\documentclass[11pt]{iopart}
\pdfoutput=1
 		            		
\usepackage{iopams}  		       
\usepackage{graphicx}									
\usepackage{caption}
\usepackage{subcaption}
\usepackage{url}
\usepackage{hyperref}
\usepackage{color,soul}
\usepackage{cite}

\begin{document}

\title[Non-equilibrium phase separation with reactions]{Non-equilibrium phase separation with reactions:\\ A canonical model and its behaviour}

\author{Yuting I. Li, 
Michael E. Cates}

\address{DAMTP, Centre for Mathematical Sciences, Wilberforce Rd, Cambridge CB3 0WA}
\ead{yuting.li@damtp.cam.ac.uk}

\begin{abstract} Materials undergoing both phase separation and chemical reactions (defined here as all processes that change particle type or number) form an important class of non-equilibrium systems. Examples range from suspensions of self-propelled bacteria with birth-death dynamics, to bio-molecular condensates, or `membraneless organelles', within cells. 
In contrast to their passive counterparts, such systems have conserved and non-conserved dynamics that do not, in general, derive from a shared free energy. This mismatch breaks time-reversal symmetry and leads to new types of dynamical competition that are absent in or near equilibrium. We construct a canonical scalar field theory to describe such systems, with conserved and non-conserved dynamics obeying Model B and Model A respectively (in the Hohenberg-Halperin classification), chosen such that the two free energies involved are incompatible. The resulting minimal model is shown to capture the various phenomenologies reported previously for more complicated models with the same physical ingredients, including microphase separation, limit cycles and droplet splitting. We find a low-dimensional subspace of parameters for which time-reversal symmetry is accidentally recovered, and show that here the dynamics of the order parameter field (but not its conserved current) is {\em exactly} the same as an equilibrium system in which microphase separation is caused by long-range attractive interactions. 
\end{abstract}



\maketitle
%
%

\section{Introduction}
The problem of phase separation in an immiscible binary liquid of molecules or polymers undergoing chemical reactions was considered in a series of papers in the mid 1990's \cite{Glotzer1, Glotzer2, GlotzerPRE,Motoyama1996, Motoyama1997, puri1994segregation, puri1998phase}. In steady state, such chemical reactions were found by simulation, and argued theoretically, to create microphase separation (such as layered phases of finite wavelength) in place of the bulk phase separations seen for immiscible molecules of constant chemistry. However it was soon clarified that can happen only under conditions where the chemical reactions are held out of equilibrium \cite{lefever1995comment}. This outcome is required by the general principle that thermodynamic equilibrium is determined solely by the free energy landscape (expressed as a function of the chemical potentials or densities of the various species) and not by kinetic details such as chemical reaction rates. This reasoning holds so long as those rates are chosen to respect the detailed balance condition with respect to the same free energy as governs the phase separation -- as must be the case for systems at, or close to, equilibrium. Only far from equilibrium can a steady-state structure emerge whose properties depend on reaction rates, as arises for the microphase separations reported in  \cite{Glotzer1, Glotzer2, GlotzerPRE,Motoyama1996, Motoyama1997}.

Recently there has been a resurgence of interest in the case of strongly non-equilibrium phase separation with chemical reactions -- which we define here as all processes that change particle type or number, including for instance the birth and death of micro-organisms such as bacteria. One major context has been cell biology, where non-membrane-bound compartments of cells, also called biomolecular condensates or membraneless organelles, can be regarded as phase separated liquid-liquid mixtures \cite{berry2018physical, Jacobs2017biophysical, hyman2014liquid}.  These structures reside in the active environment of living cells which can drive chemical reactions far from equilibrium. Models based on these principles account for multi-droplet morphologies observed in some organisms \cite{Wueseke2016bioopen, zwicker2014centrosomes} where a mechanism involving equilibrium phase separation alone would (for short-ranged interactions among species) generically lead to one single large droplet per cell. In addition, self-propelled bacteria subject to population dynamics fall into the same general class, and exhibit some similar phenomena \cite{catesPNAS2010,grafkePRL2017}. 

The general class in question is characterised by the non-equilibrium combination of a phase separating mechanism that conserves the total particle density (or composition variable in the case of liquid mixtures), and population dynamics or driven chemical reactions that allows non-conservative variations, with no detailed-balance condition linking the two sectors. Take bacteria as an example: these are known to undergo motility-induced phase separation as a result of lower swim speed in dense regions \cite{MIPS}; the total density is conserved in this process. Separately, bacteria can reproduce via cell division and die as a result of over-crowding; this is the non-conservative part of the dynamics. Similarly, in subcellular fluid mixtures, the chemical reactions among phase-separating species consumes fuel and generates waste, whose concentrations are maintained externally, driving the system far from equilibrium \cite{weberRepProgPhys2019}. 

In this paper we propose a canonical model for this class of systems via separate Landau-Ginzburg expansions of the conservative and non-conservative dynamics. The outcome is a combination of Model B and Model A, as defined by Hohenberg and Halperin \cite{HohenbergHalperin}, with chemical potentials that are mismatched. 
Note that in active systems these chemical potentials need not even stem from free energies, let alone a shared one \cite{wittkowski2014scalar, cates2018JFM}.  (This is because activity can result in gradient terms not present in either Model A or B, both of which were originally constructed for near-equilibrium systems only \cite{HohenbergHalperin}.) However, for simplicity of the analysis, and to retain a parameter space of reasonable dimensions, here we focus on the lowest-order non-equilibrium theory, which minimally disequilibrates conservative and non-conservative sectors that would each observe time-reversal symmetry in isolation. (Note that the relevant coarse-grained equations of motion in each sector can be equilibrium-like, in this sense, even if the underlying microphysics is totally irreversible \cite{MIPS}.) Time-reversal symmetry is then generically broken directly by the incompatibility of the two free energies that drive the conservative and non-conservative aspects of the dynamics.

As we shall see, however, it is possible to find a low-dimensional subspace of parameters for the coupled model in which time-reversal symmetry is accidentally restored for the dynamics of the density (albeit not of the current).  In this fine-tuned subspace, we will map the dynamics of the stochastic non-equilibrium field theory onto one with an altered combination of conserved and non-conserved dynamics that allows an effective free energy to be identified. This exact mapping supersedes one reported previously at mean-field level only (i.e., neglecting fluctuations) \cite{ Liu1989, Muratov2002, Sagui1995}; it establishes complete dynamic equivalence to an equilibrium system with certain long-range interactions. The character of these effective interactions readily explains the emergence of microphase-separated states, and captures physics that should still apply in neighbouring regions of parameter space. In more distant parameter regions, however, dynamics is seen that is clearly incompatible with any kind of equilibrium model. For example we find limit cycles, resembling those reported previously for a specific, non-minimal model of bacteria with birth and death \cite{grafkePRL2017}. In this paper we survey the phenomenology of our canonical model across various regimes, making connections with results already discovered in more specific models within the same general class. 

The paper is organised as follows. In section \ref{canonical_model}, the canonical model is introduced, and a full discussion given of the special case where an equilibrium mapping exists. Section \ref{steady_state} catalogues some stationary solutions of the dynamics, including lamellar patterns and droplet suspensions, with the key feature that the phase separation is always arrested at a fixed length-scale. Next, in section \ref{droplet_splitting}, we investigate a transient state -- where droplets stretch followed by splitting -- via linear stability analysis of the angular Fourier modes. In section \ref{limit_cycles} we address steady-state limit cycle solutions where the system oscillates between the phase separated state and the uniform state. In section \ref{conclusion}, the results are summarised and compared to more complicated, problem-specific models found in the literature. Most, but not all, of the work reported here is at mean-field level; we plan to address fluctuation effects more fully in future work. In the same future work we will also address in detail the question of steady-state entropy production whose calculation generally also requires treatment of fluctuations \cite{nardini2017entropy}. 

\section{Non-equilibrium Model AB} 
\label{canonical_model}
We shall present a canonical model for this class of systems, found as a combination of Model B for the conservative sector and Model A for the non-conservative relaxation. Model A and Model B were among those systematically catalogued in the 1970's, from conservation laws and symmetry considerations, to describe the dynamical approach to equilibrium \cite{HohenbergHalperin}. In principle, each can be separately generalized to non-equilibrium by adding terms that break time reversal symmetry in the stochastic field theory (see, e.g.,  \cite{wittkowski2014scalar, singh2019PRL, TjhungPRX2018}). We shall briefly recall their definitions before constructing the lowest-order mixed model. 

Consider a scalar field $\phi(x, t)$, which can be, say, the rescaled density of bacteria or the composition variable of binary fluid. Model A describes the dynamics of a non-conserved order parameter,  \begin{equation}
\partial_t \phi = -M_\mathrm{A} \mu + \sqrt{2 \epsilon M_\mathrm{A}} \Lambda 
\end{equation} 
where $M_\mathrm{A}$ denotes a constant relaxation rate, $\mu(\phi,\nabla\phi,\nabla^2\phi...)$ is the local chemical potential associated with the process, $\epsilon$ governs the noise strength and $\Lambda$ is a spatio-temporal white noise of unit variance: $\langle \Lambda(\boldsymbol{x}, t) \Lambda(\boldsymbol{y}, s) \rangle = \delta(\boldsymbol{x} - \boldsymbol{y}) \delta(t - s)$. If $\mu$ can be written as the derivative $\delta \mathcal{F}/\delta \phi$ of some functional $\mathcal{F}[\phi]$, the steady state solution obeys the Boltzmann distribution, with $\mathcal{F}$ playing the role of the free energy and $\epsilon$ the temperature. (A square-gradient, $\phi^4$ theory form for $\mathcal{F}$ is traditionally selected, but we defer making this choice until later.)

Model B is defined similarly but for a conserved order parameter, whose amount in any region ${\mathcal V}$, $\int_{\mathcal V} \mathrm{d} \boldsymbol{x} \, \phi(\boldsymbol{x}, t)$, only changes by virtue of a current $\boldsymbol{J}$ across the surface of that region: 
\begin{equation}
\eqalign{
\partial_t \phi =  - \bnabla \cdot \boldsymbol{J} \\
\boldsymbol{J} =  - M_\mathrm{B} \bnabla \mu + \sqrt{2 \epsilon M_\mathrm{B}} \boldsymbol{\Lambda}
}
\end{equation}
The first of these is the continuity equation enforcing the conservation law, and the second is a constitutive equation saying that the current transports mass from high to low chemical potential, with $M_\mathrm{B}$ a constant mobility. Again $\epsilon$ governs the the noise strength and $\boldsymbol{\Lambda}$ is a vector of uncorrelated unit white noises (each with the statistics of $\Lambda$ above). 

In our mixed systems, both conservative and non-conservative mechanisms are present. 
Our Model AB therefore takes the form 
\begin{equation}
\eqalign{
\partial_t \phi = - \bnabla \cdot \boldsymbol{J}- M_\mathrm{A} \mu_\mathrm{A} + \sqrt{2 \epsilon M_\mathrm{A}} \Lambda_\mathrm{A} \\
\boldsymbol{J}=  - M_\mathrm{B} \bnabla \mu_\mathrm{B} + \sqrt{2 \epsilon M_\mathrm{B}} \bLambda_\mathrm{B} \\
}
\label{eq:modelAB}
\end{equation}
As noted previously, the two chemical potentials $\mu_\mathrm{A,B}(\phi,\nabla\phi,\nabla^2\phi...)$ are not generically functional derivatives of free energies $\mathcal{F}_{\mathrm{A}, \mathrm{B}}$ in systems far from equilibrium. For simplicity we shall assume in this paper that they are, but we allow that $\mathcal{F}_{\mathrm{A}}\neq\mathcal{F}_{\mathrm{B}}$. This is enough to break time-reversal symmetry, and generically does so at lower order in $(\nabla,\phi)$ than the terms needed to break the symmetry in either sector by itself. The latter begin with contributions of order $(\nabla^2,\phi^2)$ to the chemical potentials \cite{wittkowski2014scalar} (and for Model B also include a contribution to $\boldsymbol{J}$ that is not the gradient of any local chemical potential  \cite{TjhungPRX2018, nardini2017entropy}). Note also that once $\mathcal{F}_{\mathrm{A}}\neq\mathcal{F}_{\mathrm{B}}$, unequal noise levels in the two sectors can be absorbed by rescaling either free energy, allowing us to set both noises to a common value $\epsilon$ in (\ref{eq:modelAB}).

In the next two sections, we first derive a general condition on $\mu_\mathrm{A,B}$ under which, despite the fact that $\mathcal{F}_{\mathrm{A}}\neq\mathcal{F}_{\mathrm{B}}$, an exact formal equivalence exists to a Hohenberg-Halperin-type model with a more complicated mobility operator and a single free energy $\mathcal{F}$. The resulting $\mathcal{F}$ features long range interactions of a general form that is well studied in equilibrium models and leads generically to microphase separation.  After this, we specialise to the particular forms for $\mu_\mathrm{A}$ and $\mu_\mathrm{B}$ that complete the specification of our canonical model and that we will use for the rest of the paper. 

\subsection{A special case: Accidental restoration of time-reversal symmetry for $\phi$ field} 
\label{elsp}
\begin{figure}
\includegraphics[width=\textwidth]{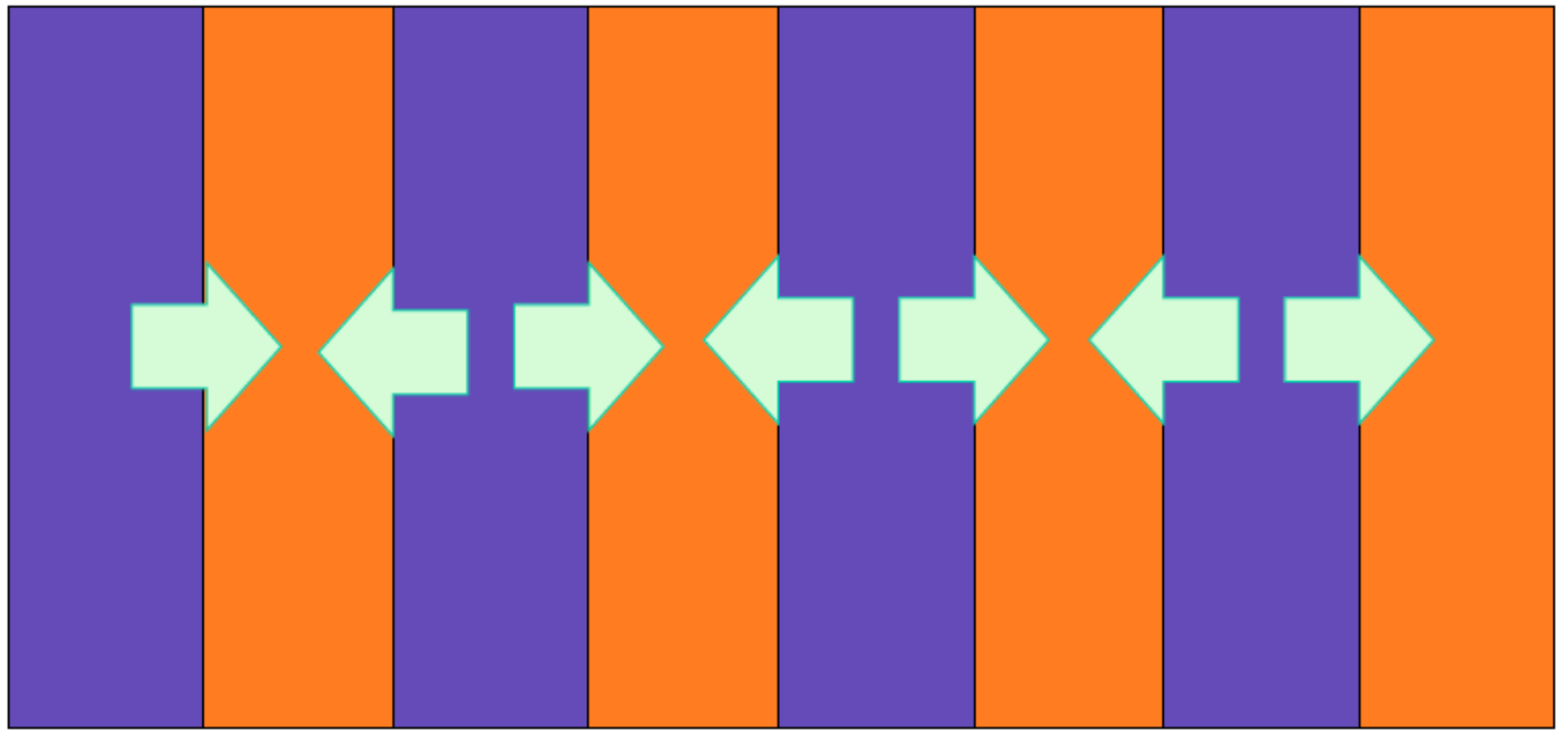}
\caption{Microphase separation in a lamellar phase. On the special parameter subspace the statistics of the $\phi$ field (orange/purple pattern) are reversible, but there are nonzero mean currents in steady state (arrows). For example in phase-separating bacteria with population dynamics, particles are born in the dilute zones, and diffuse towards the dense ones, where they die \cite{catesPNAS2010}. A movie of $\phi(\boldsymbol{x},t)$ then shows exactly reversible fluctuations whereas the currents are large and irreversible, even at deterministic order.} 
\label{fig:breeding}
\end{figure}

In general, the model defined by (\ref{eq:modelAB}) is out of equilibrium, but under certain conditions on $\mu_\mathrm{B}$ and $\mu_\mathrm{A}$, time-reversal symmetry is restored, in the sense that the steady-state probability of observing a field trajectory $\phi(\boldsymbol{x},t)$ is the same as the probability of the time-reversed trajectory. Note however that the separate contributions to the dynamics are not the same in forward and backward paths: in particular the conservative currents have a nonzero average spatial pattern in the microphase-separated steady state, and this current pattern reverses sign under time reversal. This was reported for a more specific model in \cite{catesPNAS2010}, whereby a pattern of dense and dilute bacterial domains emerges in which particles are born in the dilute regions, then move by phase separation into the dense ones, where they die off (Figure \ref{fig:breeding}).  The same physics applies here, so that time-reversal symmetry is restored only for the $\phi$ field and not for the full dynamics \cite{nardini2017entropy}. However, this is enough to allow the stationary measure for $\phi$ to be constructed as the Boltzmann distribution of an effective free energy $\mathcal{F}$. Although not solved exactly, the resulting equilibrium theory is already well studied in the context of microphase separation and its behaviour known to high accuracy \cite{TarziaPRL2006, TarziaPRE2007}.

Let $\eta$ be the sum of the Model B and Model A noise in equation (\ref{eq:modelAB}): $\eta (\boldsymbol{x}, t)  = \sqrt{2 \epsilon} \left [  - \sqrt{M_\mathrm{B}} \bnabla \cdot \bLambda_\mathrm{B} (\boldsymbol{x}, t) + \sqrt{M_\mathrm{A}} \Lambda_\mathrm{A} (\boldsymbol{x}, t) \right ] $. Using the addition property of variances for the sum of independent Gaussian distributions, 
\begin{equation}
\eqalign{
\langle \eta ( \boldsymbol{x}, t) \eta (\boldsymbol{y}, s) \rangle &= 2 \epsilon \left ( - M_\mathrm{B} \nabla_x^2 + M_\mathrm{A} \right ) \delta( \boldsymbol{x} - \boldsymbol{y} )  \delta ( t- s)\\
 & \equiv 2 \epsilon K (\boldsymbol{x}-\boldsymbol{y} ) \delta(t - s) 
}
\end{equation}
where we have formally introduced a spatial noise kernel $K(\boldsymbol{x})$. Next define a vector operator $\boldsymbol{B}_i(\boldsymbol{x}) = \sqrt{M_\mathrm{B}} \partial_{x_i}+ \sqrt{M_\mathrm{A}}$ such that\footnote{
Here $\dagger$ denotes conjugation of the operator. Let $f, g$ be functions that the operator $O$ acts on, $O^\dagger$ is defined as $\int \mathrm{d} \boldsymbol{x} f ( O g ) = \int \mathrm{d} \boldsymbol{x} (O^\dagger f) g$. For example $\bnabla^\dagger = - \bnabla$. 
} $K(\boldsymbol{x}) = \boldsymbol{B}_i (\boldsymbol{x})  \boldsymbol{B}^\dagger_i(\boldsymbol{x})$. With the above operators, we can cast equation (\ref{eq:modelAB}) in a form that closely mimics the relaxational models in \cite{HohenbergHalperin}, 
\begin{equation}
\eqalign{
\partial_t \phi = \boldsymbol{B} \cdot \left [ - \boldsymbol{B}^\dagger \mu + \sqrt{2 \epsilon} \bLambda \right ] \\
\mu = \mu_\mathrm{B} + M_\mathrm{A} K^{-1} (\mu_\mathrm{A} - \mu_\mathrm{B}) \\
\langle \Lambda_i (\boldsymbol{x}, t) \Lambda_j (\boldsymbol{y}, s) \rangle = \delta_{ij} \delta (\boldsymbol{x} - \boldsymbol{y} ) \delta ( t - s) 
} \label{eq:mimic}
\end{equation}
If there exists a $\mathcal{F}[\phi] $ that has the given $\mu$ as its functional derivative,  the steady state probabilities will follow the Boltzmann distribution with $\mathcal{F}$ as the free energy, and detailed balance will be restored. As $\mu_\mathrm{B}$ is already the functional derivative of $\mathcal{F}_\mathrm{B}$, this requires $K^{-1}(\mu_\mathrm{A} - \mu_\mathrm{B})$ to be a functional derivative. Here $K^{-1}$ is the integral operator inverse to $K(\boldsymbol{x})$: $K^{-1}\psi(\boldsymbol{x}) = \int \bar{K}(\boldsymbol{x}-\boldsymbol{x}')\psi(\boldsymbol{x}')\mathrm{d}\boldsymbol{x}'$ with $\bar K(\boldsymbol{x})$ the back Fourier transform of $1/K(\boldsymbol{q})$. Since in Fourier space $K(\boldsymbol{q}) = M_\mathrm{B}q^2+M_\mathrm{A}$, this $\bar{K}$ is a screened Coulomb (Yukawa) operator.

The simplest case is of course when $\mu_\mathrm{B} = \mu_\mathrm{A}$; then $\mathcal{F} = \mathcal{F}_\mathrm{B} = \mathcal{F}_\mathrm{A}$, meaning that the demixing and the chemical reactions between the two species are controlled by the same chemical potential. 
This is the case of full thermodynamic equilibrium, with detailed balance at the $(\phi,\boldsymbol{J})$ level, for instance a binary fluid of two immiscible passive species with a simple conversion reaction between the species \cite{weberRepProgPhys2019, lefever1995comment}. 
The conserved dynamics drives coarsening of coexisting fluid domains whose phase volumes are, however, not conserved due to the reactions. Ultimately the free energy is minimized by the complete elimination of the phase of higher bulk free energy density (or, for a symmetric system, one or other phase at random) so that no interfaces remain and the system is uniform. 

A more general and much more interesting case is when $\mu_\mathrm{B} - \mu_\mathrm{A} = Q\phi$ where $Q$ is a linear operator. (In the models of interest here, $Q$ is a differential operator, but an integral operator, symmetric under interchange of spatial arguments, is also admissible). In this case $K^{-1}(\mu_\mathrm{B} - \mu_\mathrm{A}) = K^{-1}Q\phi$ is also linear, albeit nonlocal; as a result $\mu = \mu_\mathrm{B} + M_\mathrm{A}K^{-1}Q\phi$ is the functional derivative of $\mathcal{F} = \mathcal{F}_\mathrm{B} + \frac{1}{2} M_\mathrm{A}\int \phi(\boldsymbol{x})\bar{K}(\boldsymbol{x} - \boldsymbol{x'}) (Q\phi(\boldsymbol{x'})) \mathrm{d} \boldsymbol{x} \mathrm{d} \boldsymbol{x'}  $. The second term is a nonlocal harmonic contribution whose effects on the Boltzmann distribution are readily explicable within equilibrium statistical physics.

As an example, chosen for its relevance to subsequent sections, we now take both $\mathcal{F}_\mathrm{B}, \mathcal{F}_\mathrm{A}$ to be of $\phi^4$ square-gradient form but with different linear coefficients: $\mu_\mathrm{B} = c - \alpha \phi + \beta \phi^3 - \kappa_\mathrm{B} \nabla^2 \phi$ and $\mu_\mathrm{A} = c + \alpha' \phi +  \beta \phi^3- \kappa_\mathrm{A} \nabla^2 \phi$. Note that we can add a constant to $\mu_\mathrm{B}$ without changing the equations of motion, hence the $c$ terms in $\mu_\mathrm{B}$ and $\mu_\mathrm{A}$ are chosen equal here without loss of generality. We can also set $\kappa_\mathrm{A} = 0$ since in the equations of motion this term is absorbed by a shift in $\alpha$; hereafter we choose $\kappa_\mathrm{B} = \kappa \equiv \kappa_\mathrm{B}-\kappa_\mathrm{A}$ for this reason.
We choose $\alpha >0$ to drive phase separation, but also choose $\alpha'>0$. The latter means that the non-conserved dynamics would, by itself, take the system towards a uniform target density $\phi_\mathrm{t}$ which is the unique real root of the cubic equation $c + \alpha' \phi_\mathrm{t} +  \beta \phi_\mathrm{t}^3 = 0$.
So long as this target density lies within the spinodals (resp., binodals) of the phase separation, the uniform state at $\phi_\mathrm{t}$ is locally (resp., globally) unstable. Steady Boltzmann-like states are guaranteed by the equivalence to an equilibrium system, but these must be nonuniform (Figure \ref{fig:breeding}).

We then have $\mu_\mathrm{A} - \mu_\mathrm{B} = (\alpha' + \alpha) \phi  + \kappa \nabla^2 \phi$, and, in Fourier space, with normalisation $f(\boldsymbol{q}) = \int \mathrm{d}\boldsymbol{x}  f(\boldsymbol{x}) \exp( - i \boldsymbol{q} \cdot \boldsymbol{x} ) $ and $|\boldsymbol{q}| = q$, we obtain: 
\begin{equation}
\mu(\boldsymbol{q}) = \mu_\mathrm{B}(\boldsymbol{q}) + M_\mathrm{A} \frac{(\alpha' + \alpha) - \kappa q^2  }{M_\mathrm{B}|\boldsymbol{q}|^2 + M_\mathrm{A} } \phi(\boldsymbol{q} ) 
\end{equation}
\if{The corresponding free energy is 
\begin{equation}
 \mathcal{F}[\phi] =  \mathcal{F}_\mathrm{B} [\phi] + \frac{M_\mathrm{A}}{2} \int \frac{\mathrm{d}^d k}{(2 \pi)^d } \frac{(\alpha' + \alpha) - \kappa |\boldsymbol{k}|^2  }{M_\mathrm{B}|\boldsymbol{k}|^2 + M_\mathrm{A} } |\phi(\boldsymbol{k} ) |^2 
 \end{equation} 
 }\fi
Defining $m = \sqrt{M_\mathrm{A}/M_\mathrm{B}}$ and $\alpha_\mathrm{eff} = \alpha + \kappa m^2 $, the corresponding free energy is
 \begin{equation} 
 \eqalign{
\fl \mathcal{F}[\phi]  =  \int \mathrm{d}\boldsymbol{x} \left [ c \phi  - \frac{\alpha_\mathrm{eff}}{2}  \phi^2 + \frac{\beta}{4} \phi^4  + \frac{\kappa_\mathrm{B}}{2} | \bnabla \phi |^2 \right ] +  \frac{m^2}{2} \left ( \alpha ' + \alpha_\mathrm{eff} \right )   \int \frac{\mathrm{d}\boldsymbol{q}}{(2\pi)^d} \frac{|\phi(\boldsymbol{q})|^2 }{ q^2 + m^2 } \\
\fl  = \int \mathrm{d}\boldsymbol{x}  \left [ c \phi   -  \frac{\alpha_\mathrm{eff}}{2}  \phi^2 + \frac{\beta}{4} \phi^4  + \frac{\kappa_\mathrm{B}}{2} | \bnabla \phi |^2 \right ] +  \frac{m^2}{2} \left ( \alpha ' + \alpha_\mathrm{eff} \right )   \int \phi(\boldsymbol{x}) U ( | \boldsymbol{x} - \boldsymbol{x'} | ) \phi(\boldsymbol{x'})  \mathrm{d}\boldsymbol{x}  \mathrm{d}\boldsymbol{x'}
} \label{eq:fullfree}
\end{equation}
Here $U (r)$, the D-dimensional inverse Fourier transform of $1/(q^2 + m^2)$, represents a long-range interaction of the scalar field. This is the screened Coulomb potential: $U(r) = (4 \pi r)^{-1} \exp( - m r) $ in 3D and $U(r)  = (2 \pi ) ^{-1} K_0( m r)$ in 2D, with $K_0$ the modified Bessel function of the second kind. The latter is logarithmic at short distances and decays as $(8\pi mr)^{-1/2}\exp(-mr)$ at large ones.  

The local part of the free energy favours phase separation into bulk phases whenever $\alpha_\mathrm{eff}>0$. However the non-local part is attractive for $\alpha'+\alpha_\mathrm{eff}>0$ and thereby frustrates phase separation. This is particularly easy to comprehend in the unscreened limit (recovered as $m^2 \to 0$ at constant $m^2\alpha'  \equiv \alpha''$) where the Coulomb cost of a bulk phase separation of oppositely charged species grows faster than the system volume $V$, and so cannot be overcome by any local contribution to $\mathcal{F}$. At small scales however, the local terms still dominate, creating coexisting phases which must therefore organize themselves into a microphase separated pattern on some finite length scale. 

The unscreened limit of our mapping is closely related to one studied previously for deterministic models of similar physical content (e.g., \cite{Liu1989}). In this limit, it might appear that the steady-state density $\bar\phi$ of any uniform state must vanish, so as to avoid an otherwise divergent contribution $V^2\alpha''\bar\phi^2/2$ from the long-ranged interaction in  (\ref{eq:fullfree}). However, this conclusion is invalidated by the fact that uniform states form a null space of the purely conservative dynamics (with $K(\boldsymbol{q}) = M_\mathrm{B}q^2$) that prevails in the same limit. The physically correct outcome is instead found by retaining nonzero $m$,  for which the nonlocal term in (\ref{eq:fullfree}) remains extensive, and equates to $V(\alpha'+\alpha_\mathrm{eff})/2$. Thus for homogeneous states, the free energy per unit volume, $f = \mathcal{F}/V$, becomes 
\begin{equation}
f(\bar{\phi})= c \bar{\phi} + \frac{\alpha'}{2} \bar{\phi}^2 + \frac{\beta}{4} \bar{\phi}^4 
\end{equation}
Hence for the uniform state, minimizing our $\mathcal{F}$ recovers the Model A target density, $\bar\phi = \phi_\mathrm{t}$. This is inevitable, because, as stated above, uniform states have no Model B dynamics. 

Next we consider fluctuations about such a uniform state: $\phi(\boldsymbol{x}) = \phi_\mathrm{t} + \delta \phi (\boldsymbol{x})$ for which we have 
\begin{equation}
\eqalign{
\fl F[\phi] =f(\phi_\mathrm{t}) + \frac{1}{2}   \int \frac{\mathrm{d}\boldsymbol{q}}{(2 \pi)^d}  \left [ - \alpha_\mathrm{eff} + 3 \beta \phi_\mathrm{t}^2 +  \kappa q^2 + \frac{\alpha' + \alpha_\mathrm{eff} } {m^{-2}q^2 + 1} \right ] | \delta \phi(\boldsymbol{q}) | ^2 \\
+\frac{\beta}{4} \int \mathrm{d}\boldsymbol{x} \, \left ( \delta \phi^4 + 4 \phi_\mathrm{t} \delta \phi^3 \right ) 
} 
\end{equation}
For $(\alpha - 3 \beta \phi_\mathrm{t}^2 )^2 \le 4 \kappa m^2 ( \alpha' + 3 \beta \phi_\mathrm{t}^2 )$, the coefficient of the quadratic term is non-negative for all $k$, and the homogeneous state is locally stable. Equality gives the onset threshold (or spinodal) of microphase-separated patterning at the mean field level.

\begin{figure}
\centering
\begin{subfigure}{0.45\textwidth}
\includegraphics[width=\textwidth]{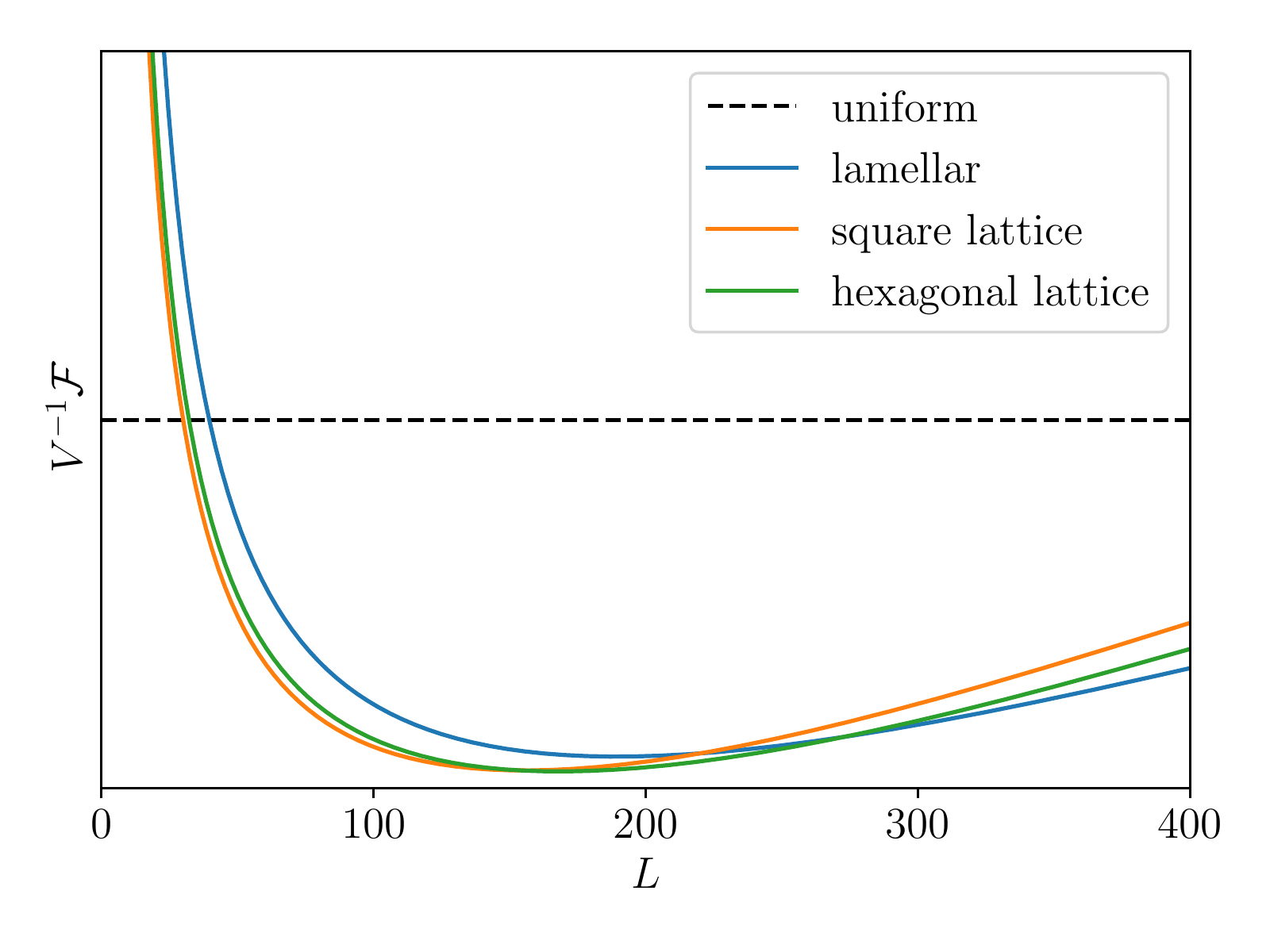}
\caption{}
\label{fig:tot_f}
\end{subfigure}
\begin{subfigure}{0.45\textwidth}
\includegraphics[width=\textwidth]{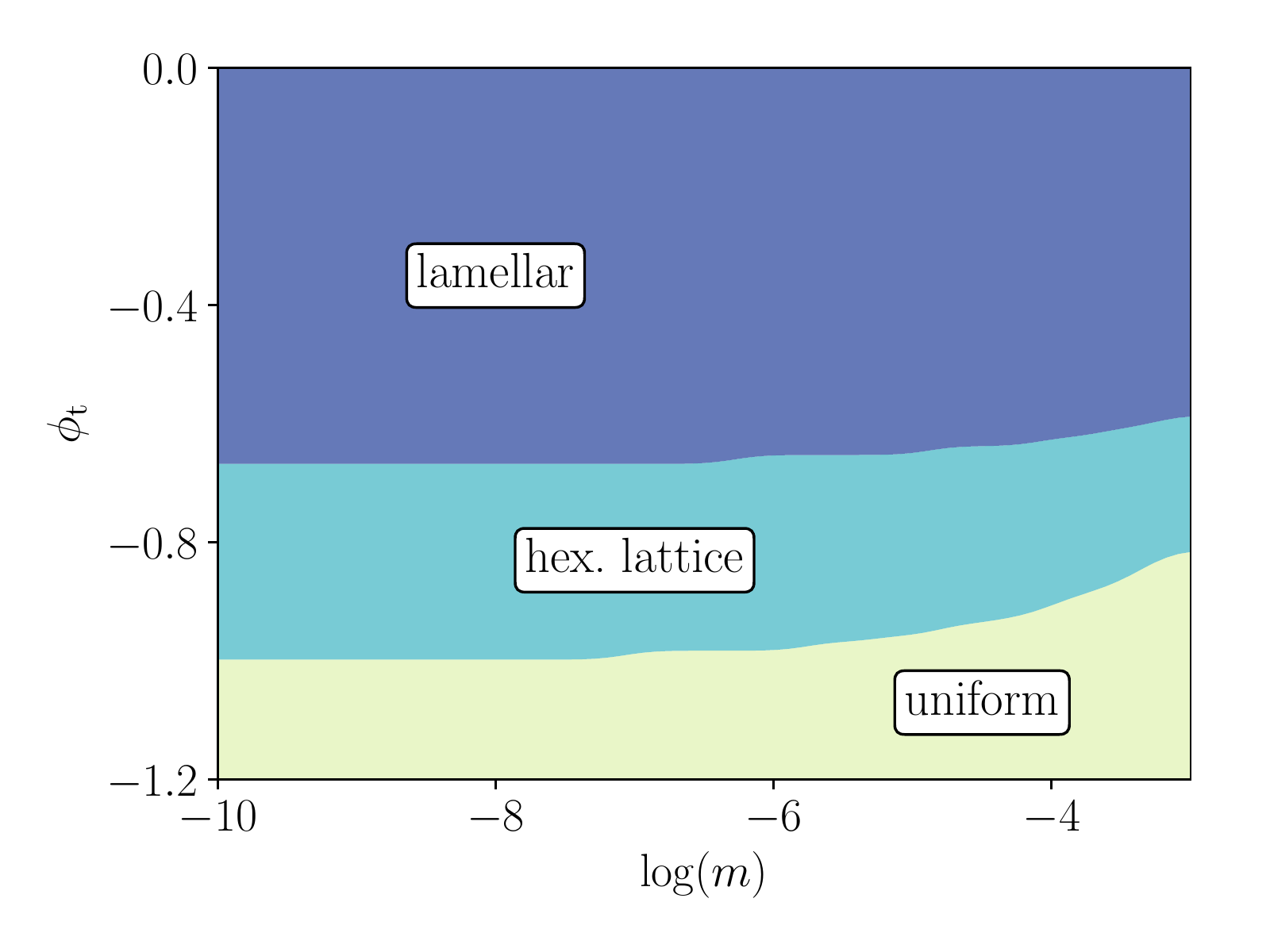}
\caption{}
\label{fig:el_phase}
\end{subfigure} 
\caption{(a) shows a plot of the total free energy density against the pattern length $L$ for the uniform state at $\phi_\mathrm{t}$ and square and hexagonal lattices of droplets, all at $(\phi_1, \phi_2)= (0.86, -0.86)$. (b) is an equilibrium phase diagram in the $(\log(m), \phi_\mathrm{t})$ plane showing the regions where the free energy is minimised by the uniform state, lamellar patterns and hexagonal droplet lattices. The square lattice phase is never the global minimizer for the range of parameters considered here.}
\end{figure}

However, it is well known that for the equilibrium microphase-separation transition, fluctuations can alter the phase boundaries, and generically drive the transition first order by suppressing the spinodal instability \cite{brazovskii}. Tarzia et al.~have presented a comprehensive study of the case with $c=0$ (which also implies $\phi_\mathrm{t}$ = 0) using the Brazovskii (Hartree) approximation \cite{TarziaPRL2006, TarziaPRE2007}. They found a first order phase transition from the uniform state to the lamellar phase as the noise strength decreases. The lamellar patterns have a domain length $L \propto ( \sqrt{\kappa^{-1}(\alpha' + \alpha) + m^2} - m)^{-1/2} m^{-1/2}$ close to the transition\footnote{The uniform state is identified as paramagnetic in these papers, whereas the ferromagnetic phase occurs for $\alpha + \alpha' < 0$, corresponding to when the Model A sector also prefers phase separation. This requires bistable chemical reactions, which goes beyond the simple class we are interested in here.}. Further away from the transition, the Hartree approximation no longer applies as the interfaces sharpen. Here it is more appropriate to substitute a pseudo-1D square wave ansatz, similar to the approach in \cite{christensen1996PRE}. \ref{ap:equilibriumAB} calculates the free energy density $V^{-1} \mathcal{F}$ for a lamellar pattern of wavelength $L$ with the two phases at $\phi_1, \phi_2$ for general values of $c$. The typical shape of the $f$ against $L$ curve is shown in Figure \ref{fig:tot_f}: for small $L$ the interfacial cost is the main contribution whereas for large $L$ the non-local screened Coulomb eventually dominates. In the limit where the reaction rates are small ($m^2 \ll \kappa^{-1} ( \alpha + \alpha') $), the pattern length obeys $L 
\propto m^{-2/3}$ for the square wave ansatz which applies further away from the transition, compared to $L \propto m^{-1/2}$ close to the transition. Interestingly, this outcome is consistent with the results in \cite{christensen1996PRE} (see also \cite{GlotzerPRE}) even though their models do not live in a parameter subspace where an equilibrium mapping holds. 

For $c \neq 0$, another possible pattern is a lattice of spherical droplets/bubbles suspended in a bath of the majority phase. We will work through the case where reactions are much slower than the conservative dynamics (so the interfaces are sharp on the droplet scale) to illustrate that the phase separation is also arrested at a finite length scale and compare the resulting minimal free energy against that of the lamellar patterns. Without loss of generality, assume c $\leq$ 0 and consider a lattice of droplets at density $\phi_1$ in a bath at density $\phi_2$. Let the lattice spacing be $L$ and the droplet radius be $R$ with the additional constraint that $R < L$ to avoid overlaps. In equilibrium, the global reaction rate must sum to zero, reducing the number of parameters to three: $\phi_1, \phi_2, L$. We refer to \ref{ap:equilibriumAB} for details of the free energy calculation and only present the results here: the typical shape of the total free energy against $L$ is shown in Figure \ref{fig:tot_f} for square and hexagonal lattices, exhibiting a clear minimum at finite $L$, similar to the lamellar case. Further minimisation over $\phi_1, \phi_2$ yields the global minimum for each lattice type, which can be compared against the minimum free energies for lamellar patterns and the uniform state to obtain the phase diagram in Figure \ref{fig:el_phase}.
There are other possible candidates for the stationary pattern, but the literature on microphase separations \cite{brazovskii} suggest that lamellar patterns and droplet suspensions are the only two found in 2D, though in 3D more exotic patterns exist, composed of structural motifs including bilayers, Y-junction cylinders, cylindrical micelles and spherical micelles \cite{hamley2005block}. 

This equilibrium subspace is special in allowing us to use equilibrium precepts to quantify the microphase-separated structure; nevertheless, we expect it to remain generic in terms of the patterns that emerge. Model parameters that are near, but not on, this subspace should therefore give qualitatively similar patterning, but they will also have steady-state entropy production \cite{nardini2017entropy}, at the level of the $\phi$ dynamics, that is absent in the subspace itself. However, if one chooses to track the Model A and Model B sectors separately, as will be explored in our future paper, the new steady-state entropy production is generically swamped by the contribution from steady-state currents, which remain large in the equilibrium limit (Figure \ref{fig:breeding}). This adds further support to our view that the qualitative behaviour in our $\phi$-reversible subspace is not exceptional, but shared by neighbouring parameter values.

\subsection{Canonical choices of $\mu_\mathrm{B}$ and $\mu_\mathrm{A}$} 

We now revert to the general case in which $\mu_\mathrm{B}$ and $\mu_\mathrm{A}$ do not differ by a constant or terms linear in $\phi$. This allows many new forms of physical behaviour to enter, so we need to reduce the parameter space to a manageable size. We therefore choose `canonical' forms for the two chemical potentials,  by writing down the simplest, leading order terms that gives rise to the physical phenomena we aim to explain. 

For the Model B sector, to lowest order in $\phi$ and $\bnabla$, we accordingly write
\begin{equation}
\mu_\mathrm{B}(\phi) = - \alpha \phi + \beta \phi^3 - \kappa \nabla^2 \phi 
\label{eq:mu1}
\end{equation}
where $\alpha, \beta, \kappa$ are positive constants. This is the functional derivative of the usual $\phi^4$ free energy, $\mathcal{F}_\mathrm{B} [\phi] = \int \mathrm{d}\boldsymbol{x} \left [ f_\mathrm{B}+ \frac{1}{2}\kappa \left | \bnabla \phi \right | ^2 \right ] $
where $f_\mathrm{B} \equiv- \frac{1}{2} \alpha \phi^2 + \frac{1}{4} \beta \phi^4$. On its own, the Model B free energy is minimised by bulk phase separation into two phases at the binodals $\pm \phi_\mathrm{B} = \pm \sqrt{\alpha/\beta}$, while the interfacial width, $\xi_0$, obeys $\xi_0 = \sqrt{2\kappa/\alpha}$ and the interfacial tension is $\sigma = (8\kappa\alpha^3/9b^2)^{1/2}$.  
Note that a linear term in $f_\mathrm{B}$ has no effect, and a cubic one can be absorbed by an additive shift of $\phi$. This shift effectively defines $\phi$ to vanish at the critical point, where $f_\mathrm{B}$ is symmetric to quartic order.

In the Model A sector, the lowest order relaxational chemical potential towards a fixed target density $\phi_\mathrm{t}$ is $\mu_\mathrm{A}(\phi) \propto ( \phi - \phi_\mathrm{t})$, as used for binary mixtures in \cite{Glotzer1, Glotzer2, GlotzerPRE}. Nevertheless, as we will see in section \ref{limit_cycles}, limit cycles cannot occur at this order due to the lack of nonlinearity. In order to allow our canonical model to capture these cycles (reported for a more elaborate model in 
\cite{grafkePRL2017}) we take the next lowest order,  
\begin{equation}
\mu_\mathrm{A}(\phi) =  u (\phi-  \phi_\mathrm{a})  (\phi - \phi_\mathrm{t}) 
\label{eq:mu2}
\end{equation}
Without loss of generality, we can assume that $\phi_\mathrm{a} < \phi_\mathrm{t}$. The non-conservative dynamics then has two fixed points -- one stable fixed point at $\phi_\mathrm{t}$ and an unstable one at $\phi_\mathrm{a}$. The latter can be used to encode the lowest physical value that the scalar field $\phi$ can take; for example, if $\phi$ is a rescaled density of bacteria, $\phi_\mathrm{a}$ can be chosen at the zero of the physical density. This reflects the situation in population dynamics where the birth-death process has a finite target density, but if the system makes an excursion to zero population it remains there forever in an absorbing state (hence the subscript `a'). To faithfully represent this behaviour requires multiplicative (and indeed non-Gaussian) noise \cite{hinrichsen} whereas in the Hohenberg-Halperin framework one always chooses additive Gaussian noise because of the vast algebraic simplification this provides 
\cite{HohenbergHalperin, tauber2014}. In combination with (\ref{eq:mu2}), this simplification is adequate here, because we are interested in phase separations whose binodals $\pm\phi_\mathrm{B}$ represent physical densities that are both positive, so that $\phi_\mathrm{a}  < - \phi_\mathrm{B}$. The absorbing state physics does not then interfere with phase separation in the slow reaction limit (which is our main interest below). Note also that in the special case when $\phi_\mathrm{a} \ll - \phi_\mathrm{B}$ (representing a narrow phase separation between two phases of nearly equal density) the reaction rate between the two binodals remains approximately linear in $\phi$ and we recover the simpler form for $\mu_\mathrm{A}$ used in \cite{Glotzer1, Glotzer2, GlotzerPRE}. 

By virtue of its being quadratic in $\phi$, equation (\ref{eq:mu2}) admits the lowest order mismatch in chemical potentials of the Model A and Model B sectors that {\em cannot} be incorporated into the equilibrium mapping of the previous section by judicious matching of higher order (i.e., cubic) terms. As stated in the introduction, it is also of lower order than the leading terms able to break time-reversal symmetry in the Model A and B sectors independently. In several ways, therefore, our choice of $\mu_\mathrm{A,B}$ offers the leading-order realization of genuinely non-equilibrium dynamics for the $\phi$ field, in systems of Model AB type.  This is why we propose the designation `canonical' for this choice.

While the arguments given above have been made on general grounds, it might help to see one concrete example. \ref{ap:lattice} starts with a microscopic lattice model for bacteria with self-propulsion, quorum sensing, and birth-death dynamics, and arrives at an approximate stochastic partial differential equation for the physical density $\rho$ by explicit coarse-graining the master equation following \cite{lefevre, gardiner, vankampen}. Then, we perturbatively expand $\rho$ around some reference density $\rho_0$ as $\rho = \rho_0 ( 1+ \phi)$, and write down the terms that are lowest order in $\phi, \bnabla$. The resulting SPDE is of the form of equation (\ref{eq:modelAB}) with the chemicals potentials in equation (\ref{eq:mu1}) and (\ref{eq:mu2}). In such a setting we can, if we wish, relate all the expansion parameters back to microscopic quantities such as the swim speed of the bacteria, and we can indeed see from \ref{ap:lattice}  that $\phi = \phi_\mathrm{a}$ corresponds to the zero of the physical density $\rho$. Note that this explicit coarse graining on a specific model yields an additional relation between $u, \phi_\mathrm{t}$ and $\phi_\mathrm{a}$: $u = (-\phi_\mathrm{a} + \phi_\mathrm{t}/2)^{-1}$. For illustrative purposes, we adopt this choice in simulations throughout the paper. 

Finally we may rescale space and other parameters to set $M_\mathrm{B}= 1$. Collecting all the terms, our canonical version of Model AB is therefore
\begin{equation}
\eqalign{
\partial_t \phi = - \bnabla \cdot \boldsymbol{J} - M_\mathrm{A} \mu_\mathrm{A} + \sqrt{2 \epsilon M_\mathrm{A}} \Lambda_\mathrm{A} \\
\boldsymbol{J} =  - \bnabla \mu_\mathrm{B} + \sqrt{2 \epsilon} \bLambda_\mathrm{B} \\
\mu_\mathrm{B} =  - \alpha \phi + \beta \phi^3 - \kappa \nabla^2 \phi  \\
\mu_\mathrm{A} = u (\phi-  \phi_\mathrm{a})  (\phi - \phi_\mathrm{t}) 
}
\label{eq:full_sto_eq}
\end{equation} 
These are the equations we will solve time and again from now on. For much of our analysis we will be able to ignore the noise terms -- guided in part by where this is possible in equilibrium systems \cite{bray_review}. Numerical evaluations are mostly performed with weak noise to avoid trapping in metastable states; where a strong effect of noise level on the stationary solution is detected, this dependence is also investigated numerically. 

All simulations are obtained via numerical integration of equation (\ref{eq:full_sto_eq}) using the pseudospectral method: computing derivatives in Fourier space and multiplications in real space \cite{boyd}. Fourier transforms are implemented using standard fast Fourier transforms (FFTs) and higher frequencies are cut off with $2/3$ dealiasing procesure \cite{orszag}. Time integration is performed using the explicit Euler–Maruyama method \cite{kloeden}. 

\section{Arrested phase separation} 
\label{steady_state}
Consider a system obeying (\ref{eq:full_sto_eq}) with target density $\phi_\mathrm{t}$ and a very small reaction coefficient $u$. This is the only uniform density sustainable by the Model A dynamics, but if it lies between the spinodals  $\pm \phi_\mathrm{S} = \pm \sqrt{\alpha/ 3\beta }$ of $\mathcal{F}_\mathrm{B}$  (meaning $d^2f_\mathrm{B}/d\phi^2<0$), the uniform state at $\phi_\mathrm{t}$ is unstable. Full bulk phase separation is then not possible in an infinite system. The phase separation is arrested at a fixed length scale, set by a balance between the flux across the interfaces and the reaction rates within the two phases. According to the equilibrium mapping of section \ref{elsp} above, the same scenario also holds outside the spinodal but within the binodals: $|\phi_\mathrm{t}|<\phi_\mathrm{B}$. Roughly speaking, matter is created in the dilute regions ($\phi<0$), pumped across the phase boundary by the Model B dynamics, and eventually destroyed in the dense regions ($\phi>0$), see Figure \ref{fig:breeding}. Qualitatively the above reasoning still holds for finite rather than very small $u$, although the binodals and spinodals do then vary with $u$ (see section \ref{droplet_sus} below). 

\begin{figure}
\includegraphics[width=\textwidth]{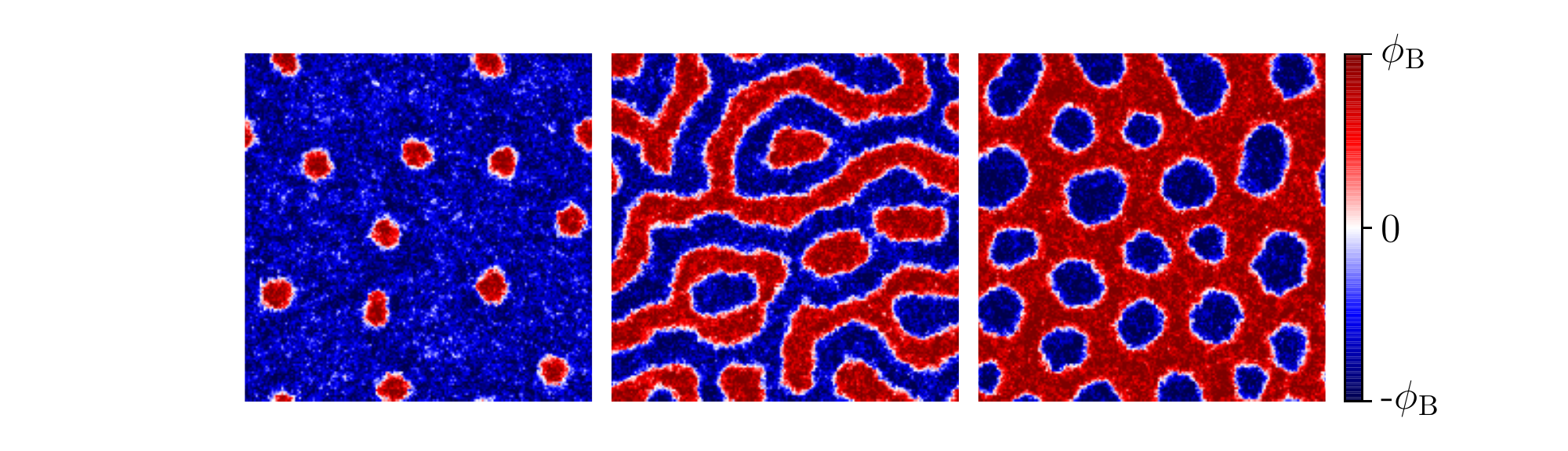}
\caption{Patterns observed for $\phi_\mathrm{t} = - 0.6, 0, 0.3$. The left panel shows a dilute suspension of droplets nucleated out of the uniform phase. The other two panels show stripe and bubble patterns that appears via spinodal decomposition. The remaining parameters are $\alpha = \beta = 0.2, \kappa = 1, u M_\mathrm{A} = 5 \times 10^{-5}, \phi_\mathrm{a} = -10, \epsilon = 0.1$. } 
\label{fig:patterns} 
\end{figure}

Some typical steady state patterns are shown in figure \ref{fig:patterns}. For $\phi_\mathrm{t}$ close to zero, the system shows lamellar patterns similar to those obtained in \cite{Glotzer1, Glotzer2, GlotzerPRE}. Otherwise, we either see either a droplet phase (disconnected regions of $\phi>0$) or its inverse, a bubble phase, depending on which binodal $\phi_\mathrm{t}$ is closer to. This echoes the findings in \cite{zwickerNatPhys2017} for a related model. Some of these steady states are reached via spinodal decomposition, others via nucleation and growth. We will discuss both in the following. 

\subsection{Spinodal decomposition} 
We consider a small perturbation about the uniform target state, $\phi(x) = \phi_\mathrm{t} + (2\pi)^{-d}\int \rmd \boldsymbol{q} \delta \phi(\boldsymbol{q}) \exp( - \rmi \boldsymbol{q} \bdot \boldsymbol{x} )$, and expand the deterministic part of equations (\ref{eq:full_sto_eq}) to linear order:  
\begin{equation} 
\eqalign{
\partial_t \delta \phi(\boldsymbol{q}) &= (\tilde{\alpha}  q^2 - \kappa q^4 - \tilde{u}) \delta \phi(\boldsymbol{q}) \\
&\equiv \sigma(q) \delta \phi (\boldsymbol{q})
} 
\end{equation} 
where $\tilde{\alpha} =  \alpha - 3 \beta \phi_\mathrm{t}^2 $,  $\tilde{u} = M_\mathrm{A} u (- \phi_\mathrm{a} +\phi_\mathrm{t})$  and we have defined the linear growth rate 
\begin{equation} 
\sigma(q)
=\left [ \left ( \frac{ \tilde{\alpha}^2}{4\kappa} - \tilde{u} \right ) -  \kappa (q^2 - q_c^2)^2 \right ] 
\label{eq:growth_rate}
\end{equation}
where  $q_c = \sqrt{\tilde{\alpha} (2\kappa)^{-1}}$. Linear stability analysis tells  us that the homogeneous state is unstable when two conditions are met: (i) $\tilde{\alpha} > 0$ so that $\phi_\mathrm{t}$ lies between the spinodals of $\mathcal{F}_\mathrm{B}$; and (ii) $\tilde{\alpha}^2/4\kappa  > \tilde{u}$, so that the conservative phase separation is fast enough to not be pulled back by the Model A relaxation. 

\subsubsection{Near the threshold: Amplitude equation.} 
\label{amplitude_eq}

\begin{figure}
\begin{subfigure}{0.47\textwidth}
\centering
\includegraphics[width=\textwidth]{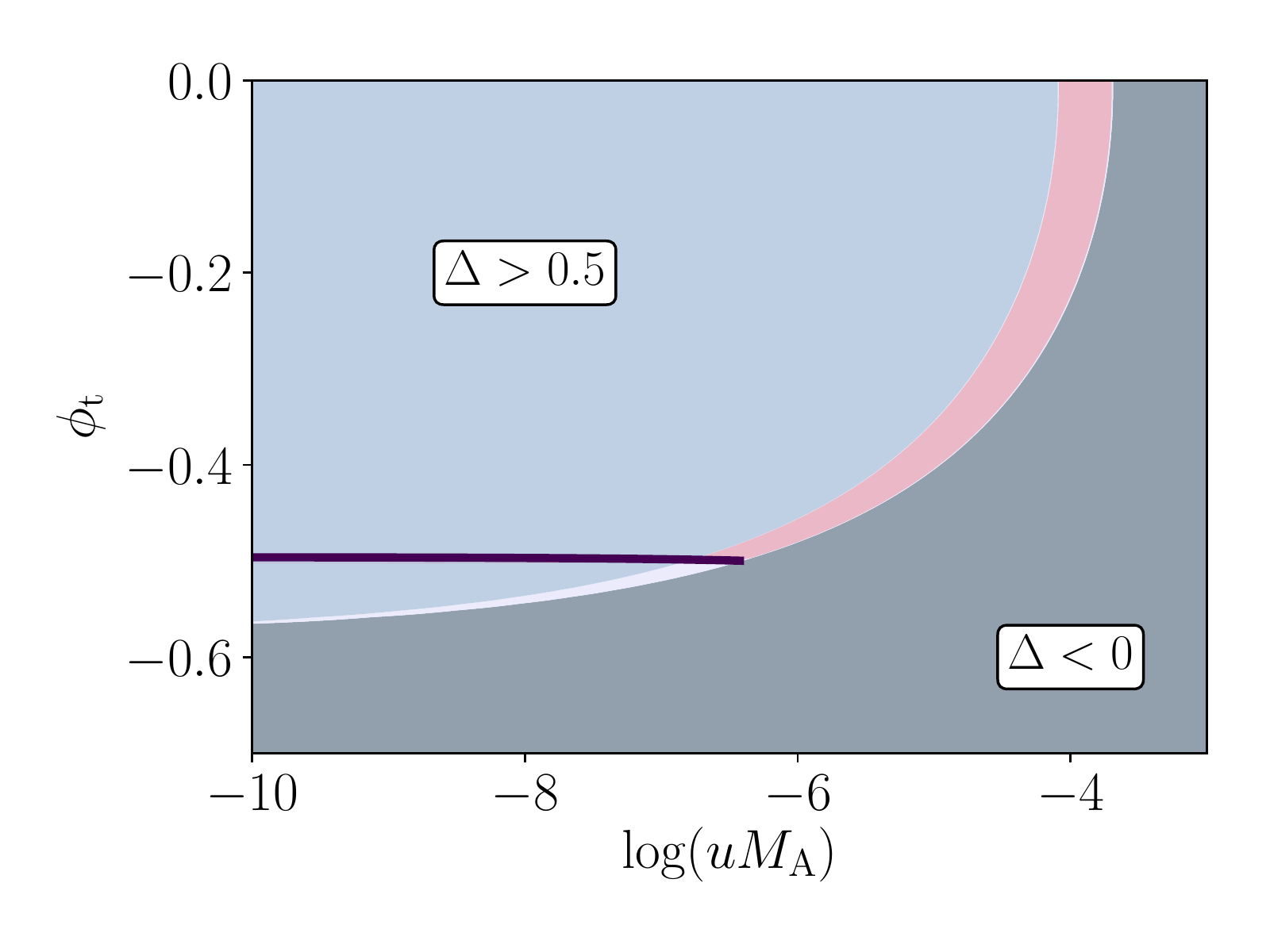}
\caption{}
\end{subfigure} 
\begin{subfigure}{0.47 \textwidth}
\centering
\includegraphics[width=\textwidth]{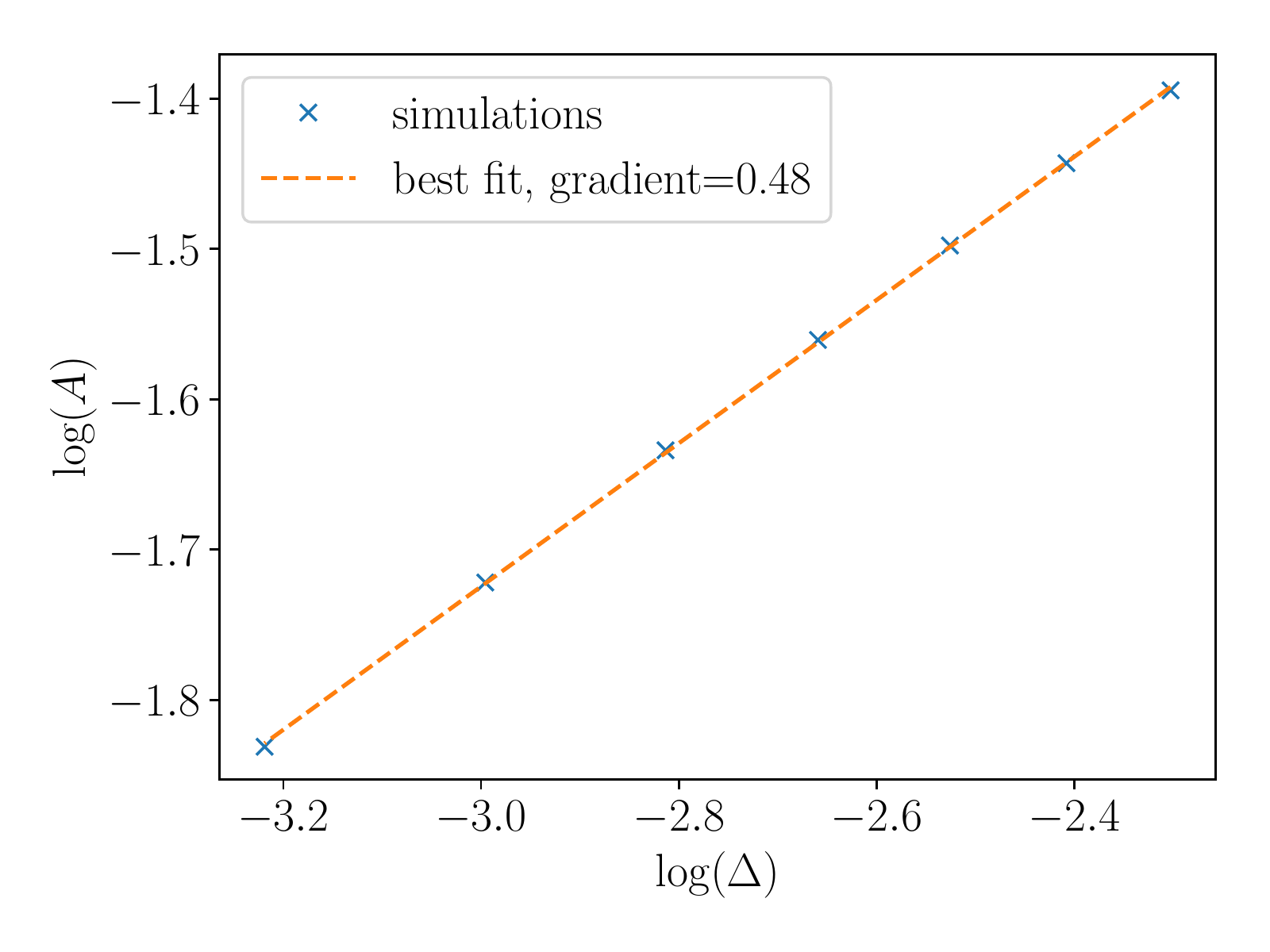}
\caption{} 
\end{subfigure}
\caption{(a) The pink region, enclosed by the $\Delta = 0$ and $\Delta = 0.5$ contours and the dark purple line, above which the expansion is self-consistent (\ref{ap: amplitude_eq}), indicates where the amplitude equation solution applies. (b)  
Log-log plot of the amplitude against $\Delta$. The simulations fall closely onto a straight line and the best fit gradient is close to $1/2$, as predicted by the amplitude equation. } 
\label{fig:amp_eq}
\end{figure}

\begin{figure}
\centering
\includegraphics[width=\textwidth]{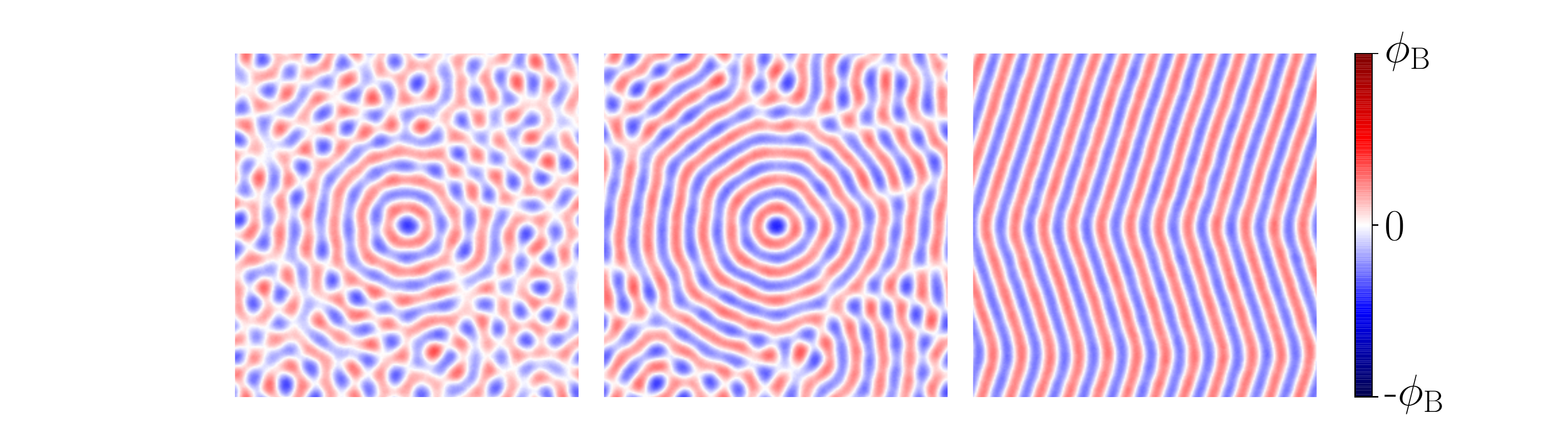}
\caption{Snapshots of the time evolution for $\Delta = 0.1, \epsilon = 1 \times 10^{-4}$ starting with a dense droplet in the middle, showing the gradual alignment of the domains after the initial spinodal decomposition. The bends in the final pattern is due to the extremely low rate of the nucleation of more domains to have the desired spacing between them. This undulation of layers whose preferred thickness times their number is less than the width of the sample is called the Helfrich instability (see \cite{DeGennes}). For even lower noise magnitudes, the pattern would be effectively arrested at the concentric rings stage (second panel) as the breaking up of rings is also noise-driven. The rest of the parameters are $\alpha = \beta = 0.2$, $\phi_\mathrm{a} = - 10$, $\phi_\mathrm{t}=0$ and $u M_\mathrm{A}$ is fixed by $\Delta$.}
\label{fig:amp_eq_patterns}
\end{figure}

We define $\Delta =\tilde{\alpha}^2(4\kappa \tilde{u})^{-1} - 1$ as a measure of the distance to the onset of the patterning instability. Also setting $\tau = 1/ \tilde{u}, \; \xi = \left (\kappa/ \tilde{u} \right )^{1/4}$, we can rewrite the growth rate for each Fourier mode as 
\begin{equation}
\sigma(q) = \frac{1}{\tau} \left [ \Delta - \xi ^4 (q^2 - q_c^2)^2 \right ] 
\label{eq:rescaled_linear}
\end{equation}
The amplitude equation is a self-consistent expansion in the small parameter $\Delta$\cite{crossRevModPhys, cross2009}. Close to the threshold ($\Delta$ small), the growth of the narrow band of linearly unstable modes is saturated by the nonlinear terms, leading to a sinusoidal pattern at $q_\mathrm{c}$ with an amplitude $\propto \sqrt{\Delta}$, within a regime of validity in parameter space as shown in Figure \ref{fig:amp_eq}. We refer to  \ref{ap: amplitude_eq} for the detailed calculations. In 1D, this simply gives $\phi(x) \propto \sqrt{\Delta} \sin (q_c x)$. In two dimensions, the pattern selection depends on the initial conditions and the amount of the noise in the simulation, as shown in Figure \ref{fig:amp_eq_patterns}. 

\subsubsection{Far away from the threshold.} 
\begin{figure}
\centering
\includegraphics[width=0.6\textwidth]{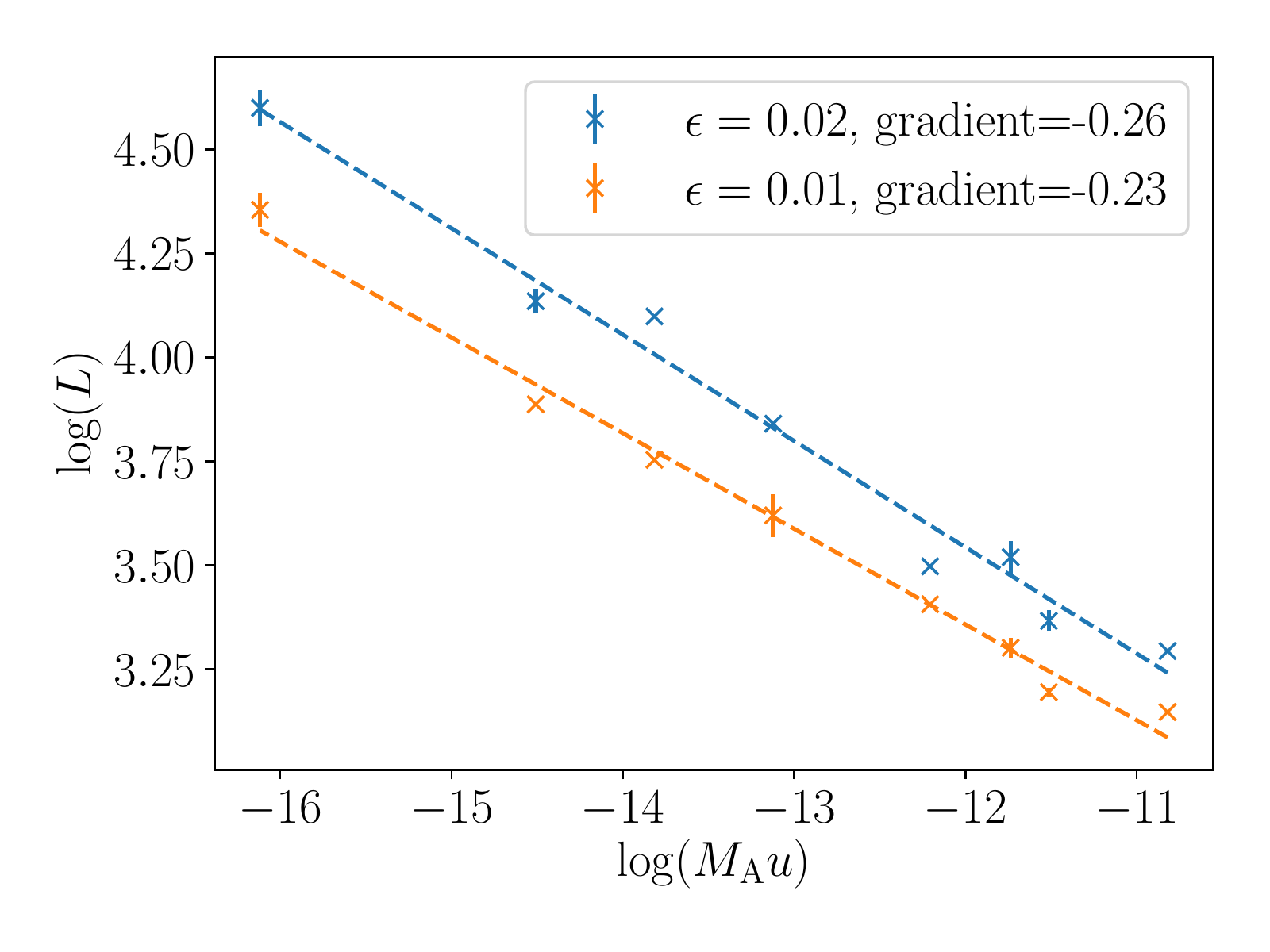}
\caption{Log-log plot of the pattern length $L$ against the birth-death rate $M_\mathrm{A} u$ with random initial conditions for $\epsilon = 0.01, 0.02$. The gradients of the best fit lines for both values of $\epsilon$ are close to $ - 1/4$, although the pattern selected depends on the noise magnitude, similar to the case of 2D droplet suspensions in Section \ref{multi_droplet}.  }  
\label{fig:pattern_length}
\end{figure}

As $M_\mathrm{A} u$ decreases, the parameter $\Delta$ is no longer small and the wavelength of the pattern deviates significantly from the fastest growing mode $2 \pi / q_\mathrm{c}$.  Numerical simulations and theoretical work have been done in 2D by Glotzer et al  \cite{Glotzer1, Glotzer2, GlotzerPRE}. They found a scaling of $L \sim (M_\mathrm{A}u)^{-1/4}$ for relative large $M_\mathrm{A}u$ by taking the dimension of the field $\phi$ to infinity (equivalent to approximating the shape by small sinusoidal perturbations around a homogeneous state), and for $M_\mathrm{A}u \rightarrow 0$, it was argued that the reactions arrest the coarsening process, leading to $L \sim (M_\mathrm{A}u) ^{-1/3}$. Curiously, Christensen et al derived the same scaling relations in 2D for the two regimes by substituting in sinusoidal and square wave ansatzes respectively, and minimising an inexactly constructed free energy with respect to the amplitude and the wavelength \cite{christensen1996PRE}. While Glotzer's argument for $M_\mathrm{A} u \rightarrow 0$ depends on the scaling law for the coarsening process and therefore the dimension of the system, Christensen's does not since the pattern is always quasi-1D. Figure \ref{fig:pattern_length} shows that our simulation results in 1D with finite noise for the domain length $L$ as a function of $M_\mathrm{A}u $ (without noise, the steady state  depends strongly on initial conditions). For each noise strength, the pattern length obeys a power law close to $L \sim (M_\mathrm{A}u) ^{-1/4}$ for all regimes of $M_\mathrm{A} u$ probed. However, for the smaller values of $M_\mathrm{A}u$, we found the patterns to be square wave shaped instead of the sinusoidal shape that the $L \sim (M_\mathrm{A} u )^{-1/4}$ scaling law of Glotzer et al and Christensen et al is supposed to fit. Thus the origin of the observed power law remains so far unexplained.  

\subsection{Nucleation and arrested growth} \label{droplet_sus}

Globally, linear stability analysis does not capture the full picture. There is a region in parameter space where although the uniform state at $\phi_\mathrm{t}$ is locally stable, there exists a competing steady state where both phases are present at a ratio that balances the total amount of reactions in the system. In the absence of reactions, $u \rightarrow 0$, this is the usual nucleation and growth regime of Model B: $\phi_\mathrm{S}<|\phi_\mathrm{t}|<\phi_\mathrm{B}$. Here droplets nucleate out of the uniform state and, once they are larger than a certain size, they grow by diffusion (and may also coalesce if the volume of the minority phase is not small) until full bulk phase separation has been reached. The diffusive growth process of droplets is commonly known as Ostwald ripening  \cite{ostwald, lifshitz1961kinetics, bray_review, cates2018JFM}; the same scaling laws also control diffusive growth of bicontinuous domains.  In our canonical Model AB, we show below that the equivalent of the Ostwald calculation reveals an additional critical radius where the growth stops, arresting the coarsening process at a finite length scale, similar to the findings in Zwicker et al.\ \cite{ZwickerPRE}. Although the kinetics differs, the steady states can again be understood in terms of microphase separation, at least close to the parameter subspace where the exact mapping of section (\ref{elsp}) holds. (Recall that this mapping shows microphase separation for $\phi_\mathrm{t}$ in the binodal region.)
The deterministic limit of our equations of motion reads (\ref{eq:full_sto_eq}), 
\begin{equation}
\eqalign{
\partial_t \phi = - \bnabla \cdot  \boldsymbol{J} + g(\phi)  \\
\boldsymbol{J}_1 = - \bnabla \left [ -\alpha \phi + \beta \phi^3 - \kappa \nabla^2 \phi \right ]  \\
g(\phi) =  - M_\mathrm{A} u (\phi - \phi_\mathrm{a}) (\phi - \phi_\mathrm{t})
}
\label{eq:det_with_current}
\end{equation}
where we have written $-M_\mathrm{A}\mu_\mathrm{A}(\phi) = g(\phi)$ to save writing later. In the analysis that follows, for tractability we will study dilute phases of well-defined droplets with sharp interfaces, although as we will see later the qualitative observations apply more generally.  
For definiteness we work in two dimensions but the generalization to 3D is straightforward and should not change qualitative outcomes for droplet suspensions. 

\subsubsection{Single droplet:}
We first consider a single high-density ($\phi> 0$) droplet of radius $R\gg\xi_0$ in an infinite bath of the dilute ($\phi<0$) phase. (Recall $\xi_0$ is the interfacial width in Model B.) Let $\phi_\pm(\boldsymbol{r})$ be the order parameter profile inside and outside the droplet respectively, and define variables $\psi_\pm(\boldsymbol{r})$ as the deviation from the binodal densities: 
\begin{equation}
\eqalign{ 
\phi_-(\boldsymbol{r} )  = - \phi_\mathrm{B} + \psi_- (\boldsymbol{r}) \qquad  &   | \boldsymbol{r} | \geq R\\
\phi_+(\boldsymbol{r} )  = \phi_\mathrm{B} + \psi_+ (\boldsymbol{r} ) 
& 0 \leq | \boldsymbol{r} | < R
} 
\end{equation}
Outside the droplet, as $r \rightarrow \infty$ the gradient terms are negligible and we must have $g(\phi_-) = 0$, implying that $\phi_- \rightarrow \phi_\mathrm{t}$ at infinity. In contrast to the standard calculation for a purely conserved order parameter \cite{cates2018JFM}, the steady-state supersaturation, $\phi(\infty)+\phi_\mathrm{B}$, is therefore not dependent on droplet size, but fixed by the non-conserved dynamics.
Assuming monotonicity of the deviation $\psi_-$ on physical grounds, we have an upper bound: $\psi_- \leq  \phi_\mathrm{t} + \phi_\mathrm{B}$. Hence for $r \geq R$, expanding equation (\ref{eq:det_with_current}) to linear order is a good approximation as long as $\phi_\mathrm{t}$ is close to $-\phi_\mathrm{B}$ and $\phi_\mathrm{a} \ll - \phi_\mathrm{B}$. Inside the droplet, to zeroth order we have $D \nabla^2 \psi_+ = g(\phi_\mathrm{B})$ where $D = 2 \alpha$, leading to a maximum deviation of approximately $g(\phi_\mathrm{B}) R^2/D$. The corresponding fractional deviation is $\psi_+/\phi_\mathrm{B} =D^{-1} R^2 / (g(\phi_\mathrm{B})^{-1} \phi_\mathrm{B}) \equiv \tau_\mathrm{diff}/ \tau_\mathrm{reac}$ where $\tau_\mathrm{diff}$ represents the time to diffusive across the droplet while $\tau_\mathrm{reac}$ is the time to deplete the dense phase by reaction. Roughly speaking,  the dynamics inside the droplet is approximately linear if reaction is much slower than diffusion. 

Provided that all the stated conditions hold, the linearised equations can be written down in the following concise form, 
\begin{equation}
\eqalign{ 
\partial_t \psi_- (\boldsymbol{r} ) = D \nabla^2 \psi_- + g^0_- + g^1_- \psi_- 
\qquad  &  | \boldsymbol{r} | \geq R  \\
\partial_t \psi_+ (\boldsymbol{r} ) = D \nabla^2 \psi_+ + g^0_+ + g^1_+ \psi_+
& 0 \leq | \boldsymbol{r} | < R
} 
\label{eq:linear}
\end{equation}
where as before $D = 2 \alpha$ and we have defined $g^0_\pm = g(\pm \phi_\mathrm{B}) , g^1_\pm = g'(\pm \phi_\mathrm{B} )$ for convenience. The linearised equations need to be solved with the appropriate boundary conditions. At $|\boldsymbol{r}| \rightarrow 0$ and $|\boldsymbol{r}| \rightarrow \infty$, we only require that $\psi_\pm$ is finite; at the boundary of the droplet, the interfacial tension $\sigma$ creates a curvature-induced offset $\delta = (d-1)\sigma/(2\phi_\mathrm{B}f''_\mathrm{B}(\phi_\mathrm{B}) R)$ to the binodal density \cite{cates2018JFM}. In two dimensions we have 
\begin{equation}
\delta = \frac{\gamma}{R},  \quad \gamma = \sqrt{\frac{\kappa}{18 \beta}}
\end{equation}

The Ostwald calculation now proceeds along standard lines \cite{cates2018JFM, ZwickerPRE}: we fix the droplet radius $R$, solve for the stationary state of the two linearised equations with $\psi_\pm(R) = \gamma R^{-1}$, then compute the current $J_\pm$ at the droplet interface using the quasi-static solutions, and finally determine the growth rate of the radius $R$ from the mismatch of currents across the interface. 

The linearised equations are the modified Helmholtz equations and a spherically symmetric solution in 2D can be written in terms of the modified Bessel functions $I_0, K_0$. Since $K_0(y) \rightarrow \infty$ as $y \rightarrow 0$ whereas $I_0(y) \rightarrow \infty$ as $y \rightarrow \infty$, finiteness of $\phi$ imposes that the solution must be of the following form, 
\begin{equation}
\eqalign{
\psi_- (\boldsymbol{r})= c_- + a_- K_0(k_- r) 
\qquad  &  | \boldsymbol{r} | \geq R \\ 
\psi_+(\boldsymbol{r}) = c_+ + a_+ I_0(k_+ r)  
& 0 \leq | \boldsymbol{r} | < R
}
\end{equation}
where $k_\pm = \sqrt{- g^1_\pm/D}$ and $c_\pm = - g_\pm^0/g_\pm^1 \approx  \phi_\mathrm{t} \mp \phi_\mathrm{B}$, and $a_\pm$  are constant coefficients whose values can be obtained by matching conditions at the droplet surface where $\psi_\pm = \gamma R^{-1}$:
\begin{equation}
\eqalign{
a_+ = (\gamma R^{-1} - c_+)/ I_0(k_+ R) \\
a_- = (\gamma R^{-1}  - c_-)/K_0(k_- R) 
}
\end{equation} 
The diffusive currents on both sides of the interface are in the radial direction due to spherical symmetry. Using the properties of the Bessel functions, 
\begin{equation}
\eqalign{
\boldsymbol{J}_+(R) = - D \bnabla \phi_+ |_{r = R} =
- D  k_+ a_+  I_1(k_+ R) \, \hat{\boldsymbol{r}} \\
\boldsymbol{J}_-(R) = - D \bnabla \phi_- |_{r = R} =
D  k_- a_-   K_1(k_- R) \, \hat{\boldsymbol{r}}
}
\label{eq:current}
\end{equation}
Whenever there is a mismatch of these currents, the interface moves accordingly. Since the interfacial width is small ($R\gg\xi_0$), reactions within the interface are negligible, and the growth rate of the droplet radius is fixed by the conservative dynamics as 
\begin{equation}
\eqalign{ 
2 \phi_\mathrm{B} \dot{R} &= J_+ - J_- \\
&= D \left [  - a_+ k_+ I_1(k_+R)  - a_- k_- K_1(k_- R)  \right ] \\
& = D \left [ \left (c_+ -   \frac{ \gamma}{R} \right ) k_+ \frac{I_1(k_+R)}{I_0(k_+R)} 
+ \left (  c_- -  \frac{\gamma }{R } \right ) k_- \frac{K_1(k_- R)}{K_0(k_- R)} \right] 
} 
\end{equation}

We will now disentangle the terms in the above equation.  Recall that $c_\pm = - g_\pm^0/g_\pm^1 \approx  \phi_\mathrm{t} \mp \phi_\mathrm{B}$, so $c_+ < 0$ and $c_- > 0$. Combined with the fact that modified Bessel functions are  positive, we can see that the first term is always negative whereas the second term can be positive.
Next, we need to find the relevant parameters. As we can always choose to rescale $\phi, x, t$ such that $\alpha = \beta = \kappa=1$, the only three remaining parameters are $M_\mathrm{A} u, \phi_\mathrm{t}, \phi_\mathrm{a}$. We are in the $\phi_\mathrm{a} \ll - \phi_\mathrm{B}$ regime where the Model A terms can be well approximated by linearity: $\mu_\mathrm{A} \approx - M_\mathrm{A} u \phi_\mathrm{a} ( \phi - \phi_\mathrm{t})$. Hence only the combinations $(-M_\mathrm{A} u \phi_\mathrm{a})$ and $\phi_\mathrm{t}$ are meaningful. 

\begin{figure}
\centering
\begin{subfigure}{0.46\textwidth}
\includegraphics[width=\textwidth]{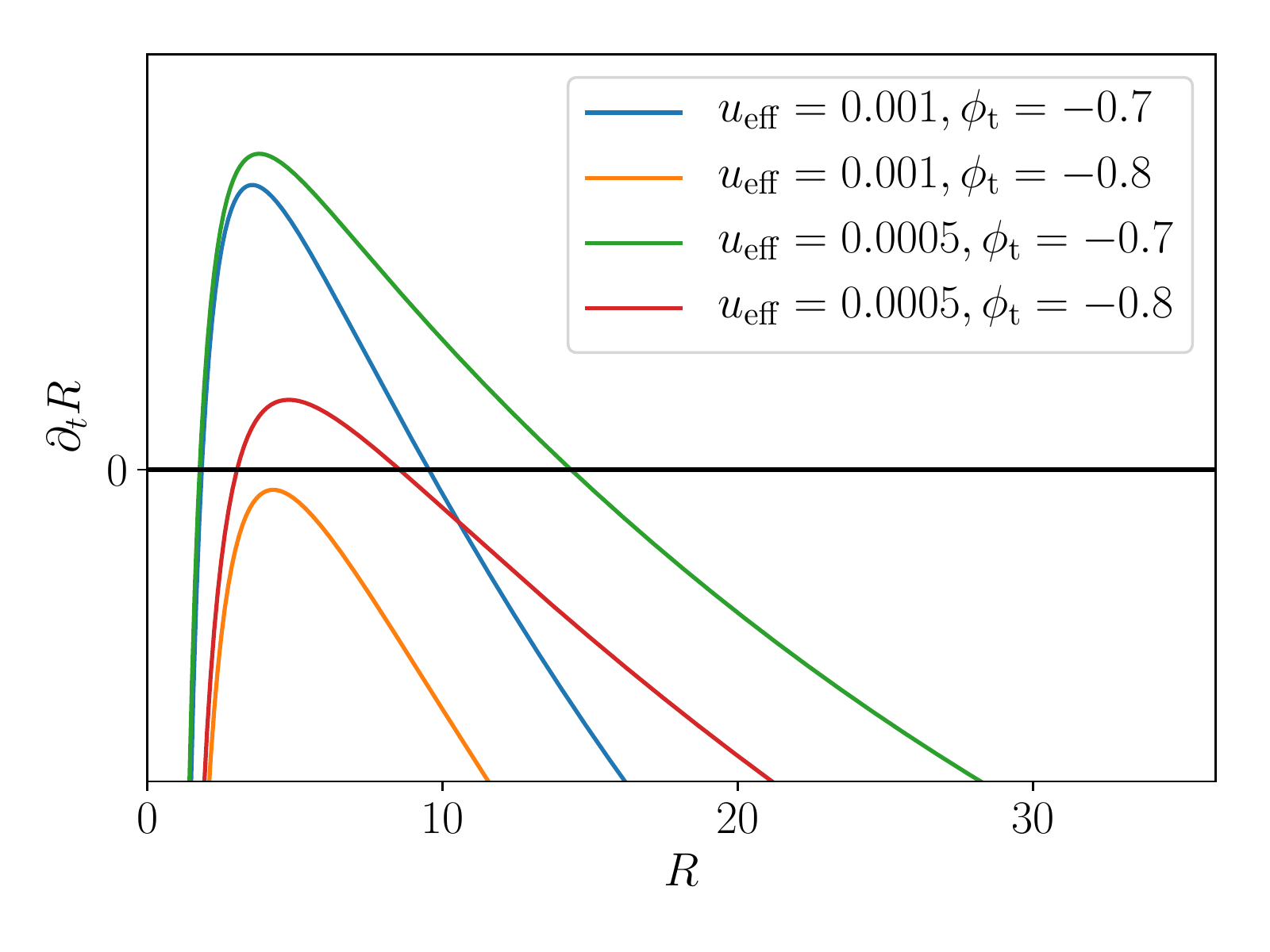}
\caption{}
\label{fig:rdot}
\end{subfigure}
\begin{subfigure}{0.46\textwidth}
\includegraphics[width=\textwidth]{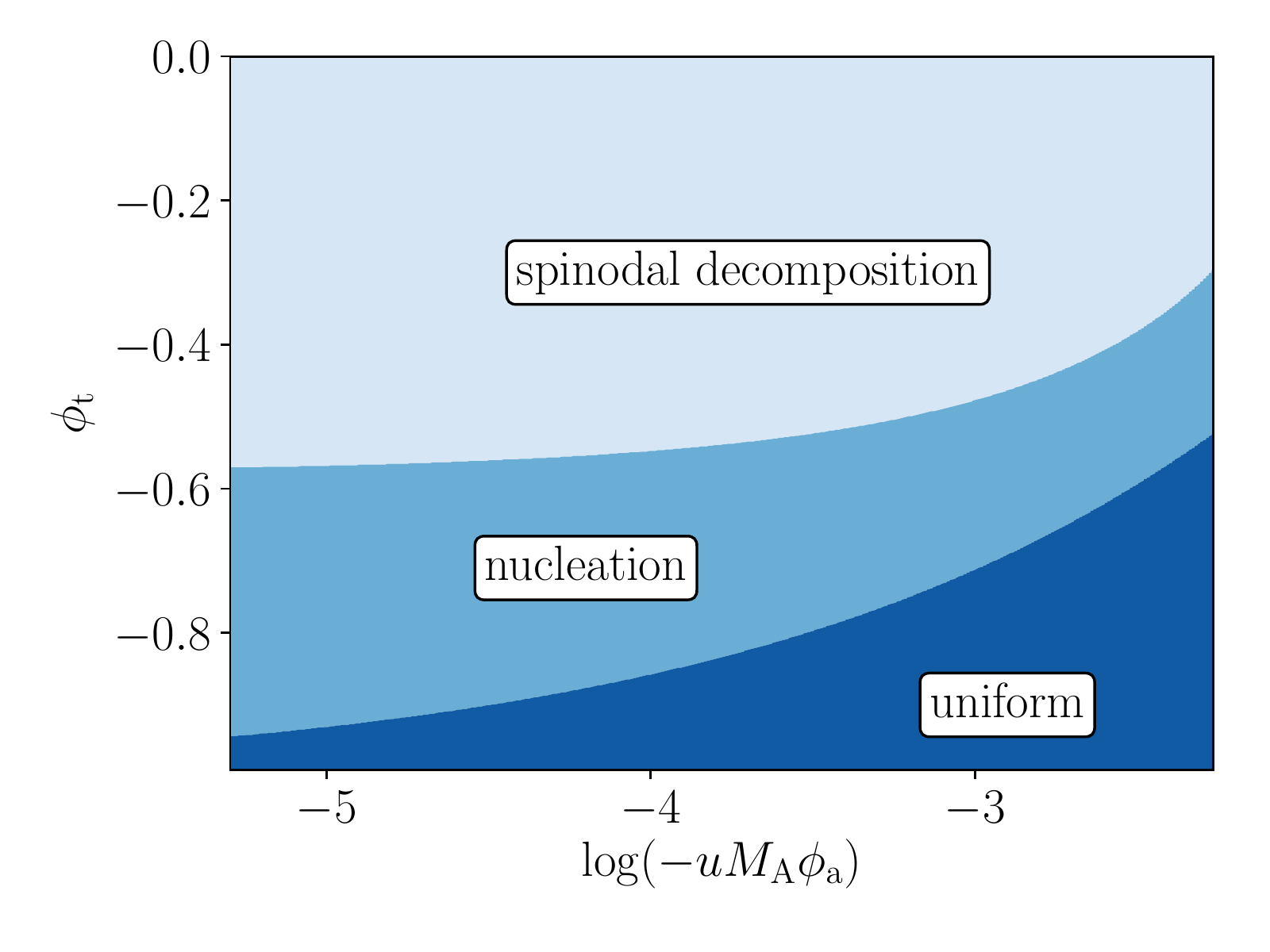}
\caption{}
\label{fig:sb}
\end{subfigure}
\caption{(a) Droplet growth rate $\dot R$ against $R$ for four sets of parameters. The orange curve is entirely below the y-axis, indicating that droplets at all radii shrink; the other three curves have two fixed points -- the one on the left is the unstable Ostwald radius $R_\mathrm{O}$ and the one on the right is the stable droplet radius $R_\mathrm{s}$. (b) Phase diagram in the $(-u M_\mathrm{A} \phi_\mathrm{a}, \phi_\mathrm{t})$ plane, showing the spinodal decomposition, nucleation and uniform phase. On the far left of the diagram, we have the Model B spinodal and binodal boundaries; these move closer together with increasing reaction rate.}
\end{figure}

Roughly speaking, $\phi_\mathrm{t}$ shifts the growth rate curve $\dot R(R)$ vertically whereas $(-M_\mathrm{A} u \phi_\mathrm{a})$ changes the shape. 
Growth rate curves for four different sets of parameters are plotted in Figure \ref{fig:rdot}. In the absence of reactions $\dot R$ is negative at small $R$, and positive and decaying towards zero at large $R$, with a single unstable fixed point at a critical radius $R_\mathrm{O}$ that depends on supersaturation. This scenario gives the familar Ostwald process, whereby large droplets grow at the expense of small ones giving a scaling $R_\mathrm{O}\sim (\sigma t/\phi_\mathrm{B}^2)^{1/3}$ \cite{cates2018JFM}. A nonzero reaction rate, however small, enforces fixed supersaturation at infinity such that the largest droplets also shrink, which changes this scenario completely. There are now two distinct cases: (i) the entire curve has negative $\dot R$, implying that droplets at all radii shrink and the uniform state is stable; (ii) there is an unstable fixed point at $R_\mathrm{O}$ as well as a new stable fixed point at a larger radius $R_\mathrm{s}$. In the second case, $R_\mathrm{O}$ is the Ostwald ripening radius; smaller droplets shrink and larger droplets grow, but with the key difference that the growth of these larger droplets stops at the second critical radius $R_\mathrm{s}$, similar to findings in \cite{weberRepProgPhys2019}.

\subsection{Multiple droplets} 
\label{multi_droplet}
Typically, within the nucleation and growth regime, multiple droplets are nucleated prior to attaining a steady state. For low phase volumes of the dispersed phase, the existence of neighbouring droplets does not alter the flux pattern close to the interface of each one \cite{ostwald, lifshitz1961kinetics}. However, we do need to account for effect of reactions in distant droplets on the supersaturation at infinity \cite{ZwickerPRE}.

\begin{figure}
\centering
\begin{subfigure}{0.52\textwidth}
\centering
\includegraphics[width=\textwidth]{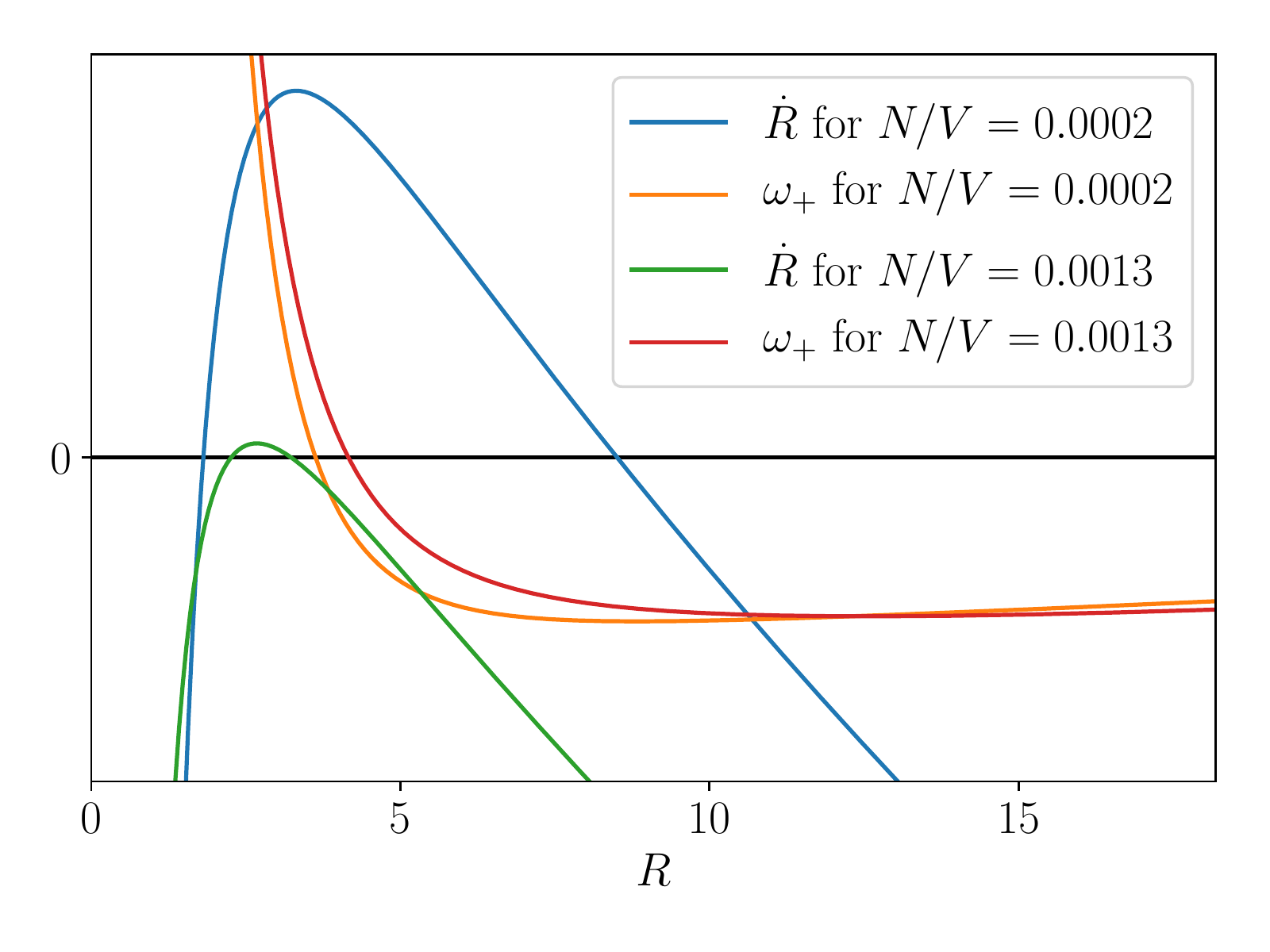}
\caption{}
\label{fig:mult_droplet}
\end{subfigure}
\begin{subfigure}{0.49\textwidth}
\centering
\includegraphics[width=\textwidth]{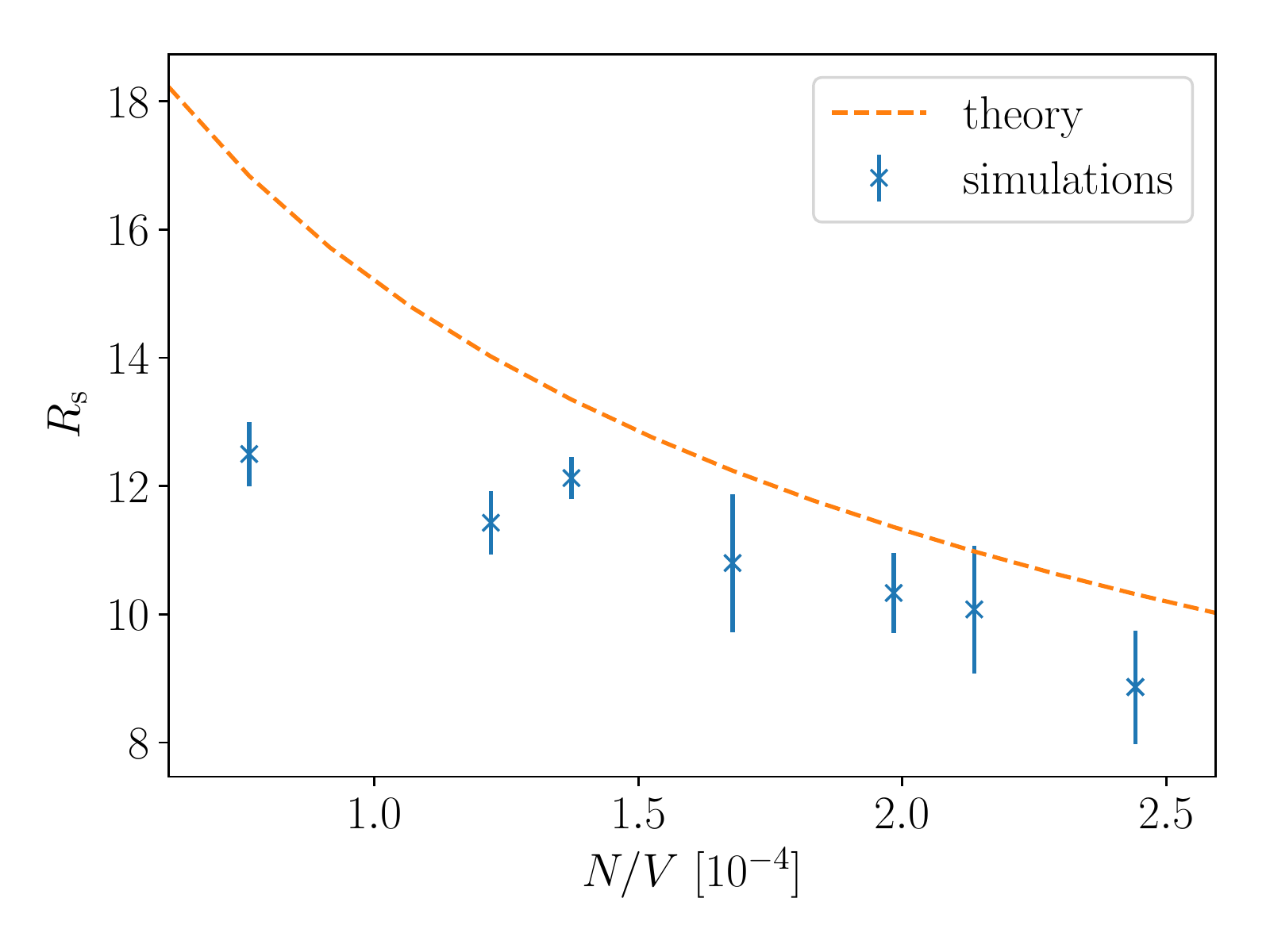}
\caption{}
\label{fig:r_against_n}
\end{subfigure}
\begin{subfigure}{0.49\textwidth}
\centering
\includegraphics[width=\textwidth]{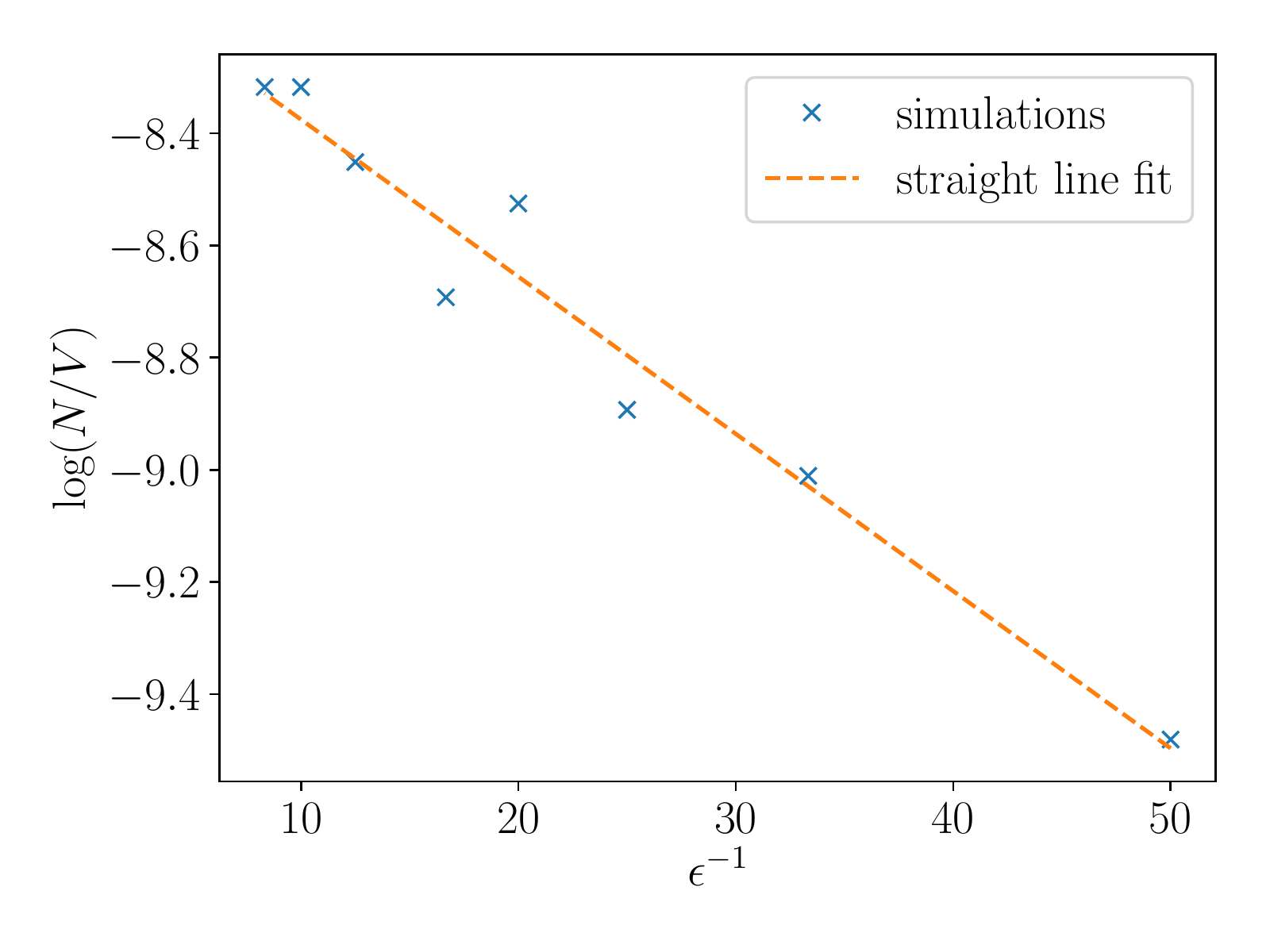}
\caption{}
\label{fig:epsilon}
\end{subfigure}
\caption{(a) Droplet growth rate $\dot R$ and exchange-mode eigenvalue $\omega_+$ plotted against $R$ for two different number densities of droplets in the suspension. The stability of the droplet suspension is indicated by the sign of $\omega_+$ at the stable radius $R_\mathrm{s}$ -- the second root of the $\partial_t R$ curve. The droplet suspension with $N/V=0.0013$ is unstable while $N/V=0.0002$ is stable. (b, c) The stable radius $R_\mathrm{s}$ plotted against the $N/V$ and $N/V$ against $\epsilon$ for several simulations with the same parameters (except $\epsilon$). Panel (b) shows a reasonable agreement between the theory and the simulations given $N/V$. Panel (c) indicates that the relation between $N/V$ and $\epsilon$ can be approximated using classical nucleation theory. }
\end{figure}

We consider a system of volume $V$ in 2D with periodic boundary conditions\footnote{This choice is made for convenience; the same calculations can be easily repeated with no flux boundary conditions at a distant surface.}. We continue to assume that $\phi_\mathrm{a} \ll - \phi_\mathrm{B}$, and consider a target density $\phi_\mathrm{t}$ close to (but above) $- \phi_\mathrm{B}$ with slow chemical reactions. This ensures (a) that droplets remain dilute because the reaction (the nonconservative part of $\dot\phi$) must integrate to zero over the whole volume in any steady state, and (b) that the linear approximation to the reaction rates remains accurate in both the dilute and the dense phase.

We now single out one droplet and treat the rest as a homogeneous background. Let $\lambda = V^{-1} \sum_j \pi R_j^2$ be the area fraction of droplets, and radius of the singled out droplet be $R_i$. As before, we denote the density inside and outside the droplet as $\phi_\pm = \pm \phi_\mathrm{B} + \psi_\pm $. The approximate linear equation is unchanged in the dense phase, but in the dilute phase, the injection of mass by other droplets plays a role, 
\begin{equation}
\partial_t \psi_- = D \nabla^2 \psi_- + \left (1 - \lambda \right )g_-^0 + g_-^1 \psi_- + V ^{-1}  \sum_{j} 2 \pi R_j J_-^j 
\end{equation}
where $J_-^j$ denotes the current at the surface of the $j$th droplet. Solving the equation quasi-statically, we obtain a self-consistent equation for the current  $J_-^i$: 
\begin{equation}
\eqalign{
J_-^i = -  D \left ( c_-^i  - \frac{\gamma}{R_i} \right ) k_- \frac{K_1(k_- R_i ) }{K_0 (k_- R_i ) } \\
c_-^i =   c_- (1 - \lambda)   -  (g_-^1 V)^{-1} \sum_{j } 2 \pi R_j J_-^j
} 
\label{eq:mult_current} 
\end{equation}
where $c_- = - g_-^0/g_-^1$ as before, and  $c_-^i$ represents the density at infinity as seen by the $i$th droplet. One solution is for all the droplets to be of the same size. Letting this size be $R$, 
the current can be obtained by substituting the second equation into the first 
\begin{equation}
J_-=  - D k_-\frac{K_1(k_- R) }{K_0(k_- R) } \left ( c_- (1 - \lambda)  - \frac{\gamma}{R} \right ) \left ( 1 -  \frac{2 \lambda D k_-}{g_-^1 R} \frac{K_1}{K_0} \right)^{-1} 
\end{equation}
Since the equation for $\psi_+$ is unchanged, we can use the result from equation (\ref{eq:current}) for $J_+$. The growth rate of the droplet radius then obeys
\begin{equation}
\fl 2 \phi_\mathrm{B} \dot{R} = 
D  k_+ \frac{I_1}{I_0}  \left (c_+ -   \frac{ \gamma}{R} \right )
+  D k_- \frac{K_1 }{K_0  } \left ( c_- (1 - \lambda)  - \frac{\gamma}{R} \right ) \left ( 1 -  \frac{2 \lambda D k_-}{g_-^1 R} \frac{K_1}{K_0} \right)^{-1} 
\label{eq:mult_droplet}
\end{equation}
The resulting growth rate $\dot R(R)$  curve is plotted in Figure \ref{fig:mult_droplet}. We see that the structure of having one unstable fixed point and one stable fixed point remains, but the position of the stable fix point $R_\mathrm{s}$ shifts to smaller value as the number of droplets increases. Crudely we can understand it in terms of balancing the production of order parameter in the bath ($g_-^0$) and its consumption in the droplets ($g_+^1$): the ratio of the volume of the dilute and the dense phase must be around $| g_+^0/g_-^0 |$ in steady state.

So far, our calculations suggest that there can be multiple steady states with different numbers of identical droplets as long as the $\dot R(R)$ curve has a stable fixed point.
However, not all these states are stable against perturbation of the droplet sizes. The linear stability analysis in \ref{ap:mult_droplet} reveals two types of modes: one corresponds to the synchronised change of all droplets, which is always stable as long as the stationary radius exists in the first place (so that equation (\ref{eq:mult_droplet}) has a stable fixed point); the other, more unstable one represents the growth of all but one droplet at the expense of the diminishing droplet. We denote by $\omega_+$ the temporal eigenvalue of the exchange mode, which is plotted against $R$ in Figure \ref{fig:mult_droplet}. If the eigenvalue is positive at the stationary radius $R_\mathrm{s}$, $\omega_+ (R_\mathrm{s}) > 0$, the droplet suspension is unstable against flux exchange and the number of droplet decreases as a result. As the number density of droplets increases, the $\omega_+$ curve shifts to the right as the stationary radius $R_\mathrm{s}$ shifts to the left, giving rise to a maximum density. This provides a more stringent upper bound on the number density of droplets in suspension than simply requiring that (\ref{eq:mult_droplet}) has a fixed point. 

Simulations show that the actual number of droplets per unit volume depends on the noise strength $\epsilon$. Crudely, classical nucleation theory for Model B predicts that the nucleation rate is $\sim \exp(- \Delta F/\epsilon) $ where $\Delta F$ represents the free energy barrier. Starting with the uniform state at the target density, once droplets have nucleated, the flux exchange happens on the much slower time-scale of the reactions. Hence a finite number of droplets can nucleate before the suspension stabilises, leading to an inverse relation between the number of droplets and the nucleation rate, $N/V \sim \exp( - \Delta F /\epsilon) \rightarrow \log (N/V) = - \Delta F / \epsilon + $const, as the linear fit in Figure \ref{fig:epsilon} shows. Once the number of droplets is determined, their common radius is fairly well predicted by our approximate linear theory (see Figure \ref{fig:r_against_n}). 

\section{Droplet splitting} 
\label{droplet_splitting} 

\begin{figure} 
\includegraphics[width=\textwidth]{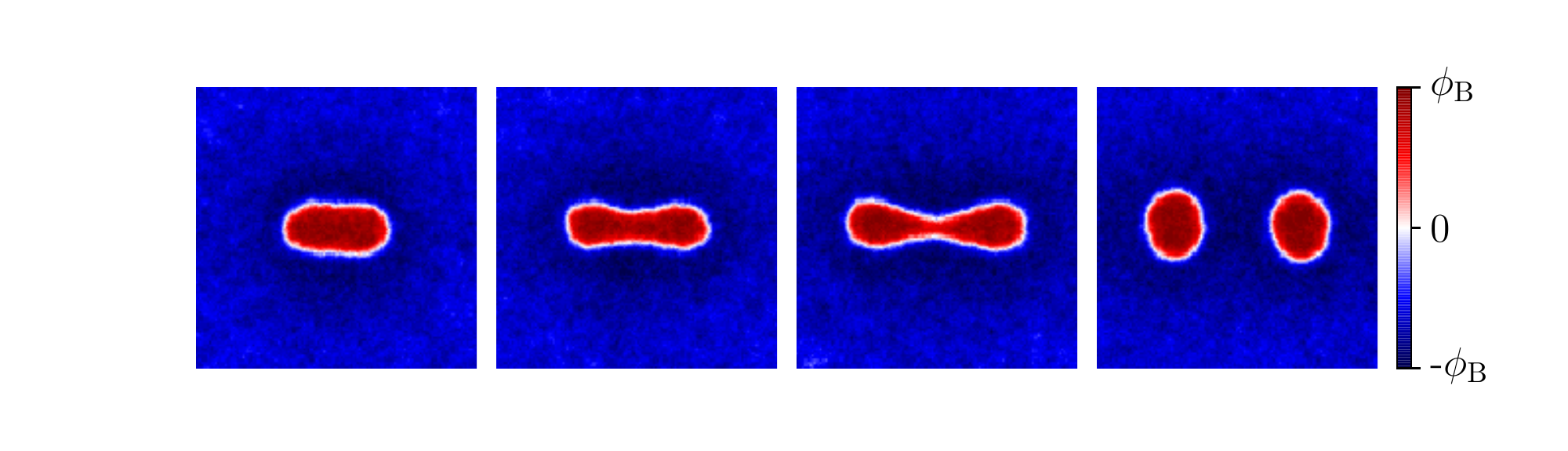}
\caption{Four snapshots of a slightly stretched droplet elongating and splitting into two. The simulation is done with parameters $ \phi_\mathrm{t} = -0.6, \phi_\mathrm{a} = -10, u M_\mathrm{A} = 4 \times 10^{-5}, \alpha = \beta = 0.2, \kappa =1, \epsilon = 0.01$ with periodic boundary conditions.  }
\label{fig:droplet_splitting}
\end{figure}

Droplets of sufficiently large radii can become unstable against shape perturbations: they can stretch and subsequently split into two smaller droplets as shown in Figure \ref{fig:droplet_splitting}. To understand this phenomenon, we consider the simple situation of a single circular droplet in an infinite bath and perturb its interface. Then, for perturbed boundary conditions we can solve equations (\ref{eq:linear}) for the scalar field inside and outside the droplet as before. Taking the quasi-static limit, we can then compute interfacial fluxes to determine the stability of the circular droplet against perturbations. Our method is inspired by a similar calculation Zwicker et al performed in 3D \cite{zwickerNatPhys2017}. 

Consider small angular perturbations about a circular droplet of radius $\bar{R}$,  
\begin{equation} 
\eqalign{
R(\theta) = \bar{R} + \sum_{l=1}^\infty \delta R_l \\
\delta R_l =\delta R^\mathrm{c}_l \cos(l \theta) + \delta R_l^\mathrm{s} \sin (l \theta) 
}
\label{eq:angular_pert}
\end{equation} 
Note that the $l=1$ mode corresponds to simple spatial translation. Thus the lowest relevant mode is the $l=2$ mode, in the shape of a dumbbell. Recall equation (\ref{eq:linear}) for $\psi_\pm$ (the excess over the binodal densities: $\phi_\pm = \pm \phi_\mathrm{B} + \psi_\pm$) 
\begin{equation}
\eqalign{ 
\partial_t \psi_+ = D \nabla^2 \psi_+ + g^0_+ + g^1_+ \psi_+\\
\partial_t \psi_- = D \nabla^2 \psi_- + g^0_- + g^1_- \psi_- 
} 
\label{eq:linear2}
\end{equation}
The values of $\psi_\pm$ at the interface are set by the local curvature $H$, which is now modified by the shape distortion, 
\begin{equation}
\eqalign{
\psi_\pm(R(\theta), \theta) &= \gamma H(R(\theta), \theta) \\
&= \frac{\gamma}{\bar{R}} \left [ 1 + \frac{1}{\bar{R}} \sum_{l=1}^\infty (l^2 - 1) \delta R_l  \right ] 
}
\label{eq:boundary}
\end{equation}
The linear equations can be solved exactly in polar coordinates in terms of the modified Bessel functions $I_l$ and $K_l$ and we refer to \ref{ap: droplet_splitting} for further details. After matching to the boundary conditions, the quasi-static solutions are 
\begin{eqnarray}
\psi_+(r, \theta) = c_+ + a_0^+ I_0(k_+ r) + \sum_{l=1}^\infty \chi_l^+ I_l (k_+ r)  \delta R_l  \\
\psi_- (r, \theta) = c_- + a_0^- K_0(k_- r) + \sum_{l=1}^\infty \chi_l^- K_l(k_- r ) \delta R_l 
\end{eqnarray}
where $k_\pm^2 = - g^1_\pm/D$ and $c_\pm = - g_\pm^0/g_\pm^1$ as before. The remaining parameters $a_0^\pm, \chi_l^\pm,$ are functions of $\bar{R}$ only and we again refer to \ref{ap: droplet_splitting} for their exact forms. 

\begin{figure}
\centering
\begin{subfigure}{0.44\textwidth}
\includegraphics[width=\textwidth]{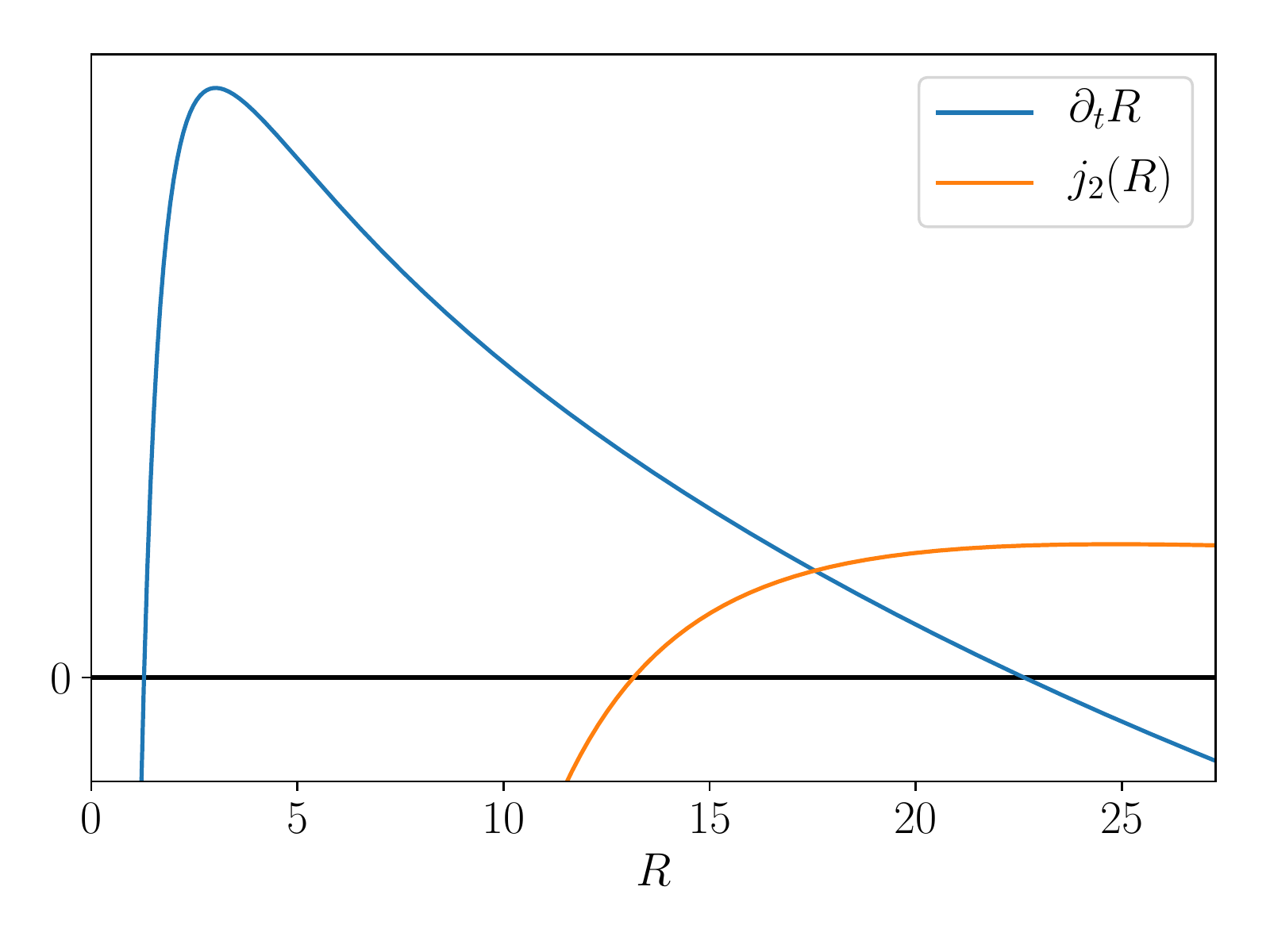}
\caption{}
\end{subfigure} 
\begin{subfigure}{0.48\textwidth}
\includegraphics[width=\textwidth]{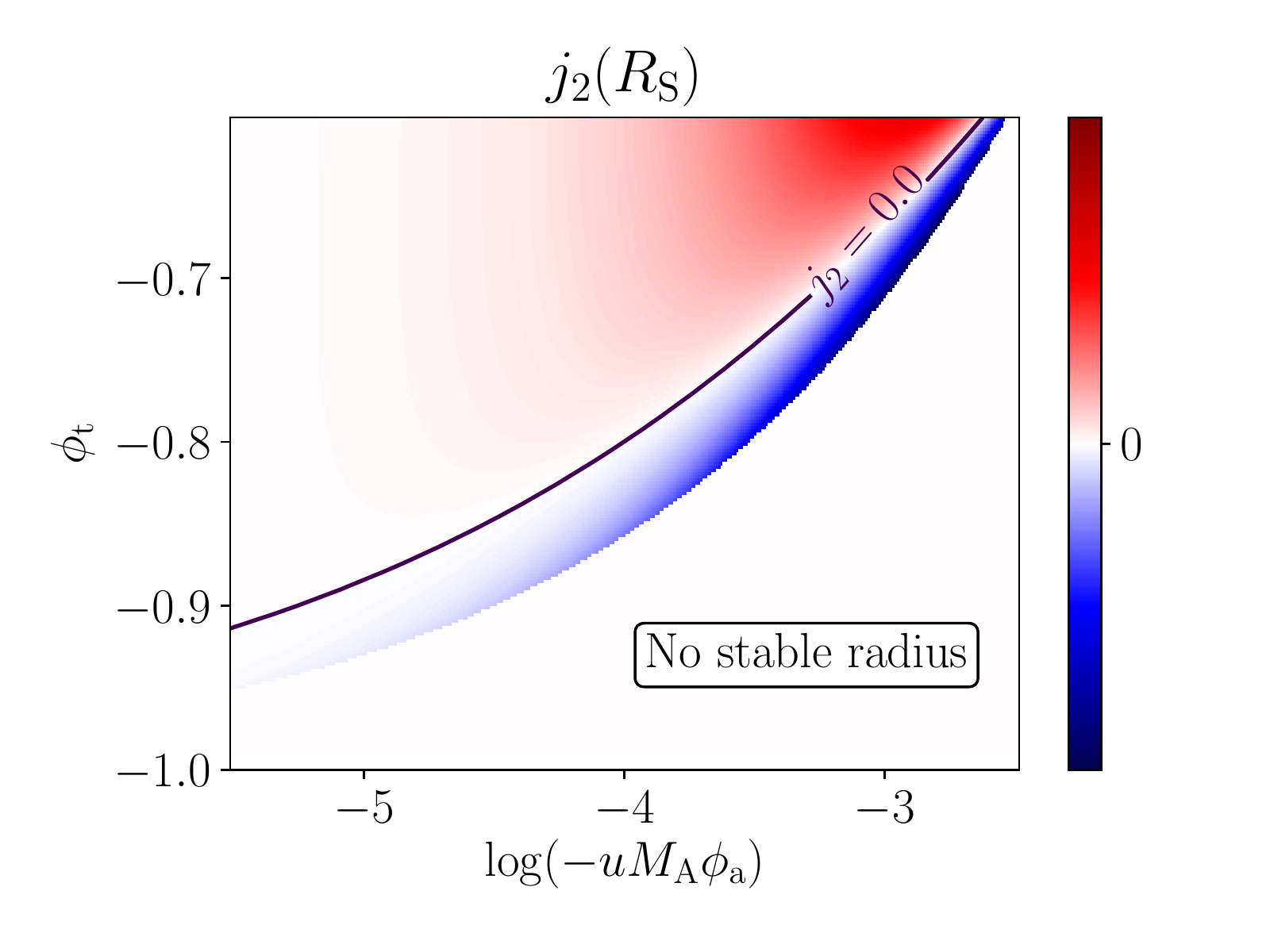}
\caption{}
\end{subfigure}
\caption{The panel on the left shows the growth rate of a spherical droplet and $j_2$ for $\phi_\mathrm{t} = -0.6, uM_\mathrm{A} = 4 \times 10^{-5}$. We can see that $j_2$ is positive at the stable radius $R_\mathrm{s}$, implying that the circular droplet is unstable against dumbbell-shaped perturbations. The right panel shows the sign of $j_2$ at the stable droplet radius for a range of $\phi_\mathrm{t}$ and $ - u M_\mathrm{A} \phi_\mathrm{a}$, and the white region on the bottom right corresponds to when no stable droplets can form. The other parameters are $\alpha = \beta = 0.2, \kappa = 1, \phi_\mathrm{a} = -10$}
\label{fig:l=2}
\end{figure}

Having obtained the quasi-static solutions for any small shape perturbation, we next calculate the diffusive currents on both sides of the interface to see which direction the interface moves. The difference between the currents can be written as
\begin{equation}
\eqalign{ 
\fl \quad \frac{1}{D} \left [  \boldsymbol{J}_+(\theta)  - \boldsymbol{J}_- (\theta) \right ]  &=  - \bnabla \psi_+ (R(\theta), \theta) + \bnabla \psi_- (R(\theta), \theta) \\
&=  j_0(\bar{R})\, \hat{\boldsymbol{r}} + \sum_{l=1}^\infty \left [ j_l (\bar{R} )  \delta R_l  \, \hat{\boldsymbol{r}}  + h_l (\bar{R}  ) (\partial_\theta \delta R_l)  \, \hat{\boldsymbol{\theta}} \right ]
}
\end{equation}

Here $j_0(\bar{R})$ is the isotropic radial flux that controls the growth of a spherical droplet, exactly the same as in section \ref{droplet_sus}. The remaining two terms give rise to  anisotropic flux as a result of the shape perturbation. Further calculations show that $h_l(\bar{R})= \bar{R}^{-1} j_0(\bar{R})$, meaning that, at the stable radius $R_\mathrm{s}$, the only remaining terms are $j_l(R_\mathrm{s})$ for $l \geq 1$. Its sign determines whether the circular shape is stable against perturbations: if it is positive, there is a net flow outwards at the protrusion and inwards at the depressions, leading to further deformation of the droplet. As shown in Figure \ref{fig:l=2},  for a certain range of parameters the $l=2$ dumbbell mode is linearly unstable, and this stretching is exactly what we see in Figure \ref{fig:droplet_splitting}. 

\begin{figure}
\includegraphics[width=\textwidth]{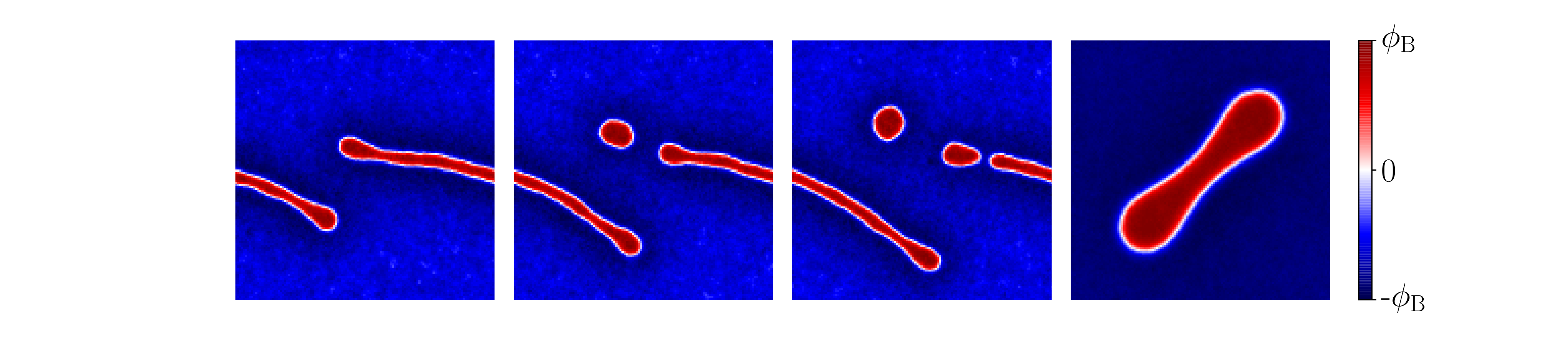}
\caption{The first three panels show another pathway to droplet formation: starting with a single droplet as before, the $l=2$ mode becomes unstable; as a result, the droplet elongates to form a long stripe whose ends can pinch off to form smaller droplets. Panel 4 shows an arrested dumbbell shape; this shape occurs when $\phi_\mathrm{a}$ is close to $-\phi_\mathrm{B}$. }
\label{fig:other_splitting_patterns}
\end{figure}

The linear stability analysis tells us whether a shape perturbation will grow, but there are several possibilities of subsequent behaviours following the initial stretch. Figure \ref{fig:droplet_splitting} shows the skewed droplet evenly splitting into two round droplets within a relatively confined volume. In a larger volume, the droplet elongates into a long stripe. This is different from the immediate splitting into smaller droplets as was reported for a similar model in 3D \cite{zwickerNatPhys2017}. This is because a cylindrical tube in 3D confined by surface tension is unstable against perturbations along the tube, known as the Plateau-Rayleigh instability, whereas such instabilities do not occur in 2D. Instead, the pinching off of smaller droplets at the end of the stripe is induced by noise (Figure \ref{fig:other_splitting_patterns}). It is also possible to arrest the stretching when the reaction rates are significantly nonlinear in the dilute phase, giving rise to a dumbbell shape as shown in the rightmost panel of Figure \ref{fig:other_splitting_patterns}. 

\section{Limit cycles} 
\label{limit_cycles}

\begin{figure}
\centering
\begin{subfigure}{0.52\textwidth}
\centering
\includegraphics[width=\textwidth]{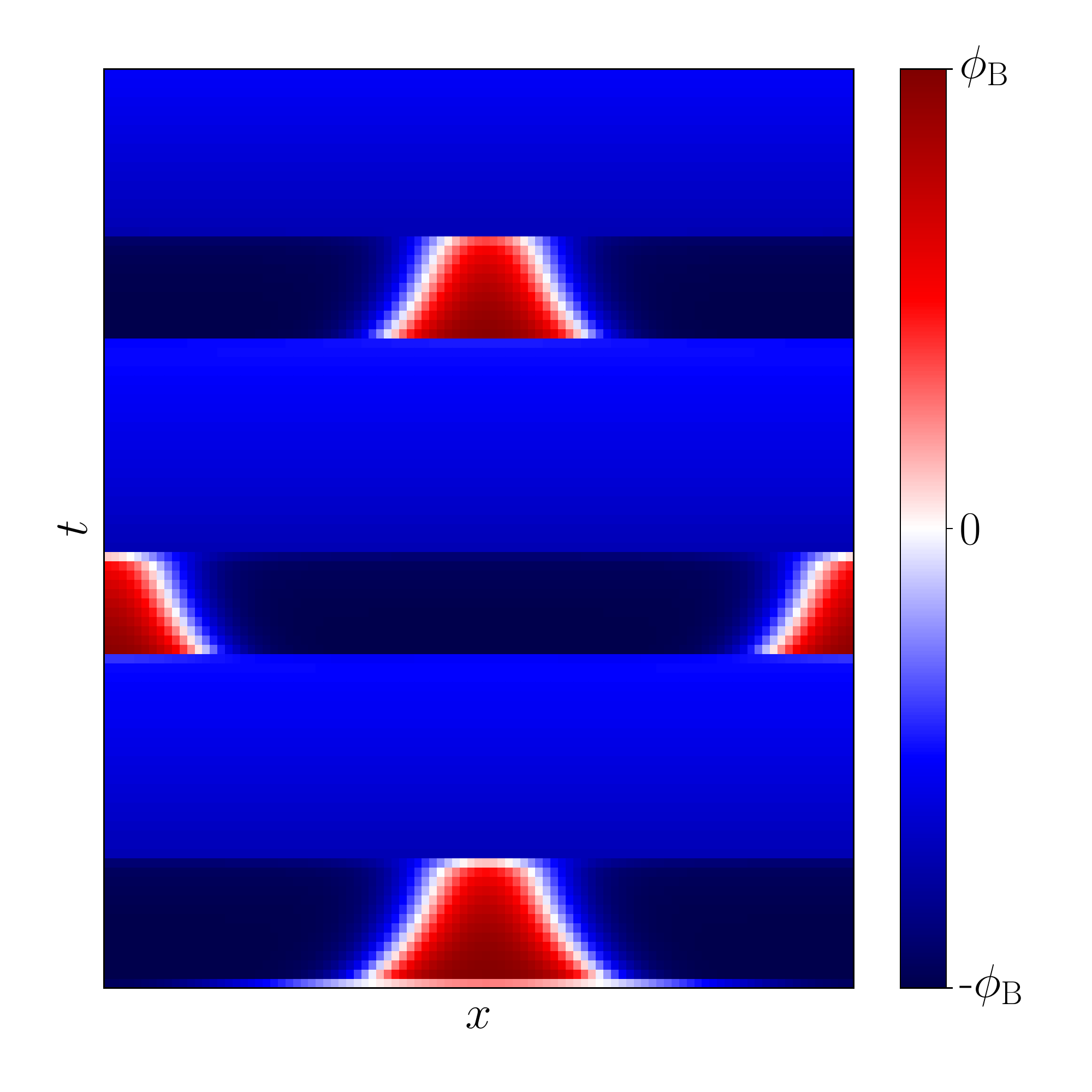}
\caption{}
\label{fig:cycle}
\end{subfigure}
\begin{subfigure}{0.45\textwidth}
\raisebox{-\height}{\includegraphics[width=\textwidth]{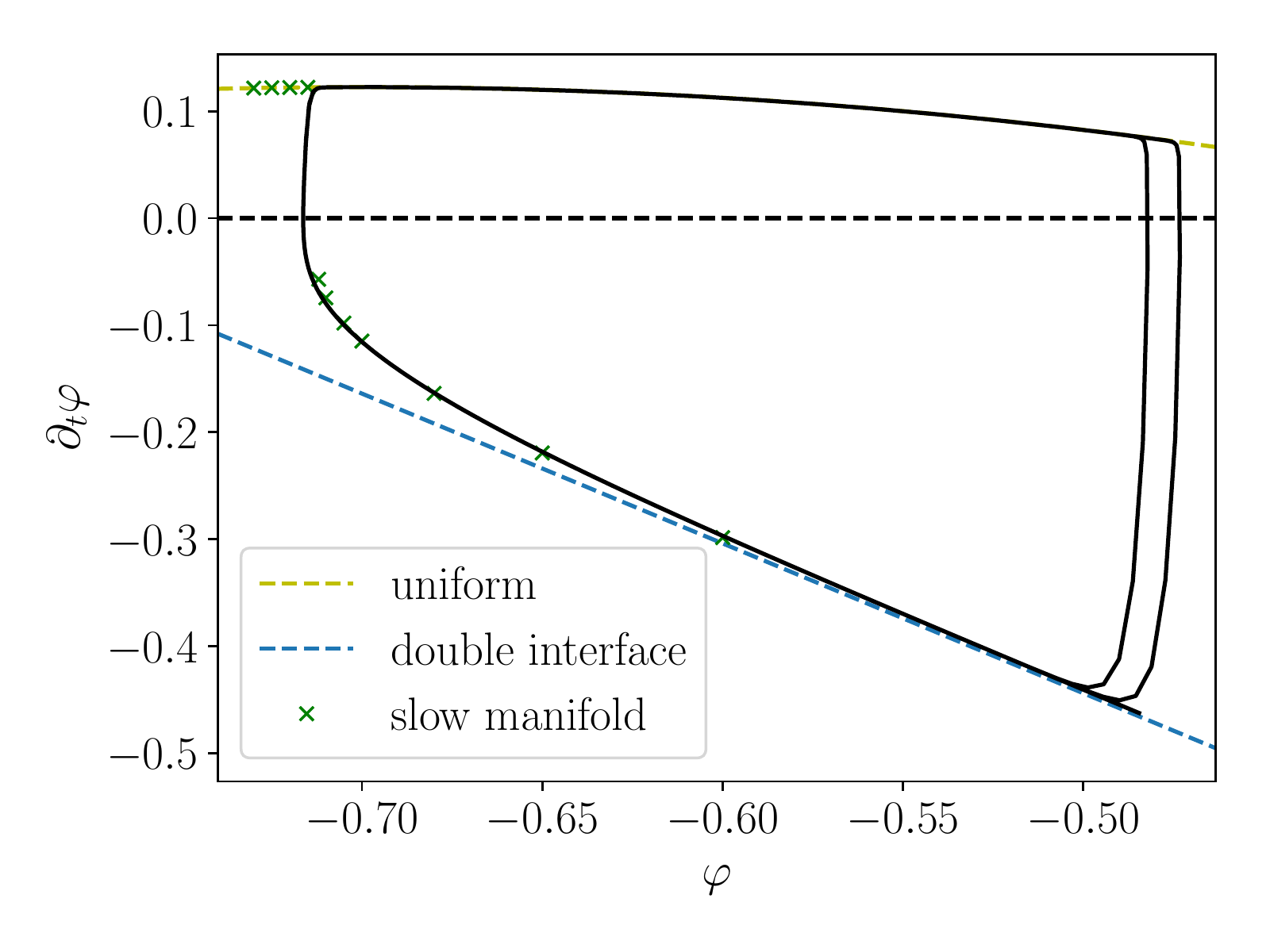}}
\vspace{-0.2cm}
\caption{}
\label{fig:phase}
\vspace{-0.2cm}
\raisebox{-\height}{\includegraphics[width=\textwidth]{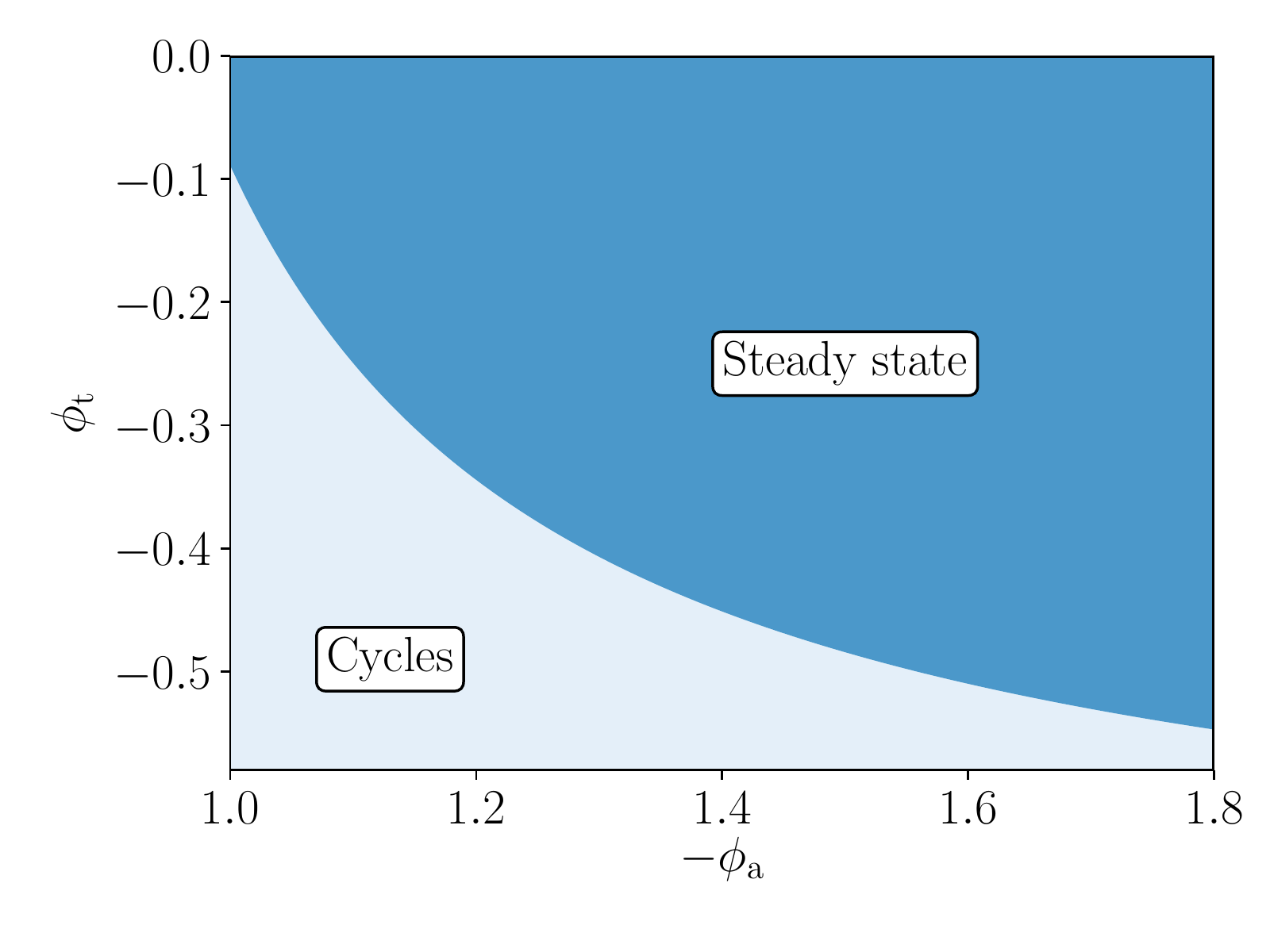}}
\vspace{-0.1cm}
\caption{}
\label{fig:phasediagram} 
\end{subfigure} 
\caption{(a) shows a space-time plot of the $\phi$ field and (b) shows the phase-space plot of the same simulation. In (b) the green crosses are numerical simulations of the slow manifold, the yellow line corresponds to equation (\ref{eq:uniform}) and the blue line corresponds to equation (\ref{eq:tanh}). The parameters for the simulation are $u M_\mathrm{A}= 5 \times 10^{-10}, \phi_\mathrm{a} = -1.05, \phi_\mathrm{t} = -0.35, \alpha = \beta = 0.001, \kappa = 1, X=800$. Panel (c) shows the phase diagram in the $(-\phi_\mathrm{a}, \phi_\mathrm{t})$ plane in the limit $ u M_\mathrm{A} \rightarrow 0$. }
\end{figure}

Above we have discussed various forms of arrested phase separation that occur for both linear and quadratic reaction rates. Next we seek to understand a curious non-equilibrium steady state that is specific to nonlinear $\mu_\mathrm{A}$, comprising limit cycles that oscillate between the homogeneous state and the phase separated state, as shown in Figure (\ref{fig:cycle}). This behaviour may be particularly relevant in bacteria as it mimics, in simplified form, a biofilm lifecycle where the bacteria alternates between swimming freely in a dilute `planktonic' phase and forming a condensed, static colony. Ref.\ \cite{grafkePRL2017} reports similar phenomena to those found below, but we will find that the use of our canonical model allows significantly greater progress in understanding the dynamics analytically. 

Such limit cycles are found in a region in parameter space where the conservative phase separation is much faster than the reactions, so that $u$ is finite but very small. Let $\varphi(t) = V^{-1}\int \mathrm{d}\boldsymbol{x}\,  \phi(\boldsymbol{x}, t)$ be the average order parameter value in the system. In a finite volume $V$, for each $\varphi$, there is at least one stable solution of the Model B dynamics. These form the slow manifold of the Model AB system. In the limit $ M_\mathrm{A} \rightarrow 0$, the reaction terms only move the system along the slow manifold \cite{grafkeJSM2017}. A sample space-time plot of the $\phi$-field in 1D on a domain of spatial extent $X$ is shown in Figure (\ref{fig:cycle}) and the corresponding circuit in $\dot{\varphi},\varphi$ shown as a phase-space plot in Figure (\ref{fig:phase}). We can see that the full dynamics indeed evolve very close to the slow manifold for the parameters chosen there, verifying the time-scale separation. The phase-space plot shows two key conditions for the limit cycle to occur:
\begin{itemize}
\item [(C1)] The rate of change of the global population, $\dot\varphi$, must be negative when there is a phase separation and remain so until $\varphi$ has been driven towards a density where the phase separated state is no longer locally stable. This leads to the turning point around $\varphi= -0.8$.
\item [(C2)] $\dot\varphi$ must be positive for the uniform state (the upper branch of Figure (\ref{fig:phase})) and remain so until local (spinodal) instabilities kick in to induce phase separation. This corresponds to the turning point around $\varphi = -0.5$.
\end{itemize}

\subsection{Uniform state}
On the upper branch in the phase space plot in Figure (\ref{fig:phase}), $\phi(x, t) = \varphi(t)$ and the uniform phase is locally stable. All conservative terms vanish and $\varphi(t)$ evolves only according to the global Model A dynamics, 
\begin{equation}
\label{eq:uniform}
\dot{\varphi} = - M_\mathrm{A} u (\varphi - \phi_\mathrm{a})(\varphi - \phi_\mathrm{t})
\end{equation}
Thus $\dot\varphi$ is positive for $\varphi < \phi_\mathrm{t}$. As long as $\phi_\mathrm{t}$ is within the spinodals, condition (C2) will be satisfied\footnote{Strictly speaking, we require $\phi_\mathrm{t}$ to lie within the range of $\varphi$ that makes the lowest Fourier mode of $\phi$ fluctuations linearly unstable; this defines the spinodal regime for a finite, as opposed to infinite, system.}.

\subsection{Phase-separated state}
The nonlinearity of the reaction rates is crucial in satisfying (C1). For suppose the reaction rates are linear: $\mu_\mathrm{A} = - u (\phi - \phi_\mathrm{t})$. Integrating over the spatial coordinates gives $\dot\varphi = - M_\mathrm{A} u (\varphi - \phi_\mathrm{t} )$, implying that $\varphi$ exponentially decays towards $\phi_\mathrm{t}$ regardless of the spatial distribution $\phi(\boldsymbol{x},t)$. Since (C2) constrains that $\phi_\mathrm{t}$ must lie within the spinodals, where a bulk phase-separated solution is always stable, (C1) can never be satisfied without nonlinearity in the Model A sector. 

To address the effects of such nonlinearity within our canonical Model AB, defined by (\ref{eq:full_sto_eq}), we will need to know the spatial distribution $\phi(\boldsymbol{x})$ for the phase-separated states that make up the slow manifold. To simplify the analysis, we restrict to 1D (or higher dimensions but with spatial variation of the density along a single axis only) where Model B can be solved analytically in an infinite system:  
\begin{equation}
\phi(x) = \phi_\mathrm{B}\tanh \left (q_c (x - x_0) \right )
\end{equation} 
Here $q_c = \sqrt{\alpha/2\kappa}$ as before, and the value of $x_0$ is such that $\int \mathrm{d}x \phi(x) = V \varphi$. In a finite system with periodic boundary conditions, we can construct an approximate solution by adding two $\tanh$ profiles together \cite{kawasaki}. Taking a domain $\Omega = [-X/2, X/2]$,  
\begin{equation}
\phi(x)/\phi_\mathrm{B} = 1 - \tanh \left (q_c (x - x_0)  \right ) + \tanh \left ( q_c (x + x_0) \right ) 
\label{eq:approx}
\end{equation}
Notice that this solution breaks down when $\varphi$ gets too close to the binodals, $\varphi = \pm \phi_\mathrm{B}+\mathcal{O}(1/q_cX)$, as the minority phase region is then smaller than the width of two $\tanh$-profiled interfaces. Otherwise, the approximate solution holds (see Figure~\ref{fig:phase}); we can then substitute (\ref{eq:approx}) into (\ref{eq:full_sto_eq}) and integrate over $\Omega$, to obtain
\begin{equation}
\eqalign{
\label{eq:tanh}
\dot\varphi &= -  M_\mathrm{A}  u (- \phi_\mathrm{a} - \phi_\mathrm{t})(\varphi -  \phi_{\mathrm{nt}} ), \\
\phi_{\mathrm{nt}} &= \frac{ \phi_\mathrm{a}\phi_\mathrm{t} + \phi_\mathrm{B}^2 (1 - 4 (q_c X)^{-1} )}{\phi_\mathrm{a} + \phi_\mathrm{t}}
}
\end{equation}
Noting that $\phi_\mathrm{a} < 0$, we thereby predict a new target (subscript ``$\mathrm{nt}$'') for the global density $\varphi$ in the phase-separated system that can be less than $\phi_\mathrm{t}$ for uniform states.

The limit-cycle arises when a negative $\dot\varphi$, caused by phase separation, remains negative until the phase-separated state becomes locally unstable under the Model B dynamics and it remixes. The new target density obtained above gives a useful guideline: cyclic behaviour is predicted for $\phi_\mathrm{nt} < -\phi_\mathrm{B}$, modulo our use of approximation (\ref{eq:approx}). To improve on this criterion requires numerical simulations of the slow manifold -- the green crosses in Figure (\ref{fig:phase}). Close to the turning point on the left, the slow manifold has deviated significantly from the tanh-profiled approximation: the dense phase occupies a region of order the interfacial width, and the peak density of this dense region is much lower than the binodal density. To obtain a more accurate estimation of the threshold between limit cycles and the steady state, we must look at the phase-separated state on the slow manifold with the smallest $\varphi$ and calculate the total reaction rate (for various $\phi_\mathrm{a}$ and $\phi_\mathrm{t}$). If the total rate is negative, (C1) is satisfied and there is a limit cycle. The resulting phase diagram in shown in Figure (\ref{fig:phasediagram}). 

Roughly speaking, limit cycles only occur for $\phi_\mathrm{a}$ close to $-\phi_\mathrm{B}$ and $\phi_\mathrm{t}$ close to $-\phi_\mathrm{S}$, so that the two phases ($\pm \phi_\mathrm{B} $) have vastly different reaction rates, and the loss of $\phi$ within in a small region of the dense phase can balance the total production of $\phi$ in the much larger region of dilute phase. This leads to a steady decline in the global density $\varphi$ within the phase-separated state, which brings the system back close to the binodal density, at which point it remixes. Having done so, there is now a net global production ($\dot\varphi > 0$) and the global density grows again until the spinodal density is reached, causing phase separation and completing the cycle.

\subsection{Limit cycles beyond the slow reaction limit}

\begin{figure}
\centering
\includegraphics[width=\textwidth]{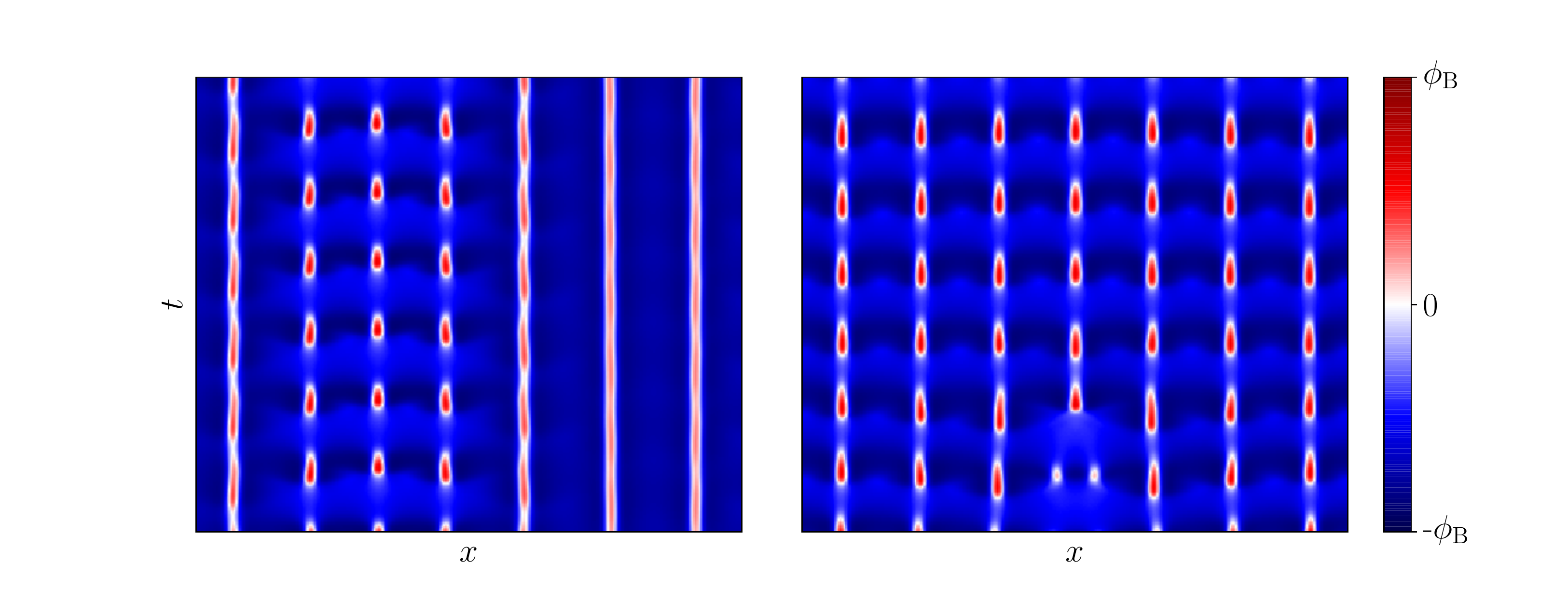}
\caption{Space-time plots of two simulation runs with the same parameters but different initial conditions. The left panel starts with a uniform state with small random noise; the right panel is initialised with a high-density bump in the middle. The parameters are $  \phi_\mathrm{a} = -1.05, \phi_\mathrm{t} = -0.45, u M_\mathrm{A} = 2 \times 10^{-4}, \alpha = \beta = 0.1, \kappa = 1, X = 600$. }
\label{fig:many_domains}
\end{figure} 

In previous sections, we studied cycles in the limit $M_\mathrm{A} \rightarrow 0$. In practice, this requires the reaction to be significantly slower than diffusion so the dynamics can be formulated in terms of evolution along a slow manifold formed by steady states of the conservative sector. More precisely, the time scale of reaction is $\sim 1/M_\mathrm{A} $ and the time scale of diffusion is characterised by the time to diffusive across the entire system, and is therefore of order $\sim X^2/D$, where $D = 2 \alpha $ is the diffusion constant in either phase. Hence there is a time scale separation if $M_\mathrm{A}   \ll D/X^2$. 

When the time scales are comparable, the system goes through more complex limit cycles where all or part of the system oscillates between a phase separated state with multiple domains and the homogeneous state, as shown in Figure \ref{fig:many_domains}. In this regime, for one unit of reaction time, the particles can only diffusive across part of the system, and this sets the domain lengths of the (transiently) phase separated state, not unrelated to the steady-state microphase separations discussed previously in which there is balance between diffusive fluxes and chemical processes of the kind shown in Figure \ref{fig:breeding}.

\subsection{Mechanism of limit cycles} 
The limit cycle is an interesting but mathematically delicate phenomenon that requires some fine-tuning of parameters within our canonical Model AB. We can imagine that by choosing a different conservative model (with non-quartic $\mathcal{F}_\mathrm{B}$, and hence different spinodal and binodal curves) we might be able to expand the size of the ``limit-cycle phase'' in parameter space. However, there are many advantages in sticking with the canonical model because of its analytic tractability.

In particular, Grafke et al.~studied the same limit cycle for a non-canonical model of bacterial phase separation with population dynamics, described by a non-polynomial $\mathcal{F}_\mathrm{B}$ with multiplicative noise \cite{grafkePRL2017}. There, the homogeneous to phase-separated limit cycle was suggested as a pre-evolutionary prototype for the planktonic-to-colonial life-cycle in bacteria. However the relative complexity of the model made it hard to fully elucidate the physical mechanism underlying the limit cycle. Addressing the same cycle within our much simpler Model AB allows the mechanism to be exposed more clearly, and with it, some potential limitations to its relevance to the bacterial life-cycle. Specifically, since the {\em global} density of motile bacteria must decrease with time throughout the lifetime of a colony, between its initial formation by phase separation and final loss by remixing, rather high rates of death (or at least of immotilization \cite{grafkePRL2017}) are required within the colony itself. Indeed, in the limit of small reaction rates, a single colony can sustain itself in coexistence with a finite planktonic reservoir but never an infinite one: in the latter case, such colony is predicted to redisperse on the fast (Model B) timescale. However, synchronised oscillations with a finite density of colonies remain possible instead (cf. Figure \ref{fig:many_domains}). The system size thus enters in a possibly unexpected way that would need to be better understood within any intended biophysical context involving an isolated colony.

\section{Conclusions} 
\label{conclusion}

In this paper, we formulated a scalar field theory (Model AB) for non-equilibrium phase-separating systems with chemical reactions. It combines Model B  for the conservative dynamics and Model A for the non-conservative reactions, with incompatibility between the free energies that drive each sector.

In general the system is far from equilibrium and no effective global free energy exists, although within a certain subspace of parameters reversibility is recovered at the level of the space-time trajectories of the order parameter $\phi$ (but not if one also monitors its conservative current $\boldsymbol{J}$). We addressed in detail the steady states that emerge from equilibrium statistical mechanics using the effective free energy so derived. The effect of reactions within this subspace is to introduce screened Coulomb interactions that frustrate bulk phase separation, replacing it with microphase separation of a form studied in the prior literature on equilibrium models. This equilibrium mapping, which is exact where it exists at all, significantly improves upon earlier works that establish a related equivalence to an (unscreened) Coulomb interaction. The latter mapping is not exact because it holds at the level of deterministic mean-field dynamics only. Our exact mapping appears to give a good qualitative guide to the physical behaviour in nearby parameter regimes where that also show stationary microphase separation.

In our studies of the more general case, we studied a `canonical' version of Model AB, retaining lowest order terms in the Landau Ginzburg expansion in each sector, plus one additional nonlinearity required to ensure that the global mean density does not dynamically decouple from other degrees of freedom. 
The canonical Model AB amounts to choosing a symmetric $\phi^4$ square-gradient free energy $\mathcal{F}_\mathrm{B}$ for the Model B sector and a nonlinear local logistic-growth type chemical potential $\mu_\mathrm{A}$ for the Model A sector. This canonical model is more general than it might at first appear, since parameter shifts can absorb both linear and cubic terms in $\mathcal{F}_\mathrm{B}$ and Laplacian terms in $\mu_\mathrm{A}$. Within the model, each sector remains separately time-reversal symmetric but with incompatible free energies driving their dynamics. (Technically our chosen form of the model has an absorbing state which could break the time reversal symmetry if entered, but we do not work in this region of very low physical density.) 

In our exploration of the resulting parameter space, we found that our canonical model shows various phenomena previously reported for diverse, more microscopically inspired, and often more complicated, models spread across the literature. These phenomena include lamellar and droplet micro-phase separated patterns; emergence of a stable radius in the single-droplet Ostwald process; droplet splitting dynamics; and steady state oscillations between a homogeneous (`planktonic')  and a phase-separated (`colonial') state \cite{Glotzer1, Glotzer2, grafkePRL2017, catesPNAS2010, weberRepProgPhys2019}. Of these phenomena, the limit cycles require oscillations of the global mean density in concert with the phase separation, so that choices of the nonlinearity parameter $\phi_\mathrm{a}$ in the Model A sector, as well as its target density $\phi_\mathrm{t}$, is crucial in that regime.

In contrast, the other phenomena listed above can, in the main, be well accounted for using only linear reaction dynamics. This gives just two free parameters in the Model A sector at the deterministic level: the target density $\phi_\mathrm{t}$, and the reaction rate $u M_\mathrm{A}$.  The time evolution, starting from a uniform state, then falls into one of three categories: spinodal decomposition, nucleation and growth, and remaining homogeneous. In the limit of the reaction rate going to zero, these regimes correspond exactly to their Model B counterparts, but with the important restriction that $\phi_\mathrm{t}$ sets both the global density and the far-field supersaturation in single-droplet growth (so that this supersaturation ceases to depend on droplet size, dramatically changing the growth law). Moreover, the spinodal and binodal lines merge towards $\phi=0$  as the reaction rate increases, until eventually the homogeneous state remains stable for all $\phi_\mathrm{t}$. 

For target densities between the binodals, the Model AB dynamics leads to the coarsening becoming arrested by the reactions at a finite length-scale, resulting in microphase separation. This comprises lamellar or droplet/bubble patterns, whose geometry depends on the phase volume ratio. This is in turn set by the global mean density $\varphi$ which, for linear reactions, evolves autonomously towards $\phi_\mathrm{t}$ in steady state.  These steady states may or may not show long range order; this depends on the noise level, and we have not explored it in detail, but see \cite{catesPNAS2010} for examples found within a more complicated non-canonical model framework. All such outcomes have natural explanations within the equilibrium mapping referred to above, but the same physics (Figure \ref{fig:breeding}) is seen far outside the parameter subspace where that mapping exists. In particular, the addition of Model A nonlinearity, introducing an extra unstable zero at $\phi_\mathrm{a}$ in $\mu_\mathrm{A}$, obviates the mapping but does not change or destroy these three regimes, although it also admits new behaviours, including a curious example where the stationary solution in a finite domain is a pronounced dumbbell-shaped droplet (Figure \ref{fig:other_splitting_patterns}).

If, instead of starting from a homogeneous phase, the system is initialized with large domains of the two phases, the system anti-coarsens until the steady-state microphase-separation pattern is recovered. This is again easily explained within the subspace of the equilibrium mapping where there is a unique Boltzmann equilibrium which the stochastic dynamics of Model AB will find eventually (modulo the possible intrusion of glass physics with long relaxation times to escape metastable states). One notable mechanism of anti-coarsening involves a large droplet stretching into a dumbbell shape and splitting into two or more smaller ones (Figure \ref {fig:droplet_splitting}). 

Within the subspace of the equilibrium mapping, the role of noise in steady states is fully addressable using the methods of finite-temperature equilibrium statistical mechanics. Away from this subspace it is relatively intractable analytically. Therefore, in addressing the more general phenomenology of Model AB, the effects of noise were only studied numerically in this work. There is room for further exploration of noise effects; for example, we found that the length-scales of the same microphase-separated morphologies depend strongly on the noise strength $\epsilon$. For droplets, the noise provides (via nucleation) a selection mechanism which decides the number density of droplets in the system; without it, there is separately stable static solution for any such number density. A very similar effect was reported recently in a purely conservative model (Active Model B+) where gradient nonlinearities, rather than non-conservative dynamics, are responsible for arrest of phase separation \cite{TjhungPRX2018}. Accordingly, we hope to return to the role of noise in Model AB in future work.

\section*{Acknowledgements:}
We thank Yongjoo Baek, Erwin Frey, Tobias Grafke, Rob Jack, Rajesh Singh, Julien Tailleur and David Zwicker for valuable discussions, and Rajesh Singh for code contributions. YIL thanks the Cambridge Trust and the Jardine Foundation for a PhD studentship. This work was funded in part by by the European Research Council under the Horizon 2020 Programme, ERC grant agreement number 740269. MEC is funded by the Royal Society. 

\begin{appendix}

\section{}
\label{ap:equilibriumAB}

In this appendix, we investigate the steady state pattern for the equilibrium subspace by calculating the free energy of the homogeneous state, lamellar patterns and droplet suspensions. Recall that the free energy can be split into a local $\phi^4$ part and a non-local screened Coulomb interaction, 
\begin{equation}
\eqalign{
\mathcal{F}[\phi] &= \mathcal{F}_\mathrm{loc} [\phi] + \mathcal{F}_\mathrm{nl} [\phi] \\
\mathcal{F}_\mathrm{loc}[\phi] &=  \int \mathrm{d}\boldsymbol{x} \left [ c \phi  - \frac{\alpha_\mathrm{eff}}{2}  \phi^2 + \frac{\beta}{4} \phi^4  + \frac{\kappa}{2} | \bnabla \phi |^2 \right ] \\
\mathcal{F}_\mathrm{nl}[\phi] &=  \frac{m^2}{2} \left ( \alpha ' + \alpha_\mathrm{eff} \right )   \int \frac{\mathrm{d}\boldsymbol{q}}{(2\pi)^d} \frac{|\phi(\boldsymbol{q})|^2 }{ q^2 + m^2 } 
}
\end{equation}
For simplicity we will work through the case where reactions are much slower than the conservative dynamics ($4 \kappa m^2 \alpha' \ll \alpha^2 $) so that the interfacial widths are small compared to the domain size. As a rule of thumb, we enforce the smallest domain to be at least $10 \xi_0$ for all numerical evaluations where $\xi_0 = \sqrt{2 \kappa / \alpha}$ is the interfacial width. 

\subsection{lamellar pattern}
We assume a quasi-1D ansatz $\phi(x) = \phi_2 - (\phi_1 - \phi_2) H(x - \zeta L / 2) + (\phi_1 - \phi_2) H(x + \zeta L/2)$ for $x \in [- L/2, L/2 ]$ where $H(x)$ is the Heaviside step function for all free energy calculations except the $\kappa$-term, where we approximate the interfaces as $\tanh$-shaped curves connecting the two domains. $\zeta$ is the ratio taken up by the $\phi_1$ phase in a period of length $L$. Note that $\zeta$ is a function of $(\phi_1, \phi_2)$, fixed by the constraint that the total amount of reaction must sum to zero.

First we will calculate the additional free energy contribution of each interface \cite{christensen1996PRE}. As the interface is narrow compared to the domain length, they can be approximate with $(\phi_1 + \phi_2)/2 + (\phi_1 - \phi_2) \tanh(x/\xi_0)/2$ where $\xi_0 = \sqrt{2 \kappa/\alpha}$ is the interfacial width \cite{cates2018JFM},  
\begin{equation}
\int_{-\infty}^{\infty} \mathrm{d} x \; \kappa \left [ \frac{\phi_1 - \phi_2}{2} \partial_x  \tanh( x/\xi_0 ) \right ] ^2  = \sqrt{\frac{\alpha \kappa}{18}} (\phi_1 - \phi_2)^2
\end{equation}
Together with the contributions from the bulk, the local part of the free energy gives a contribution per unit volume of, 
\begin{equation}
V^{-1} \mathcal{F}_\mathrm{loc}= f_\mathrm{loc}(\phi_1)\zeta + f_\mathrm{loc}(\phi_2)(1 - \zeta) + \frac{2}{L} \sqrt{\frac{\alpha \kappa}{18}} (\phi_1 - \phi_2)^2
\end{equation}
For the non-local part, let $\Psi(q) = \frac{m^2}{2} (\alpha' + \alpha_\mathrm{eff} ) (q^2 + m^2)^{-1} \phi(q) $, and $\Psi(x)$ be its Fourier transform, so that $\mathcal{F}_\mathrm{nl} = \int \mathrm{d}x \Psi(x) \phi(x)$. $\Psi(x)$ obeys the differential equation, 
\begin{equation}
\left ( - \partial_x^2 + m^2 \right ) \Psi(x) = \frac{m^2}{2} (\alpha' + \alpha_\mathrm{eff} ) \phi(x) 
\end{equation}
Substituting in for $\phi(x)$ and enforcing continuity of $\Psi(x), \partial_x \Psi$ at the boundaries, we obtain, up to an additive constant, 
\begin{equation}
\fl \Psi(x) = \frac{1}{2} \left ( \alpha' + \alpha_\mathrm{eff} \right ) 
\cases{
 a \cosh(mx) + \phi_1  & for $  | x | \leq \zeta L  / 2 $\\
b \cosh(m(x - L/2) ) + \phi_2 & for $ \zeta L /2 < | x | <  L / 2  $
}
\end{equation}
where, in this appendix, we define $z = m L /2$ and 
\begin{equation}
\eqalign{ 
a = \frac{( - \phi_1 + \phi_2 )\, \mathrm{sech}( z \zeta ) }{ 1 + \coth(z ( 1 - \zeta)  ) \tanh(z \zeta)} \\
b = \frac{( \phi_1 - \phi_2 )\, \mathrm{sech}( z ( 1 - \zeta) ) }{1 + \coth ( z \zeta  ) \tanh( z ( 1 - \zeta) ) }
}
\end{equation} 
Integrating $\Psi(x)\phi(x)$, we obtain the non-local part of free energy density, 
\begin{equation}
\eqalign{
\fl V^{-1} \mathcal{F}_\mathrm{nl}= \frac{1}{2} (\alpha' + \alpha_\mathrm{eff} ) \phi_1 \left ( \zeta \phi_1 - (\phi_1 - \phi_2)  \frac{\sinh( z ( 1 - \zeta)) \sinh( z \zeta)} {  z \sinh( z) } \right )  \\
 + \frac{1}{2} (\alpha' + \alpha_\mathrm{eff} ) \phi_2 \left ( ( 1 - \zeta) \phi_2  + (\phi_1 - \phi_2)\frac{ \sinh( z \zeta ) \sinh( z ( 1 -\zeta) }{z \sinh(z)}   \right )  
} 
\end{equation} 
The total free energy density $V^{-1} \mathcal{F}$ for a lamellar pattern is plotted against $L$ in Figure \ref{fig:tot_f}, where there is a clear minimum at a finite value of $L$, implying that the periodicity of the pattern is finite. One can then minimise the free energy by differentiating with respect to $(L, \phi_1, \phi_2)$, although the resulting systems of equations are not particularly enlightening or analytically tractable, so we will turn to numerical evaluations instead. However, when $m$ is sufficiently small, we can expand the non-local part as a power series in $z = mL /2$, and check for self-consistency, 
\begin{equation}
\fl V^{-1} \mathcal{F}_\mathrm{nl} \approx  \frac{1}{2} (\alpha' + \alpha_\mathrm{eff} ) \left [ \left ( \zeta \phi_1^2 + (1 - \zeta)^2 \phi_2^2 \right )  
+ \frac{1}{3} ( \phi_1 - \phi_2)^2 \zeta^2 ( 1 - \zeta)^2 z^2 + O(z^4) \right ] 
\end{equation}
This gives a simpler expression for the total free energy density 
\begin{equation}
\fl V^{-1} \mathcal{F}( \phi_1, \phi_2, L) =   f( \phi_1, \phi_2) + (\phi_1 - \phi_2)^2 \left [
\sqrt{ \frac{2 \alpha \kappa}{9 }} L^{-1} + \frac{m^2 ( \alpha' + \alpha_\mathrm{eff}) }{24} \zeta^2 ( 1 - \zeta^2) L^2
 \right ]  
\end{equation}
Minimising with respect to $L$ gives the length-scale of the lamellar pattern: $L \propto m^{-2/3} ( \alpha' + \alpha + \kappa m^2)^{-1/3} (\alpha \kappa)^{1/6}$. A sufficient condition for self-consistency is $m^2 \kappa \ll \alpha$  and the expression further simplifies to $L \propto m^{-2/3}$. We refer to section \ref{elsp} in the main text for further discussions of this result. 

\subsection{Droplet suspension in 2D}
\begin{figure}
\centering
\begin{subfigure}{0.45\textwidth} 
\includegraphics[width=0.9\textwidth]{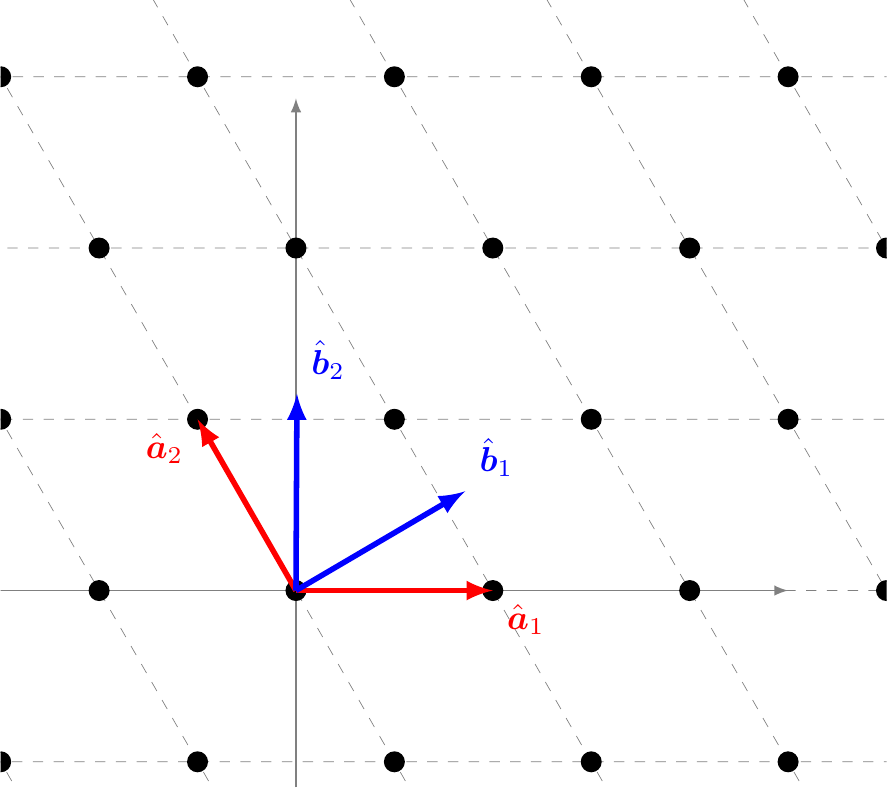}
\caption{Hexagonal lattice} 
\label{fig:lattice}
\end{subfigure} 
\hfill 
\begin{subfigure}{0.5\textwidth}
\centering
\includegraphics[width=\textwidth]{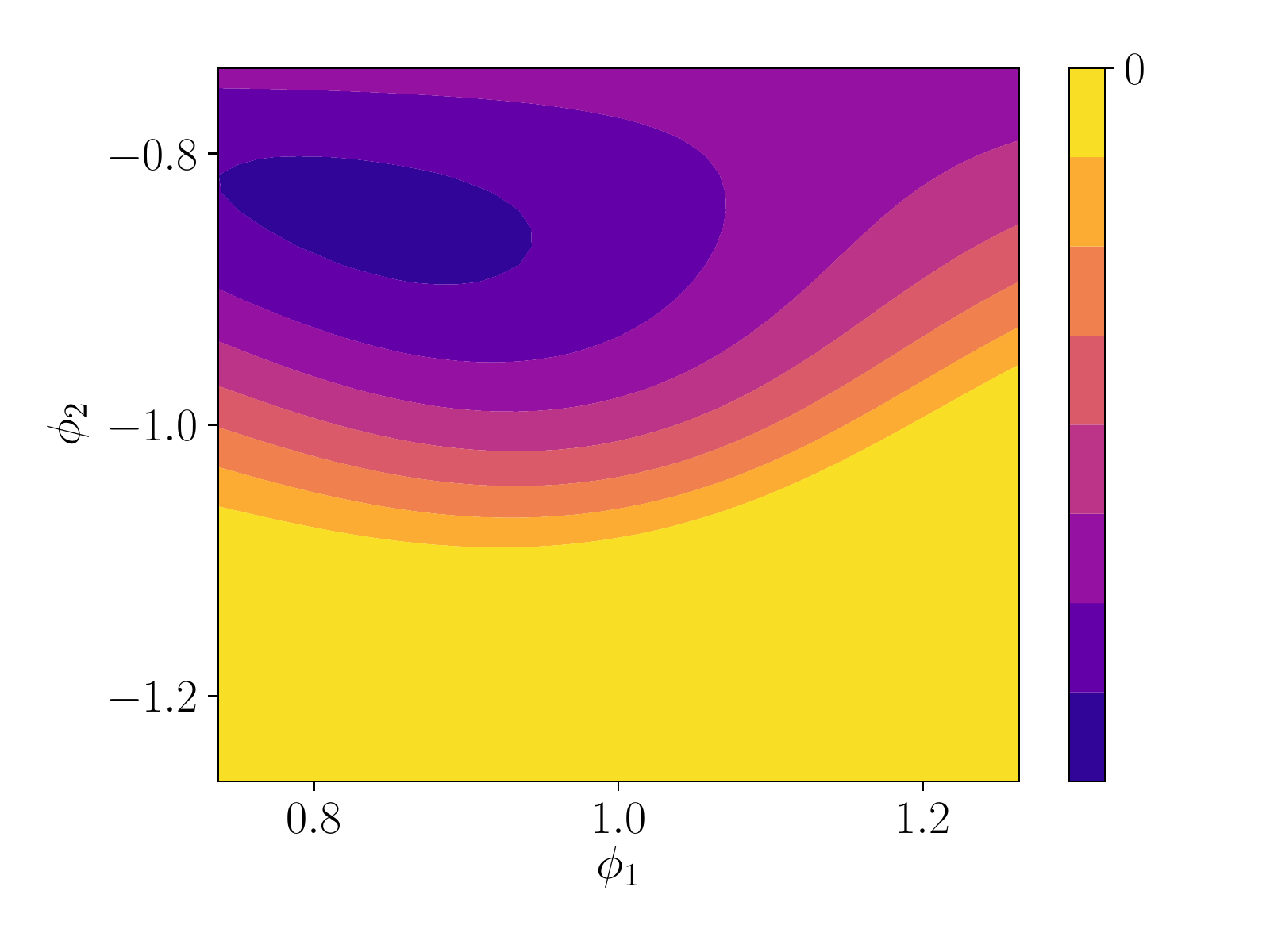}
\caption{A colormap of the partially minimised free energy for each value of $(\phi_1, \phi_2)$ to find the global minimum for a hexagonal lattice of droplets.}
\label{fig:min}
\end{subfigure} 
\end{figure} 
Without loss of generality, assume $c \leq 0$ and postulate spherical droplets at density $\phi_1$ with radius $R$ on a square or hexagonal lattice with spacing $L$ in a bath of density $\phi_2$. The scalar field $\phi$ is, up to an additive constant to obtain the correct spatial average, a convolution of a disk function $\phi_\mathrm{disk}(\boldsymbol{x}) = (\phi_1 - \phi_2 ) \left ( 1 - H(|\boldsymbol{x}| - R) \right )$, where $H$ is the same function as before, and a Dirac comb function at the lattice points, hence we can Fourier transform both functions and multiply in Fourier space using the Convolution Theorem. In 2D the Fourier transform of a disk of height $(\phi_1 - \phi_2)$ is 
\begin{equation}
\phi_\mathrm{disk}(\boldsymbol{k}) =(\phi_1 - \phi_2) \frac{2 \pi R}{| \boldsymbol{k} | } J_1(| \boldsymbol{k}| R)  
\end{equation}
where $J_1$ is the Bessel function of the first kind. A square or hexagonal lattice with unit vectors $(\hat{\boldsymbol{a}}_1, \hat{\boldsymbol{a}}_2)$ and spacing $L$ can be represented the Dirac comb function:  $\sum_{n_1, n_2} \delta( \boldsymbol{x}  - L ( n_1 \hat{\boldsymbol{a}}_1 + n_2 \hat{ \boldsymbol{a}}_2) )$. Its Fourier transform is the following sum, 
\begin{equation}
\phi_\mathrm{latt}(\boldsymbol{k}) = \sum_{n_1, n_2} \exp[ - i ( n_1 \boldsymbol{k} \cdot \hat{\boldsymbol{a}}_1 + n_2 \boldsymbol{k} \cdot \hat{\boldsymbol{a}}_2 )] 
\end{equation}
We can choose unit vectors $\hat{\boldsymbol{b}}_1, \hat{\boldsymbol{b}}_2$ such that they are respectively orthogonal to $\hat{\boldsymbol{a}}_2, \hat{\boldsymbol{a}}_1$ as shown in figure (\ref{fig:lattice}) for a hexagonal lattice. In this basis, let $\boldsymbol{k} = k_1 \hat{\boldsymbol{b}}_1 + k_2  \hat{\boldsymbol{b}}_2$, the above sum factorises to 
\begin{equation}
\phi_\mathrm{latt}(\boldsymbol{k}) = \sum_{n_1}  \exp( - i k_1 n_1 L \hat{\boldsymbol{a}}_1 \cdot \hat{\boldsymbol{b}}_1 ) \sum_{n_2} \exp( - i k_2 n_2 L \hat{\boldsymbol{a}}_2 \cdot \hat{\boldsymbol{b}}_2 )
\end{equation}
Note that $\hat{\boldsymbol{a}}_1 \cdot \hat{\boldsymbol{b}}_1 =  
\hat{\boldsymbol{a}}_2 \cdot \hat{\boldsymbol{b}}_2 \equiv s $. The Fourier series identity gives 
\begin{equation}
\phi_\mathrm{latt}(\boldsymbol{k} ) = 
\Delta_k^2 \sum_{n_1, n_2} \delta(k_1 - n_1 \Delta_k) \delta(k_2 - n_2 \Delta_k)
\end{equation}
where $\Delta_k = 2 \pi/(sL)$ is the lattice spacing in reciprocal space. Using the Convolution Theorem, the Fourier transform of the scalar field is, 
\begin{equation}
\fl 
\phi(\boldsymbol{k}) = (\phi_1 - \phi_2) 2 \pi  R \Delta_k^2 \frac{J_1(|\boldsymbol{k}| R)}{|\boldsymbol{k}|} \sum_{n_1, n_2} \delta(k_1 - n_1 \Delta_k) \delta(k_2 - n_2 \Delta_k)
 + (2 \pi)^2 \phi_2 \delta^2 ( \boldsymbol{k})
\end{equation}
where the last term gives the density $\phi_2$ of the bath. Note that due to the orthogonality relations, we also have $\hat{\boldsymbol{a}}_1 \cdot \hat{\boldsymbol{a}}_2 = \hat{\boldsymbol{b}}_1 \cdot \hat{\boldsymbol{b}}_2 = \sqrt{ 1 - s^2} $. Hence the volume element $\mathrm{d}k_x \mathrm{d}k_y = s \, \mathrm{d}k_1 \mathrm{d}k_2$, and the non-local part of the free energy density is (noting that $ [2 \pi \delta(0)]^d = V$), 
\begin{eqnarray}
\fl V^{-1} \mathcal{F}_\mathrm{nl} = \frac{1}{2} (\alpha' + \alpha_\mathrm{eff} ) \left [ \frac{ s^2 m^2 R^2 \Delta_k^4 }{  (2 \pi)^2} (\phi_1 - \phi_2 )^2 \sum_{k_1, k_2}  \frac{ J_1^2( k R  ) }{k^2 ( m^2 + k^2) }  +  \phi_2 ^2 \right ] \\
\fl \qquad \quad  = \frac{1}{2}  (\alpha' + \alpha_\mathrm{eff} ) \left [   \left ( \frac{m R }{\Delta_k L } \right )^2   \left ( \phi_1 - \phi_2 \right ) ^2 \sum_{n_1, n_2} \frac{J_1( R n \Delta_k )^2 }{n^2 \left ( m^2 \Delta_k^{-2} + n^2 \right ) } + \phi_2^2 \right ] 
\end{eqnarray}
where $k = \left (k_1^2 + k_2^2 + \sqrt{1 - s^2} k_1 k_2 \right )^{1/2}$, $n = \left ( n_1^2 + n_2^2 + \sqrt{1 - s^2}n_1 n_2 \right )^{1/2}$. The local ($\phi^4$) part of the free energy can be calculated as for the $c=0$ case, 
\begin{equation}
\fl V^{-1} \mathcal{F}_\mathrm{loc} = f_\mathrm{loc}(\phi_1)  \frac{ \pi R^2}{s L^2}  + f_\mathrm{loc}(\phi_2) \left ( 1 -  \frac{ \pi R^2}{s L^2}  \right ) + \sqrt{ \frac{\alpha \kappa}{18}} (\phi_1 - \phi_2)^2 \frac{2 \pi}{s L } \frac{R}{L} 
\end{equation}
where $f_\mathrm{loc}$ is the local free energy. We can see that there are four parameters: $\phi_1, \phi_2, L, R/L$. However since the total amount of reaction must sum to zero in equilibrium, $R/L \equiv \zeta $ is a function of $\phi_1, \phi_2$. We will also write $m/\Delta_k \equiv \chi$. The non-local part simplifies, 
\begin{equation}
 V^{-1} \mathcal{F}_\mathrm{nl} = \frac{1}{2} (\alpha' + \alpha_\mathrm{eff} ) \left [  \chi^2 \zeta ^2   \left ( \phi_1 - \phi_2 \right ) ^2 \sum_{n_x, n_y} \frac{J_1( \frac{2 \pi}{s} \zeta |n|  )^2 }{|n|^2 \left ( \chi^2 + |n|^2 \right ) } + \phi_2^2 \right ] 
\end{equation}
So does the local part of the free energy 
\begin{equation}
 V^{-1} \mathcal{F}_\mathrm{loc} = f_\mathrm{loc}(\phi_1)  \frac{\pi}{s} \zeta^2 +  f_\mathrm{loc} (\phi_2) \left ( 1- \frac{\pi}{s} \zeta^2 \right ) + \sqrt{\frac{2 \alpha \kappa}{9}} (\phi_1 - \phi_2)^2 \frac{\pi \zeta}{sL} 
\end{equation}
The total free energy for a square lattice $(s=1)$ and a hexagonal lattice $(s = \sqrt{3}/2)$ with the same $(\phi_1, \phi_2)$ are shown in figure \ref{fig:tot_f} and we can indeed see that the minimum of the free energy occurs at a finite length, where the long range screened Coulomb interaction dominates the short range interactions. This minimum for a hexagonal lattice is then plotted in figure \ref{fig:min} for various values of $(\phi_1, \phi_2)$. 

\section{Lattice model of active bacteria with birth-death}
\label{ap:lattice} 
In this section, we construct a lattice model for active bacteria with birth-death
and coarse grain into a  stochastic partial differential equation. First we perform the van Kampen system size expansion \cite{gardiner, vankampen} of the birth-death process in the well mixed limit to illustrate the method. Next, following \cite{lefevre}, we arrive at the hopping rates that give the right continuum limit for a single Active Brownian particle. Finally, we combine the hopping rates with the birth-death rates and perform a system size expansion on the entire lattice model before taking the continuum limit. 

This procedure is similar to the explicit coarse graining presented in the appendix of Grafke et al. \cite{grafkePRL2017} with the following key differences: (1) we perform the van Kampen system size expansion before taking the continuum limit and hence arrive at an approximate SPDE rather than the exact path integral with Poisson noise obtained by Grafke et al.~; (2) We take the phenomenological approach of looking at the equivalent lattice process of a single active particle and extracting the lattice hopping rates whereas Grafke et al.\ represent run-and-tumble particles with left/right-movers and coarse-grained them separately before combining into a single density. The end results are the same, although our treatment of the active diffusion extends more easily to higher dimensions. 

\subsection{Birth-death process in the well mixed limit}

The birth-death process can be represented by the following chemical equation with birth rate $\lambda_\mathrm{b}$ and death rate $\lambda_\mathrm{d}$, 
\begin{equation}
A + A \leftrightarrow A 
\end{equation}
Physically this corresponds to birth by cell division and death by overcrowding. In the well mixed limit, we assume that the diffusion of particles is sufficiently fast and uniform that the density distribution is always homogeneous. Denoting $P_n$ as the probability of having $n$ particles in the system, the master equation for the evolution of the probabilities is 
\begin{equation}
\partial_t P_n = \lambda_\mathrm{b} \left [ (n-1) P_{n-1} - n P_n \right ] + \frac{\lambda_\mathrm{d}}{N} \left [ (n+1)n P_{n+1} -  n(n-1) P_n \right ]
\end{equation}
where $N$ is an extensive variable characterising the typical number of particles in the system. We divide $\lambda_\mathrm{d}$ by this extensive parameter such that the two rate constants have the same scaling with system size -- this will become more obvious once we write down the deterministic rate equation later. We can write the master equation in an equivalent form by defining a translation operator $\mathbb{E}(a)$ that acts on any function $f(n)$ as $\mathbb{E}(a) f(n) = f(n+a)$ 
\begin{equation}
\partial_t{P_n} = \lambda_\mathrm{b} \left [ \mathbb{E}(-1) - 1\right] n P_n +\left ( \lambda_\mathrm{d}/ N \right ) \left [ \mathbb{E}(1)- 1 \right ] n(n-1) P_n  
\end{equation}
To make connection with macroscopic quantities, we will change variable to a new scaled density $\rho= n/N$. The probabilities in terms of $\rho$ evolves as follows, 
\begin{equation}
\fl \partial_t P(\rho, t) = N \lambda_\mathrm{b} \left [ \mathbb{E}(- 1/N) - 1 \right] \rho P(\rho, t) + N \lambda_\mathrm{d} \left [ \mathbb{E}(1/N)- 1 \right ]\rho \left ( \rho - \frac{1}{N} \right ) P(\rho, t) 
\end{equation}
Recall from quantum mechanics that the momentum operator is the generator of spatial translation. We can define a similar conjugate operator here as $\hat{\rho} = - i \partial_\rho$ after taking $\rho$ to be continuous \footnote{In \cite{lefevre}, they have the conjugate variable $\hat{n} = i \hat{\rho}$ and explicitly integrate over imaginary $\hat{n}$ in the path integral. }. Then the translation operator $\mathbb{E}(a) = e^{i\hat{\rho} a} = 1 + a \partial_\rho + \frac{a^2}{2} \partial^2_\rho \dots$ Now treating $1/N$ as small, we can expand the exponentials. This is the Kramers-Moyal expansion \cite{gardiner, vankampen}. To second order, the master equation becomes a Fokker-Planck equation, 
\begin{equation} 
\label{eq:birth-death}
\fl \partial_t P(\rho, t) =   -  i \hat{\rho} ( \lambda_\mathrm{b} \rho -  \lambda_\mathrm{d} \rho (\rho - 1/N ) ) P -   \frac{\hat{\rho}^2}{2N} (\lambda_\mathrm{b} \rho  + \lambda_\mathrm{d} \rho (\rho - 1/N ))  P  + O(\hat{\rho}^3 N^{-2}P)
\end{equation}
Note we cannot say that this Fokker-Planck equation is accurate to $O(N^{-2})$ without knowing the shape of $P(\rho, t)$.  For example, if $P(\rho, t)$ is a Gaussian distribution with width $\approx 1/\sqrt{N}$, $\hat{\rho} P \approx \sqrt{N} P$ and $\hat{\rho}^2 P \approx P/N$. We can self-consistently perform a perturbative expansion of $P(\rho, t)$ by assuming that $P(\rho, t)$ is Gaussian and justify \textit{a posteriori}. This yields the van Kampen system size expansion $\rho = \rho_0(t) + \rho_1(t)/\sqrt{N}$, with $\rho_0(t)$ as the deterministic trajectory and $\rho_1(t)$ the lowest order variation,  
\begin{eqnarray*}
\dot{\rho_0} = \lambda_\mathrm{b} \rho_0 - \lambda_\mathrm{d} \rho_0^2 \\
\partial_t P(\rho_1, t) = - i (\lambda_\mathrm{b} \rho_0 - 2 \lambda_\mathrm{d} \rho_0^2) \hat{\rho}_1 \rho_1 P + \frac{1}{2} (\lambda_\mathrm{b} \rho_0 + \lambda_\mathrm{d} \rho_0^2 ) \hat{\rho}_1^2 P 
\end{eqnarray*}
where, as before, $\hat{\rho}_1 = - i \partial_{\rho_1}$ is the conjugate momentum of $\rho_1$. If we wish to know higher order variations we would need to do the Kramers-Moyal expansion to higher order and perform the perturbative calculation order by order. Note that the first order van Kampen system size expansion is equivalent to the first order low noise expansion in $\sqrt{1/N}$ of the It\^o stochastic differential equation, 
\begin{equation}
\partial_t \rho = \lambda_\mathrm{b} \rho - \lambda_\mathrm{d} \rho^2 + \sqrt{ \left ( \lambda_\mathrm{b} \rho + \lambda_\mathrm{d} \rho^2 \right ) / N } \Lambda 
\end{equation}
where $\Lambda$ is a white noise. 

\subsection{Active Particles} 
Active particles can be described by the It\^o Langevin equation \cite{PRLTailleurCates}
\begin{equation}
\dot{\boldsymbol{r}} = \left ( \boldsymbol{V} + \bnabla D \right ) + \sqrt{2 D} \bLambda 
\end{equation}
where $\boldsymbol{V} = - \tau v_0 \bnabla v_0 / d $, $D = \tau v_0^2 / d $, $v_0(r)$ is the swim speed, $\tau$ is the tumbling time and $d$ is the dimension of the system. Define $\mu =  \log v_0$ such that $D \bnabla \mu = \boldsymbol{V} + \bnabla D$. We will next show that the above SDE is equivalent to the lattice diffusion model with diffusion rate $\omega_{ij} = \frac{D_i}{a^2}  \exp ( - \frac{f_{ij}}{2})$ where $f_{ij} = - ( \mu_j - \mu_i)/a$ is the force vector from site $i$ to site $j$ and $a$ is the lattice constant. 

The master equation for 1D is 
\begin{equation}
\partial_t P_i =  \omega_{i+1, i} P_{i+1} + \omega_{i-1, i} P_{i-1} - \left ( \omega_{i, i+1} + \omega_{i, i-1} \right ) P_i 
\end{equation}
Let $x = a n$. Using the translation operator $E(1) = \exp(i a p)$ where $p = - i \partial_x$, we can write the master equation as 
\begin{eqnarray}
\fl \partial_t P_x = \left ( e^{iap}  - 1 \right ) \omega_{x, x-a} P_x+ (e^{-iap} - 1) \omega_{x, x+a} P_x \\
\fl \qquad = \left [ \left ( iap - \frac{a^2}{2} p^2 + O(a^3)\right ) \omega_{x, x-a} + \left ( - i a p - \frac{a^2}{2} p^2 + O(a^3) \right ) \omega_{x, x+1} \right ] P_x \\
\fl \qquad = \left [ - i a p \left ( \omega_{x,x+a} - \omega_{x, x-a} \right ) - \frac{a^2}{2} p^2 \left ( \omega_{x,x+a} + \omega_{x, x-a} \right ) + O(a) \right ] P_x
\end{eqnarray}
Now we compute the symmetric and antisymmetric parts of the diffusion rates: 
\begin{eqnarray*}
w_{x,x+a} - w_{x, x-a} & = \frac{D_x}{ 2 a^2} (f_{x, x-a} - f_{x, x+a}) = - \frac{D_x}{a} \partial_x \mu_x + O(a^0)\\
w_{x,x+a} + w_{x, x-a} &= \frac{2 D_x}{a^2} + O(a^0)
\end{eqnarray*}
Collecting the terms and taking $a \rightarrow 0$, we obtain 
\begin{eqnarray*}
\partial_t P &= - \partial_x \left [ - ( \partial_x \mu) D P - \partial_x (D P ) \right ] \\
& = -  \partial_x \left [ f D P - \partial_x ( D P ) \right ] 
\end{eqnarray*}
This is indeed the Fokker-Planck equation corresponding to the It\^o stochastic differential equation, as promised above. 

\subsection{Minimal model} 

For multiple particles hopping on a lattice, we will use the same hopping rates $\omega_{i, j}$ as found in the previous section, but make $v_0$ depend on the density and its gradients. 
\begin{equation}
\eqalign{
\partial_t P(\{n_i\}, t) =\left (  \mathcal{L}_\mathrm{bd} + \mathcal{L}_\mathrm{diff} \right) P(\{n_i\}, t)  \\
\mathcal{L}_\mathrm{bd} = \sum_{i}  \lambda_\mathrm{b} \left [ \mathbb{E}_i(-1) - 1 \right] n_i +\left ( \lambda_\mathrm{d}/ N \right ) \left [ \mathbb{E}_i(1)- 1 \right ] n_i(n_i-1) \\
\mathcal{L}_\mathrm{diff} =   \sum_i \left [ \mathbb{E}_i(1) \mathbb{E}_{i-1}(-1) - 1 \right ] \omega_{i, i-1} n_i + \left [ \mathbb{E}_i(-1) \mathbb{E}_{i-1}(1) - 1 \right ] \omega_{i-1, i} n_{i-1}  
}
\end{equation}
Change variable to $\rho_i = n_i/N$ as before, and let $\hat{\rho}_i = - i \partial_{\rho_i}$
\begin{equation}
\eqalign{
\fl \mathcal{L}_\mathrm{diff} = \sum_i \left [  i (\hat{\rho}_i - \hat{\rho}_{i-1} ) - \frac{1}{2N} (\hat{\rho}_i - \hat{\rho}_{i-1})^2 \right ] \omega_{i, i-1} \rho_i  \\
+\sum_i  \left [ - i  (\hat{\rho}_i - \hat{\rho}_{i-1} ) - \frac{1}{2N} (\hat{\rho}_i - \hat{\rho}_{i-1})^2 \right ] \omega_{i-1, i} \rho_{i-1}  \\ 
\fl \qquad = \sum_i  i (\hat{\rho}_i - \hat{\rho}_{i-1} ) \left ( \omega_{i, i-1} \rho_i - \omega_{i-1, i} \rho_{i-1} \right ) 
- \frac{1}{2N} (\hat{\rho}_i - \hat{\rho}_{i-1} )^2 \left ( \omega_{i, i-1} \rho_i + \omega_{i-1, i} \rho_{i-1} \right ) 
}
\end{equation}
Using the diffusion rates from before, we have 
\begin{eqnarray}
\omega_{i, i-1} \rho_i - \omega_{i-1, i} \rho_{i-1} &= \frac{1}{a} \partial_x ( D \rho) + \frac{1}{a} (\partial_x \mu_i) D_i \rho_i + O(a^0) \\
\omega_{i, i-1} \rho_i + \omega_{i-1, i} \rho_{i-1} &= \frac{2}{a^2} D \rho + O(1/a) 
\end{eqnarray}
Collecting the terms 
\begin{equation}
\eqalign{
\mathcal{L}_\mathrm{diff} &= \sum_i \frac{ i (\hat{\rho}_i - \hat{\rho}_{i-1} ) }{a}  \left [  \partial_x (D_i \rho_i ) + (\partial_x \mu_i ) D_i \rho_i  + O(a) \right ] \\
&\qquad  - \sum_i \frac{(\hat{\rho}_i - \hat{\rho}_{i-1})^2}{2Na^2} \left ( 2 D \rho + O(a) \right ) \\
 &= \sum_i - i \hat{\rho}_i  \partial_x \left [  \partial_x (D_i \rho_i ) + (\partial_x \mu_i) D_i \rho_i  + O(a) \right ] \\
 &\qquad - \frac{1}{2N}  \sum_{i, j} \hat{\rho}_i \hat{\rho}_j \left [ -  \partial_x   \left ( 2 D \rho + O(a) \right ) \partial_x \right ]_{ij} 
  }
\end{equation}
Recall that the birth-death part is 
\begin{equation}
\mathcal{L}_\mathrm{bd} =- \sum_{i} \left [  i \hat{\rho}_i  ( \lambda_\mathrm{b} \rho_i -  \lambda_\mathrm{d} \rho_i (\rho_i - 1/N ) )  +  \frac{\hat{\rho}_i^2}{2N} (\lambda_\mathrm{b} \rho_i  + \lambda_\mathrm{d} \rho_i (\rho_i - 1/N ))  \right ] 
\end{equation}
As before, we perform a system size expansion, $\rho_i = \rho^0_i + \sqrt{\epsilon} \rho^1_i$ where $\epsilon = 1/N$. The zeroth order and the first order equation are the same as the low noise expansion of the following SDE 
\begin{equation}
\partial_t \rho_i =  \lambda_\mathrm{b} \rho_i - \lambda_\mathrm{d} (\rho_i)^2  + \partial_x \left [  \partial_x (D_i \rho_i ) + (\partial_x \mu_i)  D_i \rho_i  + O(a) \right ] + \Lambda_i 
\end{equation}
where $\Lambda_i$ is a Gaussian noise with variance 
\begin{equation}
\langle \Lambda_i(t) \Lambda_j(s) \rangle = \epsilon \delta(t - t') \left [  \left ( - \partial_x 2 D \rho \partial_x \right)_{ij} + \delta_{ij} \left (  \lambda_\mathrm{b} \rho_i  + \lambda_\mathrm{d} \rho^2_i \right)  + O(a) \right ] 
\end{equation}
Taking the continuum limit $\rho' = \rho/a, \lambda_\mathrm{d}' = a \lambda_\mathrm{d} $ (then relabel back) and $a \rightarrow 0$, we find
\begin{equation}
\eqalign{
\partial_t \rho = \lambda_\mathrm{b} \rho - \lambda_\mathrm{d} \rho^2  + \partial_x \left [  \partial_x (D \rho ) + (\partial_x \mu)  D \rho \right ] + \Lambda \\
 \langle \Lambda(x, t) \Lambda(y, s) \rangle = \epsilon \left [ \partial_x \partial_y 2 D \rho  + \lambda_\mathrm{d} \rho + \lambda_\mathrm{d} \rho^2  \right ] \delta(x - y) \delta(t - s) 
 }
\end{equation}
We can see that we have conservative phase separation with the chemical potential $\mu$ as well as non-conservative dynamics. The above equation can be easily generalised to higher dimensions with careful treatment of the lattice diffusion terms \cite{lefevre}. The result is the same as replacing the spatial derivatives with vector gradient operators and taking the appropriate dot products. 

Now we perturbatively expand $\rho = \rho_\mathrm{0} ( 1 + \phi) $  and only keep the lowest terms in gradients and $\phi$ for both the conservative and non-conservative part, we obtain, in vector notations, 
\begin{equation}
\eqalign{
\partial_t \phi = M \nabla^2 \mu - u (\phi + 1) (\phi - \phi_\mathrm{t}) +  \Lambda \\
\mu = - \alpha \phi + \beta \phi^3  - \kappa \nabla^2 \phi \\
\langle \Lambda(\boldsymbol{x}, t) \Lambda(\boldsymbol{y}, s) \rangle = \left [ - 2 \epsilon M \bnabla_x \boldsymbol{\cdot} \bnabla_y + \epsilon u (2 + \phi_\mathrm{t} ) \right ] \delta ( \boldsymbol{x} - \boldsymbol{y} ) \delta ( t - s) 
}
\end{equation}

We can also freely rescale $\phi$ by any factor to adjust the relative magnitude of $\alpha$ and $\beta$, and after relabelling the parameters, this gives
\begin{equation}
\eqalign{
\partial_t \phi = M \nabla^2 \mu - u (\phi - \phi_\mathrm{a}) (\phi - \phi_\mathrm{t}) +  \Lambda \\
\mu = - \alpha \phi + \beta \phi^3  - \kappa \nabla^2 \phi \\
\langle \Lambda(\boldsymbol{x}, t) \Lambda(\boldsymbol{y}, s) \rangle = \left [ - 2 \epsilon M \bnabla_x \boldsymbol{\cdot} \bnabla_y + \epsilon u (- 2\phi_\mathrm{a} + \phi_\mathrm{t} ) \right ] \delta ( \boldsymbol{x} - \boldsymbol{y} ) \delta ( t - s) 
}
\label{eq:full}
\end{equation}
This is of the form of the canonical model proposed in equation (\ref{eq:full_sto_eq}) with additional relations between the constants. 

\section{Self consistency condition for the amplitude equation}
\label{ap: amplitude_eq} 

This appendix completes the discussion on the amplitude equation in section \ref{amplitude_eq}.

Starting with the deterministic part of the full equation (\ref{eq:full_sto_eq}) and the growth rate of the linear terms (\ref{eq:rescaled_linear}), we will split the equation into linear and nonlinear parts. Note that we can still rescale both time and space as we have assumed the noise to be small and only work with the deterministic part: $t' = t/ \tau, \, x' = x/ \xi$ using $\tau, \xi$ defined in equation (\ref{eq:rescaled_linear}). For convenience we will also relabel $\delta \phi \rightarrow \phi$, 
\begin{equation}
\partial_t \phi = - \left [ \nabla^2 + \lambda^2 \right ]^2 \phi + \Delta \phi + \delta_1 \nabla^2 \phi^2 + \delta_2 \nabla^2 \phi^3  - \delta_3 \phi^2 
\label{eq:amplitude} 
\end{equation} 
where $\lambda = q_\mathrm{c} \xi$, $\delta_1 = 3 \beta \phi_t \tau \xi^{-2}, \, \delta_2 = \beta \tau \xi^{-2} $ and $\delta_3 = M_\mathrm{A} u \tau $. 

Close to the threshold ($\Delta$ small), the band of unstable modes is narrowly centred at $\lambda$, leading to a sinusoidal pattern at wavelength $\approx 1/\lambda$ with its amplitude modulated on a much larger length scale. According to linear stability analysis, the width of the band of growing modes is roughly scales as $\Delta^{1/2}$ so we expect the length scale of the slow spatial modulation to scale like $\Delta^{-1/2}$. Similarly, the growth rate scales as $\Delta$, implying that the characteristic time scale of growth is proportional to $\Delta^{-1}$. We will now separate the two scales by defining the slow variables $\boldsymbol{X}= \boldsymbol{x} \Delta^{-1/2}, T = t \Delta^{-1}$, and expanding $\phi$ perturbatively in powers of $\Delta^{1/2}$ to match the spatial and temporal scaling,  
\begin{equation}
\phi(\boldsymbol{X}, T, \boldsymbol{x}, t) = \Delta^{1/2} \phi_0 + \Delta \phi_1 + \Delta^{3/2} \phi_2
\end{equation} 
By the chain rule,  $\partial_t \rightarrow \partial_t + \Delta \partial_T$ and $\bnabla\rightarrow \bnabla_x + \Delta^{1/2} \bnabla_X$.  For convenience, we also define $L_x = \nabla_x^2 + \lambda^2 $. In terms of the slow and fast variables, $L_x \rightarrow L_x + 2\Delta^{1/2} \bnabla_x \cdot \bnabla_X + \Delta \nabla_X^2$ and $L_x^2 \rightarrow L_x^2 + 4\Delta^{1/2} \bnabla_X \cdot \bnabla_x L_x + \Delta (2L_x + 4 \nabla_x^2)\nabla_X^2$. Now we can write the entire equation in terms of the slow and fast variables. Keeping everything to at most $O\left(\Delta^{3/2}\right)$,  
\begin{equation}
\eqalign{
\fl \Delta \partial_T (\Delta^{1/2} \phi_0) =& - \left [ L_x^2 + 4 \Delta^{1/2} \bnabla_X \cdot \bnabla_x L_x + \Delta (2L_x + 4 \nabla_x^2)\nabla_X^2 \right ] (\Delta^{1/2} \phi_0 + \Delta \phi_1 + \Delta^{3/2} \phi_2) \\
& + \Delta^{3/2} \phi_0 \\
& + \delta_1 \left ( \nabla_x^2 + 2 \Delta^{1/2} \bnabla_x \cdot \bnabla_X \right ) (\Delta^{1/2} \phi_0 + \Delta \phi_1)^2 \\
& + \delta_2 \nabla_x^2 (\Delta^{1/2} \phi_0)^3 \\
& - \delta_3 (\Delta^{1/2} \phi_0 + \Delta \phi_1)^2 + O(\Delta^2)  						
}
\label{eq:amp_eq_full}
\end{equation}
To order $\Delta^{1/2}$, we get the linear equation at the threshold, 
\begin{equation}
	(\nabla_x^2 + \lambda^2) \phi_0 = 0 \Rightarrow \phi_0 = A(\boldsymbol{X}, T) \exp(\rmi \boldsymbol{\lambda} \cdot \boldsymbol{x} ) + \mathrm{c.c.}
\end{equation}
where $\boldsymbol{\lambda}$ is a vector with length $\lambda$ and consistent with the boundary conditions, and c.c. means complex conjugate. Collecting together the $O(\Delta)$ terms of equation (\ref{eq:amp_eq_full}), 
\begin{equation}
\eqalign{
	L_x^2 \phi_1 & = - 4 L_x \bnabla_x \cdot \bnabla_X \phi_0 + \delta_1 \nabla_x^2 \phi_0^2 - \delta_3 \phi_0^2 \\ 
	& =  - \left ( 4\delta_1 \lambda^2 + \delta_3 \right )  A(\boldsymbol{X}, T)^2 \exp(2 \rmi \boldsymbol{\lambda} \cdot \boldsymbol{x}) + \mathrm{c.c.}
}
\end{equation}
We can substitute in the \textit{lhs} to check that the following is a solution, 
\begin{equation}
\phi_1(\boldsymbol{X}, T, \boldsymbol{x}, t) = -  \frac{4 \delta_1 \lambda^2 + \delta_3 }{9 \lambda^4}  A(\boldsymbol{X}, T)^2 
 \exp(2 \rmi \boldsymbol{\lambda} \cdot \boldsymbol{x}) + \mathrm{c.c.} 
\end{equation}
To order $\Delta^{3/2}$, the lowest Fourier components are 
\begin{equation}
\fl \partial_T \phi_0 = - \left (2 L_x + 4 \nabla_x^2 \right ) \nabla_X^2 \phi_0 + \phi_0 + 2\delta_1 \nabla_x^2 (\phi_0\phi_1) + \delta_2 \nabla_x^2 \phi_0^3 - 2 \delta_3 \phi_0 \phi_1 - L_x^2 \phi_2 
\end{equation} 
Collecting all the $\exp(\rmi \boldsymbol{\lambda} \cdot \boldsymbol{x})$ terms and noting that $L_x^2$ annihilates any such term in $\phi_2$, the amplitude equation is
\begin{equation}
\eqalign{
\partial_T A = 4 \lambda^2 \nabla^2_X A + A - g |A|^2 A, \\
g = 3 \lambda^2 \delta_2 - \frac{(2 \delta_1 \lambda^2 + 2 \delta_3) (4 \delta_1 \lambda^2 + \delta_3) }{9 \lambda^4}
}
\end{equation}

For self-consistency, we require $g>0$ so the linear growth is suppressed by higher order terms and the amplitude has a finite stationary value, implying that
\begin{equation}
27 \lambda^6 \delta_2 > (2 \delta_1 \lambda^2 + 2 \delta_3 )(4 \delta_1 \lambda^2 + \delta_3)  
\end{equation}
Recall that $\delta_1 = 3 \phi_\mathrm{t} \delta_2$ and $\delta_2 = \beta \tau \xi^{-2} > 0$.  Substituting $ z = 3 \lambda^2 \delta_2$, we obtain 
\begin{equation}
8 \phi_\mathrm{t}^2 z^2 + (10 \phi_\mathrm{t} \delta_3 - 9 \lambda^4) z + 2 \delta_3^2 < 0 
\label{eq:inequality}
\end{equation}
Note that $z$ is by definition always positive and at $z=0$ the \textit{lhs} evaluates to $2\delta_3^2 > 0$. Treating $lhs=0$ as a quadratic equation in $z$, the \textit{lhs} has negative parts if and only if both roots of the quadratic equation are positive. This requires 
\begin{equation}
\eqalign{
10 \phi_\mathrm{t} \delta_3 - 9 \lambda^4 < 0 \\
(10 \phi_\mathrm{t}\delta_3- 9\lambda^4 )^2 - 64 \phi_\mathrm{t}^2\delta_3^2 > 0 
}
\end{equation}
Combining the two inequalities yields the first condition, 
\begin{equation}
\phi_\mathrm{t} \delta_3 < \lambda^4 / 2 
\end{equation}
Provided that the above is true, the inequality (\ref{eq:inequality}) holds for $z_- < z < z_+$, where $z_\pm$ are the two roots of the quadratic equation. Writing in terms of $\delta_2 = z / 3 \lambda^2 $, this becomes $\delta_- < \delta_2 < \delta_+$, where 
\begin{equation}
\delta_\pm = \frac{- (10 \phi_\mathrm{t} \delta_3 - 9 \lambda^4 ) \pm \sqrt{(10 \phi_\mathrm{t}\delta_3- 9\lambda^4 )^2 - 64 \phi_\mathrm{t}^2\delta_3^2}}{48 \lambda^2 \phi_\mathrm{t}^2}
\end{equation}
To grasp the physical implications, we will look at the simplified case of $\phi_\mathrm{a} \gg \phi_\mathrm{B} = 1$. The first condition is automatically true, and the second condition reduces to, 
\begin{equation}
\delta_2 < \frac{3 \lambda^2 }{8 \phi_\mathrm{t}^2 }
\end{equation}
Recall $\delta_2 = \beta \tau \xi^{-2} $ and $\lambda^2 =  q_\mathrm{c}^2 \xi^2$, the expression can be written in a rather simple form
\begin{equation}
\phi_\mathrm{t}^2< \frac{3}{25} \phi_\mathrm{B}^2
\end{equation}
where recall $\phi_\mathrm{B}$ is the binodal density defined as $\phi_\mathrm{B} = \sqrt{\alpha/\beta}$. This is a more stringent constraint than the local instability condition. Recall that the homogeneous solution is only unstable when $\tilde{\alpha} > 4 \kappa \tilde{u}$. So for small $\tilde{u}$, we have a regime where the homogeneous solution is unstable but the scheme cannot be made self-consistent, meaning that the solution is not well modelled by a sine wave with small amplitude around a homogeneous background state of density. 

\section{Stability of droplet suspension} 
\label{ap:mult_droplet} 
In this section, we will look at the stability of a suspension of $N$ droplets in volume $V$ with periodic boundary conditions. Consider a perturbation of the form $R_i = R_\mathrm{s} + \delta R_i$, where $R_\mathrm{s}$ is the stationary radius for an $N$-droplet suspension, i.e.\ the stable fixed point of equation (\ref{eq:mult_droplet}).  In addition, let the current on the outside the $i$th droplet be $J_-^i = J(R_\mathrm{s}) + \delta J_-^i$ where $\delta J_-^i$ is first order in $\{ \delta R_i \}$. Expand equation (\ref{eq:mult_current}) to first order in $\delta R_i$, 
\begin{equation}
( 1 + a) \delta J_i +  a \sum_{j \neq i } \delta J_j= b \, \delta R_i + c \sum_{j \neq i} \delta R_j 
\label{eq:vector_eq} 
\end{equation}
where in this appendix 
\begin{equation}
\eqalign{ 
\fl a = - \frac{2 \lambda D k_-}{ N g_-^1 R} \frac{K_1}{K_0} \\ 
\fl b = - D k_-^2  \left ( 1 - \frac{1}{k_-R} \frac{K_1}{K_0} + \frac{K_1^2}{K_0^2} \right )\left [ (1 - \lambda) c_- ^0 - \frac{\gamma}{R}  - \frac{2 \lambda J }{g_-^1 R } \right ] \\
- D k_-  \frac{K_1}{K_0} \left [ \frac{\gamma}{R^2}  - \frac{2 \lambda}{N R}  \left ( c_-+ \frac{2J}{g_-^1 R} \right ) \right ] \\
\fl c = \frac{2 \lambda D k_-}{N R} \frac{K_1}{K_0} \left ( c_- + \frac{J}{g_-^1 R } \right ) 
 } 
\label{eq:definitions} 
\end{equation}
Here $c_- = - g_-^0/g_-^1$ as before and $c_-^0 = c_- ( 1 - \lambda) - (g_-^1 V)^{-1} N 2 \pi R J$ is the density at infinity seen by the unperturbed droplets (see equation (\ref{eq:mult_current})). Since equation (\ref{eq:vector_eq}) is true for every $i$, we have $N$ equations with $N$ variables, which can be written in the vector equation form 
\begin{equation}
\boldsymbol{M}_1 \delta \boldsymbol{J}_- = \boldsymbol{M}_2 \delta \boldsymbol{R} 
\end{equation}
The currents on the inside of the droplets are unaffected by the presence of neighbouring droplets, implying that $\delta \boldsymbol{J}_+ = z \delta \boldsymbol{R}$ where $z$ is obtained by expanding equation (\ref{eq:current}) to first order in $\{ \delta R_i \}$, 
\begin{equation}
z =  D k_+ \left [ k_+ \left ( c_+ - \frac{\gamma}{R} \right ) \left ( 1 - \frac{1}{k_+ R} \frac{I_1}{I_0} - \frac{I_1^2}{I_0^2} \right ) + \frac{\gamma}{R^2} \frac{I_1}{I_0} \right ]
\end{equation} 
Putting the currents together for the $i$th droplet, we can get the growth rate $\partial_t \delta R_i$, 
\begin{equation}
\eqalign{
2 \phi_\mathrm{B} \partial_t \delta \boldsymbol{R} &=\delta  \boldsymbol{J}_+ - \delta \boldsymbol{J}_- \\
& = \left ( z  \boldsymbol{I} - \boldsymbol{M}_1^{-1} \boldsymbol{M}_2 \right ) \delta \boldsymbol{R} 
}
\end{equation}
Notice that both $\boldsymbol{M}_1$ and $\boldsymbol{M}_2$ are of a special form: $(\boldsymbol{M}_1)_{ij} = ( 1 + 2a)\delta_{ij} - a$ and  $(\boldsymbol{M}_2)_{ij} = (b + c) \delta_{ij} - c$. They can be simultaneously diagonalised and the vector equation can be solved separately in the eigen-space of the two matrices. The (unnormalised) eigenvalues are of two types, 
\begin{equation}
\eqalign{ 
\boldsymbol{v}_1 = \left ( 1, 1, ... , 1 \right ) \\
\boldsymbol{v}_2 = \left (-1, \frac{1}{N-1}, ... , \frac{1}{N-1} \right ) 
} 
\end{equation}  
The first one is the synchronised growth of all the droplets, and the second one is an exchange of flux: mass flows from one droplet into all the rest. The first eigenmode is always stable provided that the stationary radius exists in the first place -- in fact, its eigenvalue is the gradient of the \textit{rhs} of equation (\ref{eq:mult_droplet}) with respect to $R$, which is always negative at a stable fixed point. Therefore we are only interested in the second set of eigenvectors. Let 
\begin{equation}
\eqalign{
\delta \boldsymbol{R} = \delta R  \left ( -1, \frac{1}{N-1}, ... , \frac{1}{N-1} \right ) \\
\delta \boldsymbol{J}_- = \delta J_-  \left ( -1, \frac{1}{N-1}, ... , \frac{1}{N-1} \right ) 
} 
\end{equation}
We can solve the matrix equation to obtain that 
\begin{equation}
\delta J_- = (b - c )  \delta R 
\end{equation}
Therefore the eigenvalue of the flux exchange mode is found as
\begin{equation}
\eqalign{
\partial_t \delta R =\omega_+ \delta R \\
\omega_+ = (2 \phi_\mathrm{B})^{-1} ( z - b + c) 
}
\end{equation}
where we recall that $b, c$ are defined in (\ref{eq:definitions}). The eigenvalue $\omega_+$  is plotted as a function of $R$ along with $\dot{R}$ in Figure (\ref{fig:mult_droplet}), and we refer to section \ref{droplet_sus} for further discussions of these results. 

\section{Droplet splitting} 
\label{ap: droplet_splitting} 
Starting with equation (\ref{eq:linear2}), we will consider the stability of anisotropic perturbations around a droplet of radius $R$ in an infinite bath of the form in equation (\ref{eq:angular_pert}), with the following matching conditions at the interface set by the local curvature $H(R(\theta), \theta)$:  
\begin{equation}
\eqalign{
\psi_\pm(R(\theta) , \theta) = \gamma H(R(\theta), \theta) 
&= \gamma \frac{\left | R^2 + 2 (\partial_\theta R)^2 - R \partial_\theta^2 R \right |}{\left (R^2 + (\partial_\theta R)^2 \right )^{3/2}  } \\
&= \frac{\gamma}{\bar{R}} \left [ 1 + \frac{1}{\bar{R}} \sum_{l=1}^\infty (l^2 - 1) \delta R_l \right ] 
}
\label{eq:ap_boundary}
\end{equation}
The problem simplifies in polar coordinates because the angular degree of freedom separates from the radial one. Write $\psi_\pm(r, \theta) = \Psi_\pm(r) \Theta_\pm (\theta) $, equation (\ref{eq:linear2}) reduces to 
\begin{equation}
r^2 \Psi_\pm''/ \Psi_\pm + r \Psi_\pm' / \Psi_\pm +  r^2 \frac{g_\pm^1}{D}  = - \Theta_\pm '' / \Theta_\pm
\end{equation}
where prime denotes differentiation with respect to the single argument. Note that all the $r$-dependence is on the \textit{lhs} while the \textit{rhs} is only a function of $\theta$, hence they must both equal the same constant. Denoting the constant as $l_\pm^2$, 
\begin{equation}
\eqalign{ 
\Theta_\pm ''  = - l_\pm^2 \Theta_\pm \\
r^2 \Psi_\pm'' + r \Psi_\pm ' -   r^2 k_\pm^2  \Psi_\pm = l_\pm^2 \Psi_\pm 
} 
\end{equation}
where $k_\pm^2 = - g^1_\pm/D $. Solving the eigenvalue equations, 
\begin{equation}
\eqalign{
\Theta_\pm^l (\theta) &= a^\pm_l \cos(l \theta) + b^\pm_l \sin(l \theta) \\
\Psi_\pm^l (r) &= c^\pm_l  I_l ( k_\pm r) + d^\pm_l  K_l (k_\pm r) 
}
\end{equation}
Periodicity of $\theta$ enforces that $l \in \mathbb{N}$, and $I_l$ and $K_l$ are modified Bessel functions. Recall that $I_l(y) \rightarrow \infty$ as $y \rightarrow 0$ and $K_l(y) \rightarrow \infty$ as $y \rightarrow \infty$. As a result, the finiteness of $\psi_\pm$ in their respective domains requires that 
\begin{equation}
\eqalign{
\psi_+(r, \theta)  = c_+ + a_0^+ I_0 (k_+ r ) + \sum_{l=1}^\infty I_l (k_+ r) \left ( a_l^+ \cos(l \theta) + b_l^+ \sin (l \theta) \right ) \\
\psi_- (r, \theta) = c_- + a_0^- K_0 ( k_- r) + \sum_{l=1}^\infty K_l(k_- r) \left (  a_l^+ \cos(l \theta) + b_l^+ \sin (l \theta) \right)
}
\end{equation}
The constant term in equation (\ref{eq:linear2}) fixes $c_\pm = - g_\pm^0/g_\pm^1$. At the interface $R(\theta)$, to first order in $\delta R_l$, we have,
\begin{equation}
\eqalign{ 
\fl \psi_+(\theta) = c_+ + a_0^+ \left (  I_0 (k_+ \bar{R}) + k_+  I_0' (k_+ \bar{R}) 
\sum_{l=1}^\infty \delta R_l   \right )
 + \sum_{l=1}^\infty  I_l (k_+ \bar{R} )  \left ( a_l^+ \cos(l \theta) + b_l^+ \sin (l \theta) \right ) \\
\fl \psi_-(\theta) = c_- + a_0^- \left (  K_0 (k_- \bar{R}) + k_-  K_0' (k_- \bar{R}) \sum_{l=1}^\infty \delta R_l   \right ) + \sum_{l=1}^\infty  K_l (k_- \bar{R} ) \left ( a_l^- \cos(l \theta) + b_l^- \sin (l \theta) \right ) 
 } 
\end{equation}
After a rather long and tedious calculation, omitted here for brevity, the matching conditions in equation (\ref{eq:ap_boundary}) yields the following expressions for the coefficients, 
\begin{equation}
\eqalign{
a_0^+ = \frac{ \gamma / \bar{R} - c_+}{I_0 (k_+ \bar{R} ) } , \; a_0^- =  \frac{ \gamma / \bar{R} - c_-}{K_0 (k_- \bar{R} ) }   \\
a_l^\pm = \chi_l^\pm \delta R_l^\mathrm{c}, \; b_l^\pm = \chi_l^\pm \delta R_l^\mathrm{s} \\
\chi_l^+ = \frac{\gamma (l^2 - 1) / \bar{R}^2 - a_0^+ k_+ I_0' ( k_+ \bar{R} ) }{I_l (k_+ \bar{R} ) } , \; 
\chi_l^- = \frac{\gamma (l^2 - 1) / \bar{R}^2 - a_0^- k_- K_0' ( k_- \bar{R} ) }{K_l (k_- \bar{R} ) }  
}
\end{equation}
Collecting the terms together, to $O(\delta R)$, the fields inside and outside the droplet are directly related to the angular perturbations $\delta R_l$, 
\begin{equation}
\eqalign{
\psi_+(r, \theta) = c_+ + a_0^+ I_0(k_+ r) + \sum_{l=1}^\infty \chi_l^+ I_l (k_+ r)\delta R_l  \\
\psi_- (r, \theta) = c_- + a_0^- K_0(k_- r) + \sum_{l=1}^\infty \chi_l^- K_l(k_- r ) \delta R_l 
}
\end{equation}
Next we will look at the current flow $ \boldsymbol{J}_\pm = - D \bnabla \psi_\pm$ across the droplet interface to see whether the perturbation grows. To lowest order in $\delta R$, the difference between the currents on the two sides of the interface is 
\begin{equation}
\eqalign{
\fl \frac{1}{D} \left [  \boldsymbol{J}_+(\theta) - \boldsymbol{J}_- (\theta) \right ]
= j_0(\bar{R})\, \hat{\boldsymbol{r}} + \sum_l \left [ j_l \left (\bar{R} \right )  \delta R_l  \, \hat{\boldsymbol{r}}  + h_l \left (\bar{R} \right ) (\partial_\theta \delta R_l)  \, \hat{\boldsymbol{\theta}} \right ] \\
j_0(\bar{R}) =  - \left ( \frac{\gamma}{\bar{R}} - c_+  \right )  k_+ \frac{I_1}{I_0} - \left ( \frac{\gamma}{\bar{R}} - c_- \right ) k_- \frac{K_1}{K_0} \\
j_l(\bar{R}) = f'(\bar{R}) - \chi_l^+ k_+ I_l'(k_+ \bar{R}) + \chi_l^- k_- K_l' (k_- \bar{R} ) \\
h_l(\bar{R}) = \frac{1}{\bar{R}}\left [ -  \chi_l^+ I_l (k_+ \bar{R}) + \chi_l^- K_l (k_- \bar{R} )   \right ]  =  \bar{R}^{-1} j(\bar{R}) 
}
\end{equation}
Recall that $j_0(\bar{R})$ governs the growth of the spherical droplet as in section \ref{droplet_sus}. In addition, the various terms cancel out in such a way that $h_l(\bar{R})$ is proportional to $j_0(\bar{R})$. Thus at the stable radius (the stable fixed point of $2 \phi_\mathrm{B} \partial_t R = j_0(R)$), the only remaining term is $j_l(\bar{R})$ and its sign determines whether the shape is stable. Using the properties of Bessel functions $I_l '(z) = I_{l-1}(z) - \frac{l}{z} I_l (z)$, $I_0' = I_1, K_0' = - K_1$, $j_l(\bar{R})$ simplifies to
\begin{equation}
\eqalign{ 
j_l = & -  k_+^2  \left ( \frac{\gamma}{\bar{R} } - a_+ \right ) \left ( 1 - \frac{I_1}{ k_+ R I_0 } \right )  -  k_-^2 \left ( \frac{\gamma}{\bar{R} } - a_- \right )   \left ( 1 - \frac{K_1}{k_- R K_0 } \right )   \\
& - \left ( k_+ \frac{I_{l-1}}{I_l }  - \frac{l}{R} \right ) \left [ \frac{\gamma (l^2 - 1)}{\bar{R}^2 }  -  \left ( \frac{\gamma}{\bar{R} } - a_+ \right )  \frac{k_+ I_1 }{I_0 } \right ] \\
& + \left ( k_- \frac{K_{l-1}}{K_l }  - \frac{l}{R} \right )\left [ \frac{\gamma (l^2 - 1)}{\bar{R}^2 }  + \left ( \frac{\gamma}{\bar{R} } - a_- \right )  \frac{k_- K_1}{K_0 } \right ]
} 
\end{equation}
We refer back to section \ref{droplet_splitting} for plots of $j_l$ and further discussions of the result. 

\end{appendix}

\section*{Bibliography}

\bibliographystyle{unsrt}
\bibliography{JStatPaper}

\begin{thebibliography}{10}

\bibitem{Glotzer1}
S.~C. Glotzer, D.~Stauffer, and N.~Jan.
\newblock {M}onte {C}arlo simulations of phase separation in chemically
  reactive binary mixtures.
\newblock {\em Phys. Rev. Lett.}, 72:4109--4112, 1994.

\bibitem{Glotzer2}
S.~C. Glotzer, E.~A. Di~Marzio, and M.~Muthukumar.
\newblock Reaction-controlled morphology of phase-separating mixtures.
\newblock {\em Phys. Rev. Lett.}, 74:2034--2037, 1995.

\bibitem{GlotzerPRE}
S.~C. Glotzer and A.~Coniglio.
\newblock Self-consistent solution of phase separation with competing
  interactions.
\newblock {\em Phys. Rev. E}, 50:4241--4244, 1994.

\bibitem{Motoyama1996}
M.~Motoyama.
\newblock {Morphology of binary mixtures which undergo phase separation during
  chemical reactions}.
\newblock {\em J. Phys. Soc. Jpn.}, 65:1894--1897, 1996.

\bibitem{Motoyama1997}
M.~Motoyama and T.~Ohta.
\newblock {Morphology of phase-separating binary mixtures with chemical
  reaction}.
\newblock {\em J. Phys. Soc. Jpn.}, 66:2715--2725, 1997.

\bibitem{puri1994segregation}
S.~Puri and H.~L. Frisch.
\newblock Segregation dynamics of binary mixtures with simple chemical
  reactions.
\newblock {\em J. Phys. A}, 27:6027, 1994.

\bibitem{puri1998phase}
S.~Puri and H.~L. Frisch.
\newblock Phase separation in binary mixtures with chemical reactions.
\newblock {\em Int. J. Mod. Phys. B}, 12:1623--1641, 1998.

\bibitem{lefever1995comment}
R.~Lefever, D.~Carati, and N.~Hassani.
\newblock Comment on “{M}onte {C}arlo simulations of phase separation in
  chemically reactive binary mixtures”.
\newblock {\em Phys. Rev. Lett.}, 75:1674, 1995.

\bibitem{berry2018physical}
J.~Berry, C.~P. Brangwynne, and M.~Haataja.
\newblock Physical principles of intracellular organization via active and
  passive phase transitions.
\newblock {\em Rep. Prog. Phys.}, 81:046601, 2018.

\bibitem{Jacobs2017biophysical}
W.~M. Jacobs and D.~Frenkel.
\newblock Phase transitions in biological systems with many components.
\newblock {\em Biophys. J.}, 112:683 -- 691, 2017.

\bibitem{hyman2014liquid}
A.~A. Hyman, C.~A. Weber, and F.~J{\"u}licher.
\newblock Liquid-liquid phase separation in biology.
\newblock {\em Ann. Rev. Cell Dev. Biol.}, 30:39--58, 2014.

\bibitem{Wueseke2016bioopen}
O.~Wueseke, D.~Zwicker, A.~Schwager, Y.~L. Wong, K.~Oegema, F.~J{\"u}licher,
  A.~A. Hyman, and J.~B. Woodruff.
\newblock Polo-like kinase phosphorylation determines {C}aenorhabditis elegans
  centrosome size and density by biasing {SPD}-5 toward an assembly-competent
  conformation.
\newblock {\em Biol. Open}, 5:1431--1440, 2016.

\bibitem{zwicker2014centrosomes}
D.~Zwicker, M.~Decker, S.~Jaensch, A.~A. Hyman, and F.~J{\"u}licher.
\newblock Centrosomes are autocatalytic droplets of pericentriolar material
  organized by centrioles.
\newblock {\em Proc. Natl Acad. Sci. USA}, 111:E2636--E2645, 2014.

\bibitem{catesPNAS2010}
M.~E. Cates, D.~Marenduzzo, I.~Pagonabarraga, and J.~Tailleur.
\newblock Arrested phase separation in reproducing bacteria creates a generic
  route to pattern formation.
\newblock {\em Proc. Natl Acad. Sci. USA}, 107:11715--11720, 2010.

\bibitem{grafkePRL2017}
T~Grafke, M.~E. Cates, and E~Vanden-Eijnden.
\newblock Spatiotemporal self-organization of fluctuating bacterial colonies.
\newblock {\em Phys. Rev. Lett.}, 119:188003, 2017.

\bibitem{MIPS}
M.~E. Cates and J~Tailleur.
\newblock Motility-induced phase separation.
\newblock {\em Annu. Rev. Condens. Matter Phys.}, 6(1):219--244, 2015.

\bibitem{weberRepProgPhys2019}
C.~A. Weber, D.~Zwicker, F.~J{\"u}licher, and C.~F. Lee.
\newblock Physics of active emulsions.
\newblock {\em Rep. Prog. Phys.}, 82:064601, 2019.

\bibitem{HohenbergHalperin}
P.~C. Hohenberg and B.~I. Halperin.
\newblock Theory of dynamic critical phenomena.
\newblock {\em Rev. Mod. Phys.}, 49:435--479, 1977.

\bibitem{wittkowski2014scalar}
R.~Wittkowski, A.~Tiribocchi, J.~Stenhammar, R.~J. Allen, D.~Marenduzzo, and
  M.~E. Cates.
\newblock Scalar $\phi^4$ field theory for active-particle phase separation.
\newblock {\em Nat. Commun.}, 5:4351, 2014.

\bibitem{cates2018JFM}
M.~E. Cates and E.~Tjhung.
\newblock Theories of binary fluid mixtures: from phase-separation kinetics to
  active emulsions.
\newblock {\em J. Fluid Mech.}, 836:1, 2018.

\bibitem{Liu1989}
F.~Liu and N.~Goldenfeld.
\newblock {Dynamics of phase separation in block copolymer melts}.
\newblock {\em Phys. Rev. A}, 39:4805--4810, 1989.

\bibitem{Muratov2002}
C.~B. Muratov.
\newblock {Theory of domain patterns in systems with long-range interactions of
  Coulomb type}.
\newblock {\em Phys. Rev. E}, 66:066108, 2002.

\bibitem{Sagui1995}
C.~Sagui and R.~C. Desai.
\newblock {Ostwald ripening in systems with competing interactions}.
\newblock {\em Phys. Rev. Lett.}, 74:1119--1122, 1995.

\bibitem{nardini2017entropy}
C.~Nardini, {\'E}.~Fodor, E.~Tjhung, F.~Van~Wijland, J.~Tailleur, and M.~E.
  Cates.
\newblock Entropy production in field theories without time-reversal symmetry:
  {Q}uantifying the non-equilibrium character of active matter.
\newblock {\em Phys. Rev. X}, 7:021007, 2017.

\bibitem{singh2019PRL}
R.~Singh and M.~E. Cates.
\newblock Hydrodynamically interrupted droplet growth in scalar active matter.
\newblock {\em Phys. Rev. Lett.}, 123:148005, 2019.

\bibitem{TjhungPRX2018}
E.~Tjhung, C.~Nardini, and M.~E. Cates.
\newblock Cluster phases and bubbly phase separation in active fluids: reversal
  of the {O}stwald process.
\newblock {\em Phys. Rev. X}, 8:031080, 2018.

\bibitem{TarziaPRL2006}
M.~Tarzia and A.~Coniglio.
\newblock Pattern formation and glassy phase in the ${\ensuremath{\phi}}^{4}$
  theory with a screened electrostatic repulsion.
\newblock {\em Phys. Rev. Lett.}, 96:075702, 2006.

\bibitem{TarziaPRE2007}
M.~Tarzia and A.~Coniglio.
\newblock Lamellar order, microphase structures, and glassy phase in a field
  theoretic model for charged colloids.
\newblock {\em Phys. Rev. E}, 75:011410, 2007.

\bibitem{brazovskii}
S.~A. {Brazovski{\v{i}}}.
\newblock {Phase transition of an isotropic system to a nonuniform state}.
\newblock {\em J. Exp. Theor. Phys.}, 41:85, 1975.

\bibitem{christensen1996PRE}
J.~J. Christensen, K.~Elder, and H.~C. Fogedby.
\newblock Phase segregation dynamics of a chemically reactive binary mixture.
\newblock {\em Phys. Rev. E}, 54:R2212, 1996.

\bibitem{hamley2005block}
I.~W. Hamley.
\newblock {\em Block Copolymers in Solution: Fundamentals and Applications}.
\newblock John Wiley \& Sons, 2005.

\bibitem{hinrichsen}
H.~Hinrichsen.
\newblock Non-equilibrium critical phenomena and phase transitions into
  absorbing states.
\newblock {\em Adv. Phys.}, 49:815--958, 2000.

\bibitem{tauber2014}
U.~C. T{\"a}uber.
\newblock {\em Critical Dynamics: a Field Theory Approach to Equilibrium and
  Non-equilibrium Scaling Behavior}.
\newblock Cambridge University Press, 2014.

\bibitem{lefevre}
A.~Lefevre and G.~Biroli.
\newblock Dynamics of interacting particle systems: stochastic process and
  field theory.
\newblock {\em J. Stat. Mech.: Theory Exp.}, 2007:P07024, 2007.

\bibitem{gardiner}
C.~Gardiner.
\newblock {\em Stochastic Methods}.
\newblock Springer Berlin, 2009.

\bibitem{vankampen}
N.~G. Van~Kampen.
\newblock {\em Stochastic Processes in Physics and Chemistry}.
\newblock Elsevier, 1992.

\bibitem{bray_review}
A.~J. Bray.
\newblock Theory of phase-ordering kinetics.
\newblock {\em Adv. Phys.}, 51(2):481--587, 2002.

\bibitem{boyd}
J.~P. Boyd.
\newblock {\em Chebyshev and Fourier spectral methods}.
\newblock Berlin: Springer, 2001.

\bibitem{orszag}
S.~A. Orszag.
\newblock On the elimination of aliasing in finite-difference schemes by
  filtering high-wavenumber components.
\newblock {\em J. Atmos. Sci.}, 28:1074--1074, 1971.

\bibitem{kloeden}
P.~E. Kloeden and E.~Platen.
\newblock {\em Numerical Solution of Stochastic Differential Equations}.
\newblock Springer Science \& Business Media, 2013.

\bibitem{zwickerNatPhys2017}
D~Zwicker, R~Seyboldt, C.~A Weber, A.~A Hyman, and F~J{\"u}licher.
\newblock Growth and division of active droplets provides a model for
  protocells.
\newblock {\em Nat. Phys.}, 13:408, 2017.

\bibitem{crossRevModPhys}
M.~C. Cross and P.~C. Hohenberg.
\newblock Pattern formation outside of equilibrium.
\newblock {\em Rev. Mod. Phys.}, 65:851--1112, 1993.

\bibitem{cross2009}
M.~Cross and H.~Greenside.
\newblock {\em Pattern Formation and Dynamics in Nonequilibrium Systems}.
\newblock Cambridge University Press, 2009.

\bibitem{DeGennes}
P-G. De~Gennes and J.~Prost.
\newblock {\em The Physics of Liquid Crystals}.
\newblock Oxford University Press, 1993.

\bibitem{ostwald}
W.~Ostwald.
\newblock Studien {\"u}ber die bildung und umwandlung fester k{\"o}rper.
\newblock {\em Z. Phys. Chem.}, 22:289--330, 1897.

\bibitem{lifshitz1961kinetics}
I.~M. Lifshitz and V.~V. Slyozov.
\newblock The kinetics of precipitation from supersaturated solid solutions.
\newblock {\em J. Phys. Chem. Solids}, 19:35--50, 1961.

\bibitem{ZwickerPRE}
D.~Zwicker, A.~A. Hyman, and F.~J\"ulicher.
\newblock Suppression of ostwald ripening in active emulsions.
\newblock {\em Phys. Rev. E}, 92:012317, 2015.

\bibitem{grafkeJSM2017}
T~Grafke and E~Vanden-Eijnden.
\newblock Non-equilibrium transitions in multiscale systems with a bifurcating
  slow manifold.
\newblock {\em J. Stat. Mech.: Theory Exp.}, 2017:093208, 2017.

\bibitem{kawasaki}
Kyozi K. and Takao O.
\newblock Kink dynamics in one-dimensional nonlinear systems.
\newblock {\em Physica A}, 116:573 -- 593, 1982.

\bibitem{PRLTailleurCates}
J.~Tailleur and M.~E. Cates.
\newblock Statistical mechanics of interacting run-and-tumble bacteria.
\newblock {\em Phys. Rev. Lett.}, 100:218103, 2008.

\end{thebibliography}

\end{document}